\documentclass[aps,prl,superscriptaddress, nofootinbib,reprint,nolongbibliography]{revtex4-2}

\usepackage{graphicx}
\usepackage{makecell}
\usepackage{lineno}
\usepackage{amsmath,amssymb}
\usepackage[table]{xcolor}
\usepackage{ulem}
\usepackage{physics}

\newcommand{\Eeff}{\mathcal{E}_\mathrm{eff}}

\begin{document}

\title{Quantum-Enhanced Metrology for Molecular Symmetry Violation\\ using Decoherence-Free Subspaces}

\author{Chi Zhang}
\email[]{chizhang@caltech.edu}
\author{Phelan Yu}
\author{Arian Jadbabaie}
\author{Nicholas R. Hutzler}
\affiliation{California Institute of Technology, Division of Physics, Mathematics, and Astronomy.  Pasadena, CA 91125}

\date{\today}

\begin{abstract}

We propose a method to measure time-reversal symmetry violation in molecules that overcomes the standard quantum limit while leveraging decoherence-free subspaces to mitigate sensitivity to classical noise.  The protocol does not require an external electric field, and the entangled states have no first-order sensitivity to static electromagnetic fields as they involve superpositions with zero average lab-frame projection of spins and dipoles.  This protocol can be applied with trapped neutral or ionic species, and can be implemented using methods which have been demonstrated experimentally.

\end{abstract}

\maketitle

Precision measurements of time-reversal (T) symmetry violation in molecular systems provide stringent tests of new physics beyond the Standard Model \cite{Safronova2018}. For example, searches for the electron's electric dipole moment (eEDM) have excluded a broad parameter space of T violating leptonic physics at energy scales up to $\sim 50~\mathrm{TeV}$ \cite{Roussy2023,Andreev2018,Cesarotti2019,Alarcon2022}. 
Experiments aiming to laser cool and trap eEDM-sensitive neutral molecules \cite{Alauze2021,Fitch2020b,Augenbraun2020,Kozyryev2017b,isaev2010,Lasner2022SrOHBR} are currently under construction and promise significantly improved measurement precision. The immediate impact of cooling and trapping is the substantially longer coherence time compared to beam experiments, a result of both long trapping time and easier field control for quasi-stationary molecules confined in a small volume. Furthermore, quantum metrology techniques \cite{Pezze2018,Pezze2009}, such as entanglement and squeezing, promise routes to additional enhancement of eEDM sensitivity. However, a specific scheme providing metrological gain without added susceptibility to classical noise from electromagnetic fields has, to our knowledge, not yet been conceived.

Additionally, contemporary eEDM searches with molecular ions are conducted in non-stationary rotating traps \cite{Cairncross2017,Roussy2023}, since an external electric field is used to polarize the molecules. Although various improvements will be implemented for near-future experiments \cite{Ng2022,Taylor2022}, molecule motion in the rotating trap during spin precession remains a challenge for implementing entanglement-enhanced metrology.

In this manuscript, we show that the eEDM can be observed as a coupling between two entangled molecules within a decoherence-free subspace. The eEDM sensitivity scales linearly with the entangled molecule number, thereby offering Heisenberg-limited sensitivity beyond the standard quantum limit, while the susceptibility to electromagnetic fields remains mitigated. In addition, the two molecules do not have to be aligned in the lab frame by an external electric field; instead, they are prepared in orthogonal superpositions of opposite parity states. As a result, the scheme is applicable to neutral molecules in optical lattices or tweezer arrays \cite{Holland2022,Bao2022} as well as molecular ions in quasi-stationary traps \cite{Lin2020,Fan2021}, which enable entanglement generation and are a well-established platform for precision measurement \cite{Brewer2019,Sanner2019}. Importantly, the entangled molecular states involved are experimentally achievable using existing entanglement protocols \cite{Hughes2020,Hudson2018,Ni2018,Yelin2006,Zhang2022b,Wang2022,Omran2019}, some of which have been demonstrated recently \cite{Holland2022,Bao2022,Lin2020}, together with single molecule operations \cite{Park2017,Gregory2021}. Our discussion here focuses on the eEDM as an example, but the method can be straightforwardly extended to measure other T violating moments, including the nuclear Schiff moment \cite{Graner2016} and nuclear magnetic quadrupole moment \cite{Flambaum2014}.

The energy shift of the eEDM ($d_e$) in an effective internal molecular electric field ($\Eeff$) is $d_e \cdot \Eeff$. The internal field points along the molecule axis ($\hat{n}$) and its amplitude is determined by the electronic structure of the molecule, while the eEDM is collinear with the total electron spin ($S$). Conventional eEDM experiments \cite{Safronova2018,Hudson2011,Baron2014,Roussy2023,Andreev2018} orient the molecule axis in the lab frame by mixing opposite parity states with an external electric field, and subsequently polarize the electron spin in the lab frame as well. The eEDM interaction then manifests as a small spin-dependent energy shift, measured by performing spin precession in the polarized molecules. However, the polarized molecular dipoles and electron spins also make these experiments sensitive to uncontrolled external fields. As a consequence, the most common quantum metrology methods, such as spin squeezing \cite{Wineland1992,Wineland1994,Perlin2020}, increase sensitivity to external electromagnetic fields by the same amount as the gain in eEDM sensitivity.  The resulting increased susceptibility to decoherence and systematic errors from these fields, which are a main concern for eEDM experiments, can counteract the eEDM sensitivity boost.

\begin{figure*}
	\includegraphics[width=0.8\textwidth]{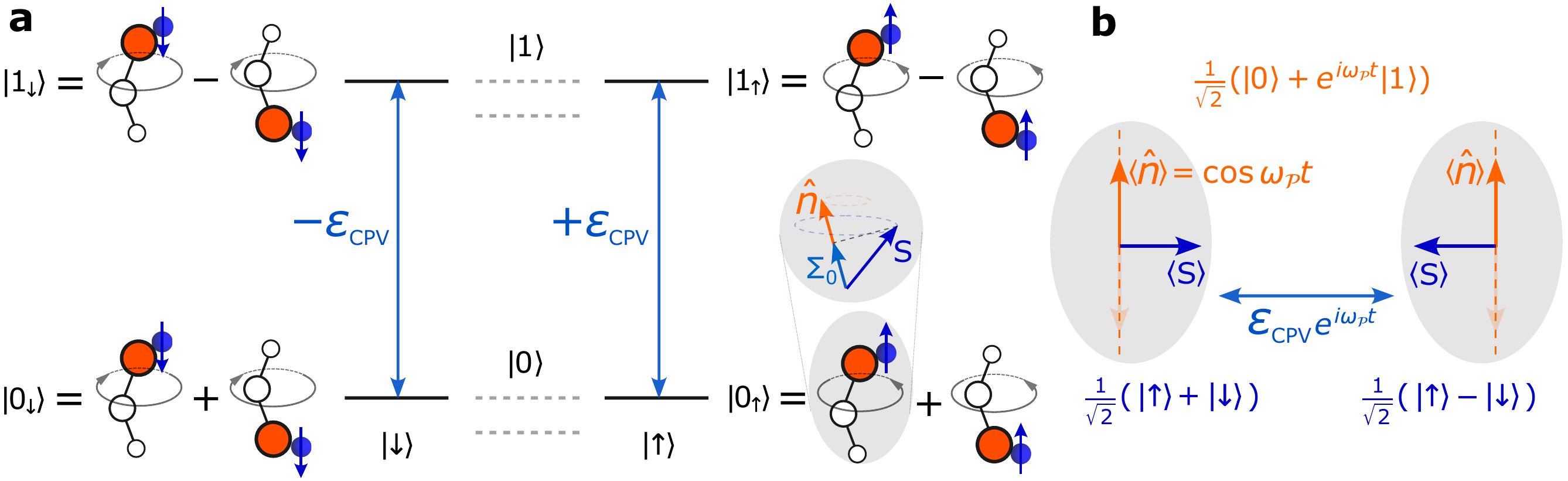}
	\caption{(a) A typical level diagram of a parity-doubled molecule, for example a triatomic bending mode~\cite{Kozyryev2017b}. Molecule eigenstates $\ket{0}$ and $\ket{1}$ are superpositions of molecular dipole orientation. They have magnetic sublevels, the stretched states (thick levels) $\ket{\uparrow}$ and $\ket{\downarrow}$ represent electron spin up and down in the lab frame. The dashed levels in the middle indicate magnetic sublevels resulting from electron spin coupling with other angular momenta, which are not needed in our scheme. The insets show the spin $S$ (dark blue arrow) projection  $\Sigma$ on the molecule axis $\hat{n}$ (light blue arrow) in the molecule frame. The eEDM gives a coupling $\pm \varepsilon_\mathrm{CPV}$ between $\ket{0} \leftrightarrow \ket{1}$, and the sign of the coupling depends on the spin. (b) The effective electric field along the molecule axis couples spin states. In the lab frame, due to the energy difference $\omega_\mathcal{P}$ between $\ket{0}$ and $\ket{1}$, the orientation of the molecule axis is oscillating, thus the coupling, the spin-precession direction, is also oscillating.}
	\label{Fig1}
\end{figure*}

Here we instead probe the eEDM as a \textit{coupling} between two opposite-parity states in a molecule. We first consider the effects of this coupling in a single molecule to build understanding of the system, and then discuss how we can engineer entangled states in a two (or more) molecule system which have Heisenberg-limited sensitivity $(\propto N)$ to the eEDM but without concurrent increases in collective electric or magnetic field sensitivity.  Again, we consider the eEDM as it provides the simplest possible system, but the methods are applicable to symmetry violating nuclear moments as well.

In Fig.~\ref{Fig1}, we provide an example of a single molecule in the parity-doubled bending mode of a $^2\Sigma$ triatomic molecule~\cite{Kozyryev2017b}, though the method should be generalizable to other types of parity-doubled states. The opposite-parity states are labeled as $\ket{0}$ and $\ket{1}$, and the spin states in the lab basis are labeled by $\ket{\uparrow}$ and $\ket{\downarrow}$. The eEDM causes a spin-dependent coupling between $\ket{0} \leftrightarrow \ket{1}$ with a coupling strength $\varepsilon_\mathrm{CPV} = \bra{0_\uparrow} \Eeff d_e\Sigma \ket{1_\uparrow} = 2\Eeff d_e \Sigma_0$, where $\Sigma = S\cdot \hat{n}$ is the projection of spin on the molecule axis and $\Sigma_0$ is the expectation value of $\Sigma$ when averaged over other angular momentum quantum numbers of the molecule wavefunction \cite{Petrov2022}. The coupling changes sign to $-\varepsilon_\mathrm{CPV}$ for the time-reversed state $\ket{\downarrow}$.

In a superposition state such as $\frac{1}{2}(\ket{0}+\ket{1})(\ket{\uparrow} + e^{i\theta} \ket{\downarrow})$, which corresponds to an orientation of $\Eeff$ perpendicular to the electron spin, the eEDM interaction causes spin precession that changes the phase $\theta$ of the spin superposition. Note that this is conceptually similar to the usual idea of creating a superposition of $\ket{0},\ket{1}$ by polarizing the molecule with a static external electric field. However, here we consider creating a superposition of these states without static applied fields, meaning that the orientation of the molecular dipole, and therefore $\Eeff$, will be oscillating in the lab frame at a frequency given by the parity splitting $\omega_\mathcal{P}$ (typically $\sim 2\pi\times100~\mathrm{kHz}$ to $\sim 2\pi\times100~\mathrm{MHz}$) between $\ket{0}$ and $\ket{1}$ \cite{Kozyryev2017b}. Thus, the eEDM spin precession ($\lesssim 100~\mathrm{\mu Hz}$) can only accumulate phase in the frame rotating at $\omega_\mathcal{P}$; in the lab frame, the direction of spin precession oscillates rapidly and averages to zero, so there is no eEDM-induced energy shift or spin precession.

\begin{figure*}
	\includegraphics[width=\textwidth]{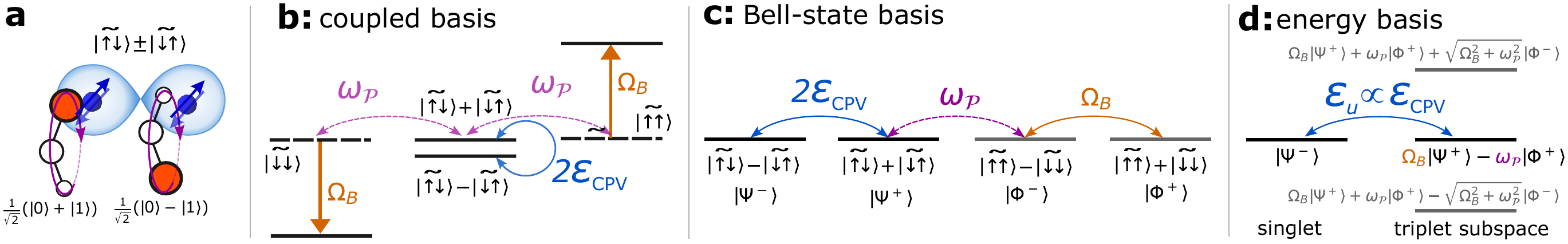}
	\caption{(a) eEDM interaction for two molecules in opposite molecular dipole superposition states $\widetilde{\ket{\Uparrow}}$ and $\widetilde{\ket{\Downarrow}}$ and entangled spin state $\widetilde{\ket{\uparrow \downarrow}}\pm\widetilde{\ket{\downarrow \uparrow}}$. (b) In the rotating frame, an eEDM couples the degenerate singlet and triplet pair states with zero spin projections. The triplet states are coupled by rotation of the frame. A magnetic field shifts $\widetilde{\ket{\uparrow \uparrow}}$ and $\widetilde{\ket{\downarrow\downarrow}}$ oppositely and suppresses the coupling of the rotation. This is equivalent to (c) in the Bell-state basis, where $\ket{\Psi^\pm} = \frac{1}{\sqrt{2}} (\widetilde{\ket{\uparrow\downarrow}} \pm \widetilde{\ket{\downarrow\uparrow}})$ and $\ket{\Phi^\pm} = \frac{1}{\sqrt{2}} (\widetilde{\ket{\uparrow\uparrow}} \pm \widetilde{\ket{\downarrow\downarrow}})$. The eEDM interaction couples $\ket{\Psi^-} \leftrightarrow \ket{\Psi^+}$, which is separated from other couplings by an external magnetic field. The rotation couples $\ket{\Psi^+} \leftrightarrow \ket{\Phi^-}$, and an rf magnetic field in phase with the molecule rotation couples $\ket{\Phi^-} \leftrightarrow \ket{\Phi^+}$. As a result, in (d), the eEDM interaction effectively couples $\ket{\Psi^-}$ to the unshifted state of the three-level system ($\ket{\Psi^+} \leftrightarrow \ket{\Phi^-} \leftrightarrow \ket{\Phi^+}$) with a reduced coupling strength of $4 \varepsilon_\mathrm{CPV} \frac{\Omega_B}{\sqrt{\Omega_B^2 + \omega_\mathcal{P}^2}}$, which reaches 90\% of the maximum $4\varepsilon_\mathrm{CPV}$ for $\Omega_B \gtrsim 2\omega_\mathcal{P}$ and 98\% of the maximum for $\Omega_B \gtrsim 5\omega_\mathcal{P}$.}
	\label{Fig2}
\end{figure*}

\begin{figure*}
	\includegraphics[width=0.9\textwidth]{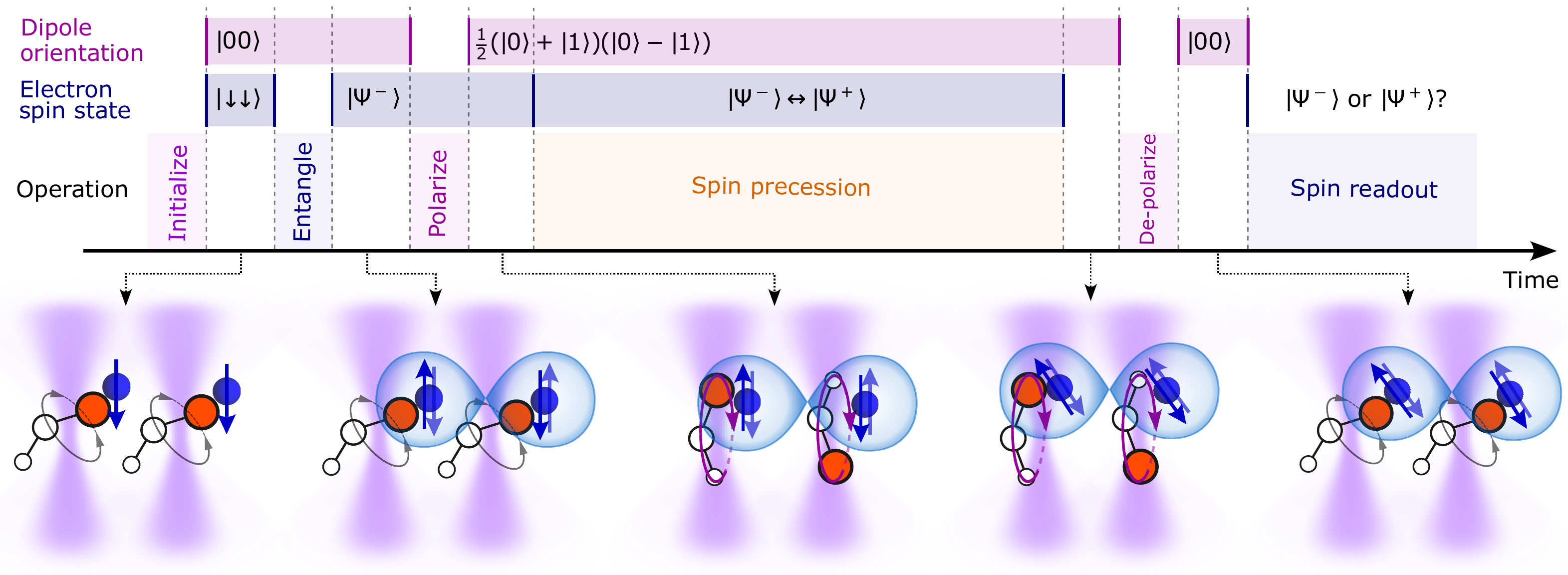}
	\caption{Experimental sequence for the eEDM measurement. The steps are indicated by the boxes on the time line. The purple boxes represent operations on molecule orientation, and the blue boxes mostly act on the spin degree of freedom. The spin precession enabled by an rf magnetic field is represented by the orange box. The molecule orientation and spin state are specified above the sequence boxes, and shown schematically in the illustrations below it. See more details in text.}
	\label{Fig3}
\end{figure*} 

However, with \textit{two} (or more) molecules, we can engineer states where eEDM precession does not average to zero, yet the oscillation in the lab frame makes the molecules highly insensitive to external fields. Furthermore, we shall see that these states have a metrological gain in sensitivity due to entanglement. We denote the superpositions $\frac{1}{\sqrt{2}}(\ket{0}+e^{i\omega_\mathcal{P}t}\ket{1})=\widetilde{\ket{\Uparrow}}$ and $\frac{1}{\sqrt{2}}(\ket{0}-e^{i\omega_\mathcal{P}t}\ket{1})=\widetilde{\ket{\Downarrow}}$, suggestive of the fact that these states have opposite orientation of the (rotating) molecular dipole.
Consider two molecules in the state $\widetilde{\ket{\Uparrow\Downarrow}}$, as shown in Fig.~\ref{Fig2}, where we label rotating frame spin states using $\widetilde{\ket{\uparrow}}$ and $\widetilde{\ket{\downarrow}}$. The rotation of the frame is described as $H_\mathrm{rot} = \hbar\omega_\mathcal{P} \widetilde{\sigma}_x$ in the rotating frame basis \cite{Loh2013} (also see Supplemental Material). An eEDM shifts $\widetilde{\ket{\uparrow\downarrow}}$ and $\widetilde{\ket{\downarrow\uparrow}}$ oppositely, as they have opposite relative orientations of electron spins and molecular dipoles. Therefore, an eEDM couples the degenerate singlet and triplet pair states with zero total spin projections. These states constitute a decoherence-free subspace as the molecular electric and magnetic dipole moments have zero average projection on the laboratory fields and are therefore insensitive to them to first order. This is conceptually similar to the eEDM coupling in a hyperfine clock transition~\cite{Verma2020}. 

Similar to the single molecule case, the eEDM has little effect on the eigenstates of $H_\mathrm{rot}$. However, now we can switch on and off the eEDM spin precession by applying a radio-frequency (rf) magnetic field $B$ in phase with the rotating frame (this is challenging for a single molecule; see Supplemental Material). The rf magnetic field is described by $H_B = \Omega_B \widetilde{\sigma}_z$, with $\Omega_B$ the interaction strength ($\Omega_B \approx \mu_B B$ for $^2\Sigma_{1/2}$ electronic states), and it shifts $\widetilde{\ket{\uparrow \uparrow}}$ and $\widetilde{\ket{\downarrow\downarrow}}$ oppositely, as they have different orientations relative to the rf field. The couplings of $H_\mathrm{rot}$, $H_B$, and eEDM are shown in Fig.~\ref{Fig2}(c) in the Bell state basis ($\ket{\Psi^\pm} = \frac{1}{\sqrt{2}} (\widetilde{\ket{\uparrow\downarrow}} \pm \widetilde{\ket{\downarrow\uparrow}})$, $\ket{\Phi^\pm} = \frac{1}{\sqrt{2}} (\widetilde{\ket{\uparrow\uparrow}} \pm \widetilde{\ket{\downarrow\downarrow}})$).

$H_\mathrm{rot}$ and $H_B$ couple $\ket{\Psi^+} \leftrightarrow \ket{\Phi^-}$ and $\ket{\Phi^-} \leftrightarrow \ket{\Phi^+}$, respectively. The resulting eigenstates are shown in Fig.~\ref{Fig2}(d); the middle state, whose eigenenergy is not shifted, is $\ket{u} = \sin{\theta} \ket{\Psi^+} - \cos{\theta}\ket{\Phi^+}$, with the mixing angle $\theta$ given by $\tan{\theta} = \Omega_B/\omega_\mathcal{P}$. Note that these interactions do not couple to $\ket{\Psi^-}$. However, the eEDM interaction couples $\ket{\Psi^-} \leftrightarrow \ket{\Psi^+}$ but with coupling strength much smaller than $H_B$ or $H_\mathrm{rot}$. The eEDM therefore induces a resonant coupling $\ket{\Psi^-}\leftrightarrow\ket{u}$ with a reduced coupling strength of $\varepsilon_u = 4\varepsilon_\mathrm{CPV} \frac{\Omega_B}{\sqrt{\Omega_B^2 + \omega_\mathcal{P}^2}}$, which reaches $\sim 90\%$ of the maximum ($4\varepsilon_\mathrm{CPV}$) when $\Omega_B \gtrsim 2\omega_\mathcal{P}$. 
Note that this is twice the coupling of a fully polarized single molecule, thereby beating the standard quantum limit. A static magnetic field, or more generally, a magnetic field at a different frequency, causes the phase on the $\ket{\Psi^+}$ part of $\ket{u}$ to oscillate and thus the eEDM coupling averages to zero. Consequently, the eEDM spin precession is turned on only when the magnetic field is in-phase. The eEDM spin-precession subspace is also known as a decoherence-free subspace \cite{Monz2009}; it is robust to noise since the total spin and dipole projections, and therefore the expectation of electric and magnetic dipole moments, is zero. 

The experimental sequence for two molecules, as an example, is illustrated in Fig.~\ref{Fig3}. Molecules are initialized in $\ket{0_\downarrow 0_\downarrow}$ by optical pumping. Then the spins are entangled in $\ket{\Psi^-}_\mathrm{lab}$ -- this can be realized by direct dipole-dipole \cite{Holland2022,Bao2022} or Rydberg atom mediated interactions \cite{Zhang2022b,Wang2022}, or, for trapped ions, the spin-dependent force gate \cite{Lin2020} or the M\o{}lmer-S\o{}renson interaction \cite{Sorensen1999}. For more than two molecules, the entangled singlet state can be generated by a set of gate operations, adiabatic sweeping to the ground state of the many-body system \cite{Omran2019}, or extracting from cluster states \cite{Briegel2001,Verresen2022,Lee2022,Tscherbul2023}. Subsequently, the molecule orientation is prepared in $\widetilde{\ket{\Uparrow\Downarrow}}$. This can be done in two sub-steps: first drive a global $\pi/2$-pulse between $\ket{0}\leftrightarrow\ket{1}$, and then apply an AC-Stark shift using a far-detuned laser focused on one of the molecules and imprint a $\pi$ phase on $\ket{1}$. By addressing different molecules, or by changing the detuning of the laser, the direction of eEDM spin precession can be controlled, thus providing ``switches'' to observe the eEDM \cite{Baron2017}.  Note that for multiple pairs of molecules trapped simultaneously, this could be performed in parallel across different pairs to mitigate imperfections in the laser pulses. The initial spin state $\ket{\Psi^-}_\mathrm{lab}$ is invariant under rotations and thus is equal to $\ket{\Psi^-}$ in the rotating frame. Next, when the rf magnetic field is turned on, eEDM spin precession $\ket{\Psi^-} \leftrightarrow \ket{u}$ starts in the rotating frame. After eEDM spin precession, the magnetic field is turned off and then the orientation of the molecules is rotated back to $\ket{00}$. In the lab frame, $\ket{u}$ is an rf-dressed state, which is oscillating in the triplet subspace of $\{\ket{\Psi^+}_\mathrm{lab}, \ket{\Phi^-}_\mathrm{lab}, \ket{\Phi^+}_\mathrm{lab}\}$. After turning off the magnetic field, the population in $\ket{u}$ is distributed in the triplet subspace but mostly mapped to $\ket{\Psi^+}_\mathrm{lab}$. Finally, the eEDM spin precession phase, i.e. the phase between $\ket{\uparrow\downarrow}$ and $\ket{\downarrow\uparrow}$ components, is measured by a projection measurement in the $\frac{1}{\sqrt{2}} (\ket{\uparrow \downarrow} \pm i \ket{\downarrow \uparrow})$ basis by the parity oscillation measurement \cite{leibfried2005,Holland2022,Bao2022} as described further in the Supplemental Material.

Our scheme has many advantages. First, the spin precession rate in the entangled basis is two times faster than in a fully polarized single molecule, and it scales linearly with molecule number for the anti-ferromagnetic spin states (i.e. between $\frac{1}{\sqrt{2}} (\ket{\uparrow \downarrow \uparrow ... \uparrow \downarrow} + \ket{\downarrow \uparrow \downarrow ... \downarrow \uparrow}) \leftrightarrow \frac{1}{\sqrt{2}} (\ket{\uparrow \downarrow \uparrow ... \uparrow \downarrow} - \ket{\downarrow \uparrow \downarrow ... \downarrow \uparrow})$ for molecule orientation $\ket{\Uparrow\Downarrow\Uparrow ... \Uparrow\Downarrow}$), thus realizing a metrological gain from entanglement. More importantly, the eEDM spin precession subspace is decoupled from various environmental noise sources, including magnetic fields, vector and tensor light shifts, etc., since the total spin and dipole projections are zero and the spin precession takes place in a rotating frame where slow noise is averaged out. This is unlike conventional eEDM protocols using polarized molecules in the lab frame, where the eEDM-enhanced entangled states, such as squeezed states or the GHZ state $\frac{1}{\sqrt{2}} (\ket{\uparrow \uparrow ... \uparrow} + \ket{\downarrow \downarrow ... \downarrow})$, normally require spins aligned collectively in the lab frame and thus are also increasingly sensitive to magnetic field noise, AC Stark shifts, etc. Magnetic field gradients at the same frequency may cause spin precession in the same subspace; however, this effect can be disentangled from an eEDM by switching the sign of eEDM interaction, which is controlled by the phase of the rf magnetic field and the phase of the molecule orientation. For example, the spin precession directions in $\widetilde{\ket{\Uparrow \Downarrow}}$ and $\widetilde{\ket{\Downarrow \Uparrow}}$ are opposite, and the spin does not precess in $\widetilde{\ket{\Uparrow \Uparrow}}$ or $\widetilde{\ket{\Downarrow \Downarrow}}$. Other couplings, including $H_\mathrm{rot}$ and $H_B$, are insensitive to the $\pm$ phase between $\ket{0}$ and $\ket{1}$.

Furthermore, our scheme is robust to various experimental imperfections. For example, the fidelity of entanglement generation does not have a lower threshold; the population that is not initialized in $\ket{\Psi^-}$ is not coupled by the eEDM and only contributes a constant background. Many possible sources may cause imperfect initialization of the molecule orientation; they include, for instance, fluctuations in the $\pi/2$-pulse power, Stark shifts, imperfect single molecule addressing light shift, or small difference in the $g$-factors of $\ket{0}$ and $\ket{1}$ states (resulting from perturbations of other electronic states), etc. If a molecule is not in equal superposition of $\ket{0}$ and $\ket{1}$ the eEDM interaction ($\Sigma_0$) is slightly reduced. If two molecules are not in exact opposite phases of $\ket{0}$ and $\ket{1}$ superpositions, the splitting between $\widetilde{\ket{\uparrow\downarrow}}$ and $\widetilde{\ket{\downarrow\uparrow}}$ is reduced (this can be used as a switch to tune the spin precession rate). If two molecules have different $\ket{0}$ and $\ket{1}$ populations, their eEDM interactions ($\Sigma_0$) are different and thus $\ket{\Psi^-}$ is also coupled to the $\ket{\Phi^\pm}$ states. However, this additional coupling does not cause spin precession since the $\ket{\Phi^\pm}$ states are strongly coupled by the magnetic field (see Fig.~\ref{Fig2}[c]). Importantly, all the fields are applied independently and they do not have correlation with the eEDM switch (AC Stark shift from the addressing beam). As a consequence, these imperfections do not lead to systematic effects directly, but instead to contrast reduction and increased statistical noise.

Magnetic field correlated rf electric fields, stray electric fields, and black-body radiation (BBR) have detrimental effects on the state of molecule orientation and need to be shielded. Our scheme does not require a DC electric field, and shielding electric fields is straightforward, especially without the need for electric field plates nearby. The effects of the residual fields include near-resonant couplings between $\ket{0} \leftrightarrow \ket{1}$ and off-resonant effects, such as energy shifts on $\ket{0}$ and $\ket{1}$. The coupling effect is suppressed by the dipole-dipole interaction between two molecules when the residual-field coupling strength is weaker than the dipole-dipole interaction (typically $\sim\mathrm{kHz}$ at $\sim \mathrm{\mu m}$ separation), and it can also be mitigated by applying a stronger electric field in phase with the molecule oscillation.

Stray electric fields or off-resonance BBR can cause an energy shift between $\ket{0}$ and $\ket{1}$. This alters the oscillating frequency of the rotating molecules, which may affect coherent control of the molecule orientation and may interfere with the eEDM spin precession by shifting the oscillation out of phase with the magnetic rf field. Nevertheless, stray electric fields can be actively measured and cancelled, especially since the molecules needed for this protocol will be trapped in a small volume $\sim$mm$^3$; for example, in trapped ions a residual electric field lower than $0.1~\mathrm{mV/cm}$ has been achieved \cite{Higgins2021,Nadlinger2021}. A $0.1~\mathrm{mV/cm}$ fluctuation corresponds to a maximum $\sim 50~\mathrm{mHz}$ dephasing rate for a molecule of $d_0 \approx 2~\mathrm{D}$ dipole moment and $\omega_\mathcal{P} \approx 100~\mathrm{kHz}$ parity splitting. This leads to a coherence time of $\sim 10~\mathrm{s}$, and the coherence time is inversely proportional to parity splitting. On the other hand, we need $\Omega_B \approx \mu_B g B \gtrsim 2\omega_\mathcal{P}$, where $g$ is the electron magnetic $g$-factor. To avoid using high magnetic field (a few Gauss, using a similar magnetic field coil setup in ref.~\cite{Anderegg2017}), our scheme is most suitable for molecules with $\omega_\mathcal{P} \lesssim 10~\mathrm{MHz}$, which is a typical range for parity doubling. In addition, for trapped ions, $\omega_\mathcal{P}$ needs to be much lower than the trap rf frequency ($\sim 20~\mathrm{MHz}$). Some examples of suitable neutral and ion species are listed in the Supplemental Material.

In summary, we have presented a quantum metrology scheme to probe T-violating effects in molecular systems. The Heisenberg scaling is particularly important for the future experiments where the molecules are well-controlled but do not necessarily have large molecule numbers, such as molecules in tweezer arrays and ion traps, as well as rare radioactive molecules \cite{Arrowsmithkron2023}. The T-violating interaction causes spin precession in an entangled, decoherence-free subspace in a rotating frame, where the slow noise in the lab frame is averaged out, and the molecules do not need to be polarized by an external electric field. As a result, the scheme is compatible with stationary ion traps, such as the linear Paul trap, in which a powerful toolbox of precision spectroscopy and quantum metrology has been developed, including sympathetic cooling \cite{Barrett2003}, quantum logic spectroscopy \cite{Tan2015,Taylor2022}, ion shuttling \cite{Kielpinski2002}, micromotion compensation \cite{Keller2015}, entanglement generation, etc. Furthermore, the direction of spin precession is controlled by the phase of the applied magnetic rf field and the phase of the oscillation of the molecule orientation. In T-violation measurements, systematic effects normally arise from imperfections correlated with the switch of the sign of the T-violating interaction, such as parity state or external electric field. Our eEDM switch is an AC-Stark shift by the far-detuned addressing beam on one of the molecules, which has little correlation with other imperfections, and can be performed in parallel across multiple pairs of molecules. In addition, because of the magnetic field insensitivity, this scheme will also improve the coherence in a shot-noise limited measurement using magnetic molecules, including all laser coolable neutral molecules and certain T-sensitive molecular ions whose ground states are magnetic. These advantages will significantly improve the precision of T-violating new physics searches in the near future.

\begin{acknowledgements}

We acknowledge helpful discussions with Andreas Elben, Manuel Endres, Ran Finkelstein, Andrew Jayich, Dietrich Leibfried, Christopher Pattison, John Preskill, Tim Steimle, Yuiki Takahashi, Michael Tarbutt, Fabian Wolf, Xing Wu and the PolyEDM Collaboration. This work was supported by Gordon and Betty Moore Foundation Award GBMF7947, Alfred P. Sloan Foundation Award G-2019-12502, and NSF CAREER Award PHY-1847550. C.Z. acknowledges support from the David and Ellen Lee Postdoctoral Fellowship at Caltech.  P.Y. acknowledges support from the Eddleman Graduate Fellowship through the Institute for Quantum Information and Matter (IQIM).

\end{acknowledgements}
\vspace{1cm}





\bibliography{references}

\begin{thebibliography}{57}%
\makeatletter
\providecommand \@ifxundefined [1]{%
 \@ifx{#1\undefined}
}%
\providecommand \@ifnum [1]{%
 \ifnum #1\expandafter \@firstoftwo
 \else \expandafter \@secondoftwo
 \fi
}%
\providecommand \@ifx [1]{%
 \ifx #1\expandafter \@firstoftwo
 \else \expandafter \@secondoftwo
 \fi
}%
\providecommand \natexlab [1]{#1}%
\providecommand \enquote  [1]{``#1''}%
\providecommand \bibnamefont  [1]{#1}%
\providecommand \bibfnamefont [1]{#1}%
\providecommand \citenamefont [1]{#1}%
\providecommand \href@noop [0]{\@secondoftwo}%
\providecommand \href [0]{\begingroup \@sanitize@url \@href}%
\providecommand \@href[1]{\@@startlink{#1}\@@href}%
\providecommand \@@href[1]{\endgroup#1\@@endlink}%
\providecommand \@sanitize@url [0]{\catcode `\\12\catcode `\$12\catcode
  `\&12\catcode `\#12\catcode `\^12\catcode `\_12\catcode `\%12\relax}%
\providecommand \@@startlink[1]{}%
\providecommand \@@endlink[0]{}%
\providecommand \url  [0]{\begingroup\@sanitize@url \@url }%
\providecommand \@url [1]{\endgroup\@href {#1}{\urlprefix }}%
\providecommand \urlprefix  [0]{URL }%
\providecommand \Eprint [0]{\href }%
\providecommand \doibase [0]{https://doi.org/}%
\providecommand \selectlanguage [0]{\@gobble}%
\providecommand \bibinfo  [0]{\@secondoftwo}%
\providecommand \bibfield  [0]{\@secondoftwo}%
\providecommand \translation [1]{[#1]}%
\providecommand \BibitemOpen [0]{}%
\providecommand \bibitemStop [0]{}%
\providecommand \bibitemNoStop [0]{.\EOS\space}%
\providecommand \EOS [0]{\spacefactor3000\relax}%
\providecommand \BibitemShut  [1]{\csname bibitem#1\endcsname}%
\let\auto@bib@innerbib\@empty
\bibitem [{\citenamefont {Safronova}\ \emph {et~al.}(2018)\citenamefont
  {Safronova}, \citenamefont {Budker}, \citenamefont {DeMille}, \citenamefont
  {Kimball}, \citenamefont {Derevianko},\ and\ \citenamefont
  {Clark}}]{Safronova2018}%
  \BibitemOpen
  \bibfield  {author} {\bibinfo {author} {\bibfnamefont {M.~S.}\ \bibnamefont
  {Safronova}}, \bibinfo {author} {\bibfnamefont {D.}~\bibnamefont {Budker}},
  \bibinfo {author} {\bibfnamefont {D.}~\bibnamefont {DeMille}}, \bibinfo
  {author} {\bibfnamefont {D.~F.~J.}\ \bibnamefont {Kimball}}, \bibinfo
  {author} {\bibfnamefont {A.}~\bibnamefont {Derevianko}},\ and\ \bibinfo
  {author} {\bibfnamefont {C.~W.}\ \bibnamefont {Clark}},\ }\href
  {https://doi.org/10.1103/RevModPhys.90.025008} {\bibfield  {journal}
  {\bibinfo  {journal} {Rev. Mod. Phys.}\ }\textbf {\bibinfo {volume} {90}},\
  \bibinfo {pages} {025008} (\bibinfo {year} {2018})}\BibitemShut {NoStop}%
\bibitem [{\citenamefont {Roussy}\ \emph {et~al.}(2023)\citenamefont {Roussy},
  \citenamefont {Caldwell}, \citenamefont {Wright}, \citenamefont {Cairncross},
  \citenamefont {Shagam}, \citenamefont {Ng}, \citenamefont {Schlossberger},
  \citenamefont {Park}, \citenamefont {Wang}, \citenamefont {Ye},\ and\
  \citenamefont {Cornell}}]{Roussy2023}%
  \BibitemOpen
  \bibfield  {author} {\bibinfo {author} {\bibfnamefont {T.~S.}\ \bibnamefont
  {Roussy}}, \bibinfo {author} {\bibfnamefont {L.}~\bibnamefont {Caldwell}},
  \bibinfo {author} {\bibfnamefont {T.}~\bibnamefont {Wright}}, \bibinfo
  {author} {\bibfnamefont {W.~B.}\ \bibnamefont {Cairncross}}, \bibinfo
  {author} {\bibfnamefont {Y.}~\bibnamefont {Shagam}}, \bibinfo {author}
  {\bibfnamefont {K.~B.}\ \bibnamefont {Ng}}, \bibinfo {author} {\bibfnamefont
  {N.}~\bibnamefont {Schlossberger}}, \bibinfo {author} {\bibfnamefont {S.~Y.}\
  \bibnamefont {Park}}, \bibinfo {author} {\bibfnamefont {A.}~\bibnamefont
  {Wang}}, \bibinfo {author} {\bibfnamefont {J.}~\bibnamefont {Ye}},\ and\
  \bibinfo {author} {\bibfnamefont {E.~A.}\ \bibnamefont {Cornell}},\ }\href
  {https://doi.org/10.1126/science.adg4084} {\bibinfo {title} {An improved
  bound on the electron’s electric dipole moment}} (\bibinfo {year}
  {2023})\BibitemShut {NoStop}%
\bibitem [{\citenamefont {Andreev}\ \emph {et~al.}(2018)\citenamefont
  {Andreev}, \citenamefont {Ang}, \citenamefont {DeMille}, \citenamefont
  {Doyle}, \citenamefont {Gabrielse}, \citenamefont {Haefner}, \citenamefont
  {Hutzler}, \citenamefont {Lasner}, \citenamefont {Meisenhelder},
  \citenamefont {O’Leary}, \citenamefont {Panda}, \citenamefont {West},
  \citenamefont {West},\ and\ \citenamefont {Wu}}]{Andreev2018}%
  \BibitemOpen
  \bibfield  {author} {\bibinfo {author} {\bibfnamefont {V.}~\bibnamefont
  {Andreev}}, \bibinfo {author} {\bibfnamefont {D.~G.}\ \bibnamefont {Ang}},
  \bibinfo {author} {\bibfnamefont {D.}~\bibnamefont {DeMille}}, \bibinfo
  {author} {\bibfnamefont {J.~M.}\ \bibnamefont {Doyle}}, \bibinfo {author}
  {\bibfnamefont {G.}~\bibnamefont {Gabrielse}}, \bibinfo {author}
  {\bibfnamefont {J.}~\bibnamefont {Haefner}}, \bibinfo {author} {\bibfnamefont
  {N.~R.}\ \bibnamefont {Hutzler}}, \bibinfo {author} {\bibfnamefont
  {Z.}~\bibnamefont {Lasner}}, \bibinfo {author} {\bibfnamefont
  {C.}~\bibnamefont {Meisenhelder}}, \bibinfo {author} {\bibfnamefont {B.~R.}\
  \bibnamefont {O’Leary}}, \bibinfo {author} {\bibfnamefont {C.~D.}\
  \bibnamefont {Panda}}, \bibinfo {author} {\bibfnamefont {A.~D.}\ \bibnamefont
  {West}}, \bibinfo {author} {\bibfnamefont {E.~P.}\ \bibnamefont {West}},\
  and\ \bibinfo {author} {\bibfnamefont {X.}~\bibnamefont {Wu}},\ }\href
  {https://doi.org/10.1038/s41586-018-0599-8} {\bibfield  {journal} {\bibinfo
  {journal} {Nature}\ }\textbf {\bibinfo {volume} {562}},\ \bibinfo {pages}
  {355} (\bibinfo {year} {2018})}\BibitemShut {NoStop}%
\bibitem [{\citenamefont {Cesarotti}\ \emph {et~al.}(2019)\citenamefont
  {Cesarotti}, \citenamefont {Lu}, \citenamefont {Nakai}, \citenamefont
  {Parikh},\ and\ \citenamefont {Reece}}]{Cesarotti2019}%
  \BibitemOpen
  \bibfield  {author} {\bibinfo {author} {\bibfnamefont {C.}~\bibnamefont
  {Cesarotti}}, \bibinfo {author} {\bibfnamefont {Q.}~\bibnamefont {Lu}},
  \bibinfo {author} {\bibfnamefont {Y.}~\bibnamefont {Nakai}}, \bibinfo
  {author} {\bibfnamefont {A.}~\bibnamefont {Parikh}},\ and\ \bibinfo {author}
  {\bibfnamefont {M.}~\bibnamefont {Reece}},\ }\href
  {https://doi.org/10.1007/JHEP05(2019)059} {\bibfield  {journal} {\bibinfo
  {journal} {Journal of High Energy Physics}\ }\textbf {\bibinfo {volume}
  {2019}},\ \bibinfo {pages} {59} (\bibinfo {year} {2019})}\BibitemShut
  {NoStop}%
\bibitem [{\citenamefont {Alarcon}\ \emph {et~al.}(2022)\citenamefont {Alarcon}
  \emph {et~al.}}]{Alarcon2022}%
  \BibitemOpen
  \bibfield  {author} {\bibinfo {author} {\bibfnamefont {R.}~\bibnamefont
  {Alarcon}} \emph {et~al.},\ }\href
  {https://doi.org/10.48550/ARXIV.2203.08103} {\bibinfo {title} {Electric
  dipole moments and the search for new physics}} (\bibinfo {year} {2022}),\
  \Eprint {https://arxiv.org/abs/2203.08103} {arXiv:2203.08103} \BibitemShut
  {NoStop}%
\bibitem [{\citenamefont {Alauze}\ \emph {et~al.}(2021)\citenamefont {Alauze},
  \citenamefont {Lim}, \citenamefont {Trigatzis}, \citenamefont {Swarbrick},
  \citenamefont {Collings}, \citenamefont {Fitch}, \citenamefont {Sauer},\ and\
  \citenamefont {Tarbutt}}]{Alauze2021}%
  \BibitemOpen
  \bibfield  {author} {\bibinfo {author} {\bibfnamefont {X.}~\bibnamefont
  {Alauze}}, \bibinfo {author} {\bibfnamefont {J.}~\bibnamefont {Lim}},
  \bibinfo {author} {\bibfnamefont {M.~A.}\ \bibnamefont {Trigatzis}}, \bibinfo
  {author} {\bibfnamefont {S.}~\bibnamefont {Swarbrick}}, \bibinfo {author}
  {\bibfnamefont {F.~J.}\ \bibnamefont {Collings}}, \bibinfo {author}
  {\bibfnamefont {N.~J.}\ \bibnamefont {Fitch}}, \bibinfo {author}
  {\bibfnamefont {B.~E.}\ \bibnamefont {Sauer}},\ and\ \bibinfo {author}
  {\bibfnamefont {M.~R.}\ \bibnamefont {Tarbutt}},\ }\href
  {https://doi.org/https://doi.org/10.1088/2058-9565/ac107e} {\bibfield
  {journal} {\bibinfo  {journal} {Quantum Sci. Technol.}\ }\textbf {\bibinfo
  {volume} {6}},\ \bibinfo {pages} {044005} (\bibinfo {year}
  {2021})}\BibitemShut {NoStop}%
\bibitem [{\citenamefont {Fitch}\ \emph {et~al.}(2021)\citenamefont {Fitch},
  \citenamefont {Lim}, \citenamefont {Hinds}, \citenamefont {Sauer},\ and\
  \citenamefont {Tarbutt}}]{Fitch2020b}%
  \BibitemOpen
  \bibfield  {author} {\bibinfo {author} {\bibfnamefont {N.~J.}\ \bibnamefont
  {Fitch}}, \bibinfo {author} {\bibfnamefont {J.}~\bibnamefont {Lim}}, \bibinfo
  {author} {\bibfnamefont {E.~A.}\ \bibnamefont {Hinds}}, \bibinfo {author}
  {\bibfnamefont {B.~E.}\ \bibnamefont {Sauer}},\ and\ \bibinfo {author}
  {\bibfnamefont {M.~R.}\ \bibnamefont {Tarbutt}},\ }\href
  {https://doi.org/10.1088/2058-9565/abc931} {\bibfield  {journal} {\bibinfo
  {journal} {Quantum Sci. Technol.}\ }\textbf {\bibinfo {volume} {6}},\
  \bibinfo {pages} {014006} (\bibinfo {year} {2021})}\BibitemShut {NoStop}%
\bibitem [{\citenamefont {Augenbraun}\ \emph {et~al.}(2020)\citenamefont
  {Augenbraun}, \citenamefont {Lasner}, \citenamefont {Frenett}, \citenamefont
  {Sawaoka}, \citenamefont {Miller}, \citenamefont {Steimle},\ and\
  \citenamefont {Doyle}}]{Augenbraun2020}%
  \BibitemOpen
  \bibfield  {author} {\bibinfo {author} {\bibfnamefont {B.~L.}\ \bibnamefont
  {Augenbraun}}, \bibinfo {author} {\bibfnamefont {Z.~D.}\ \bibnamefont
  {Lasner}}, \bibinfo {author} {\bibfnamefont {A.}~\bibnamefont {Frenett}},
  \bibinfo {author} {\bibfnamefont {H.}~\bibnamefont {Sawaoka}}, \bibinfo
  {author} {\bibfnamefont {C.}~\bibnamefont {Miller}}, \bibinfo {author}
  {\bibfnamefont {T.~C.}\ \bibnamefont {Steimle}},\ and\ \bibinfo {author}
  {\bibfnamefont {J.~M.}\ \bibnamefont {Doyle}},\ }\href
  {https://doi.org/10.1088/1367-2630/ab687b} {\bibfield  {journal} {\bibinfo
  {journal} {New J. Phys.}\ }\textbf {\bibinfo {volume} {22}},\ \bibinfo
  {pages} {022003} (\bibinfo {year} {2020})}\BibitemShut {NoStop}%
\bibitem [{\citenamefont {Kozyryev}\ and\ \citenamefont
  {Hutzler}(2017)}]{Kozyryev2017b}%
  \BibitemOpen
  \bibfield  {author} {\bibinfo {author} {\bibfnamefont {I.}~\bibnamefont
  {Kozyryev}}\ and\ \bibinfo {author} {\bibfnamefont {N.~R.}\ \bibnamefont
  {Hutzler}},\ }\href {https://doi.org/10.1103/PhysRevLett.119.133002}
  {\bibfield  {journal} {\bibinfo  {journal} {Phys. Rev. Lett.}\ }\textbf
  {\bibinfo {volume} {119}},\ \bibinfo {pages} {133002} (\bibinfo {year}
  {2017})}\BibitemShut {NoStop}%
\bibitem [{\citenamefont {Isaev}\ \emph {et~al.}(2010)\citenamefont {Isaev},
  \citenamefont {Hoekstra},\ and\ \citenamefont {Berger}}]{isaev2010}%
  \BibitemOpen
  \bibfield  {author} {\bibinfo {author} {\bibfnamefont {T.~A.}\ \bibnamefont
  {Isaev}}, \bibinfo {author} {\bibfnamefont {S.}~\bibnamefont {Hoekstra}},\
  and\ \bibinfo {author} {\bibfnamefont {R.}~\bibnamefont {Berger}},\ }\href
  {https://doi.org/10.1103/PhysRevA.82.052521} {\bibfield  {journal} {\bibinfo
  {journal} {Phys. Rev. A}\ }\textbf {\bibinfo {volume} {82}},\ \bibinfo
  {pages} {052521} (\bibinfo {year} {2010})}\BibitemShut {NoStop}%
\bibitem [{\citenamefont {Lasner}\ \emph {et~al.}(2022)\citenamefont {Lasner},
  \citenamefont {Lunstad}, \citenamefont {Zhang}, \citenamefont {Cheng},\ and\
  \citenamefont {Doyle}}]{Lasner2022SrOHBR}%
  \BibitemOpen
  \bibfield  {author} {\bibinfo {author} {\bibfnamefont {Z.}~\bibnamefont
  {Lasner}}, \bibinfo {author} {\bibfnamefont {A.}~\bibnamefont {Lunstad}},
  \bibinfo {author} {\bibfnamefont {C.}~\bibnamefont {Zhang}}, \bibinfo
  {author} {\bibfnamefont {L.}~\bibnamefont {Cheng}},\ and\ \bibinfo {author}
  {\bibfnamefont {J.~M.}\ \bibnamefont {Doyle}},\ }\href
  {https://doi.org/10.1103/PhysRevA.106.L020801} {\bibfield  {journal}
  {\bibinfo  {journal} {Physical Review A}\ }\textbf {\bibinfo {volume}
  {106}},\ \bibinfo {pages} {L020801} (\bibinfo {year} {2022})}\BibitemShut
  {NoStop}%
\bibitem [{\citenamefont {Pezz\`e}\ \emph {et~al.}(2018)\citenamefont
  {Pezz\`e}, \citenamefont {Smerzi}, \citenamefont {Oberthaler}, \citenamefont
  {Schmied},\ and\ \citenamefont {Treutlein}}]{Pezze2018}%
  \BibitemOpen
  \bibfield  {author} {\bibinfo {author} {\bibfnamefont {L.}~\bibnamefont
  {Pezz\`e}}, \bibinfo {author} {\bibfnamefont {A.}~\bibnamefont {Smerzi}},
  \bibinfo {author} {\bibfnamefont {M.~K.}\ \bibnamefont {Oberthaler}},
  \bibinfo {author} {\bibfnamefont {R.}~\bibnamefont {Schmied}},\ and\ \bibinfo
  {author} {\bibfnamefont {P.}~\bibnamefont {Treutlein}},\ }\href
  {https://doi.org/10.1103/RevModPhys.90.035005} {\bibfield  {journal}
  {\bibinfo  {journal} {Rev. Mod. Phys.}\ }\textbf {\bibinfo {volume} {90}},\
  \bibinfo {pages} {035005} (\bibinfo {year} {2018})}\BibitemShut {NoStop}%
\bibitem [{\citenamefont {Pezz\'e}\ and\ \citenamefont
  {Smerzi}(2009)}]{Pezze2009}%
  \BibitemOpen
  \bibfield  {author} {\bibinfo {author} {\bibfnamefont {L.}~\bibnamefont
  {Pezz\'e}}\ and\ \bibinfo {author} {\bibfnamefont {A.}~\bibnamefont
  {Smerzi}},\ }\href {https://doi.org/10.1103/PhysRevLett.102.100401}
  {\bibfield  {journal} {\bibinfo  {journal} {Phys. Rev. Lett.}\ }\textbf
  {\bibinfo {volume} {102}},\ \bibinfo {pages} {100401} (\bibinfo {year}
  {2009})}\BibitemShut {NoStop}%
\bibitem [{\citenamefont {Cairncross}\ \emph {et~al.}(2017)\citenamefont
  {Cairncross}, \citenamefont {Gresh}, \citenamefont {Grau}, \citenamefont
  {Cossel}, \citenamefont {Roussy}, \citenamefont {Ni}, \citenamefont {Zhou},
  \citenamefont {Ye},\ and\ \citenamefont {Cornell}}]{Cairncross2017}%
  \BibitemOpen
  \bibfield  {author} {\bibinfo {author} {\bibfnamefont {W.~B.}\ \bibnamefont
  {Cairncross}}, \bibinfo {author} {\bibfnamefont {D.~N.}\ \bibnamefont
  {Gresh}}, \bibinfo {author} {\bibfnamefont {M.}~\bibnamefont {Grau}},
  \bibinfo {author} {\bibfnamefont {K.~C.}\ \bibnamefont {Cossel}}, \bibinfo
  {author} {\bibfnamefont {T.~S.}\ \bibnamefont {Roussy}}, \bibinfo {author}
  {\bibfnamefont {Y.}~\bibnamefont {Ni}}, \bibinfo {author} {\bibfnamefont
  {Y.}~\bibnamefont {Zhou}}, \bibinfo {author} {\bibfnamefont {J.}~\bibnamefont
  {Ye}},\ and\ \bibinfo {author} {\bibfnamefont {E.~A.}\ \bibnamefont
  {Cornell}},\ }\href {https://doi.org/10.1103/PhysRevLett.119.153001}
  {\bibfield  {journal} {\bibinfo  {journal} {Phys. Rev. Lett.}\ }\textbf
  {\bibinfo {volume} {119}},\ \bibinfo {pages} {153001} (\bibinfo {year}
  {2017})}\BibitemShut {NoStop}%
\bibitem [{\citenamefont {Ng}\ \emph {et~al.}(2022)\citenamefont {Ng},
  \citenamefont {Zhou}, \citenamefont {Cheng}, \citenamefont {Schlossberger},
  \citenamefont {Park}, \citenamefont {Roussy}, \citenamefont {Caldwell},
  \citenamefont {Shagam}, \citenamefont {Vigil}, \citenamefont {Cornell},\ and\
  \citenamefont {Ye}}]{Ng2022}%
  \BibitemOpen
  \bibfield  {author} {\bibinfo {author} {\bibfnamefont {K.~B.}\ \bibnamefont
  {Ng}}, \bibinfo {author} {\bibfnamefont {Y.}~\bibnamefont {Zhou}}, \bibinfo
  {author} {\bibfnamefont {L.}~\bibnamefont {Cheng}}, \bibinfo {author}
  {\bibfnamefont {N.}~\bibnamefont {Schlossberger}}, \bibinfo {author}
  {\bibfnamefont {S.~Y.}\ \bibnamefont {Park}}, \bibinfo {author}
  {\bibfnamefont {T.~S.}\ \bibnamefont {Roussy}}, \bibinfo {author}
  {\bibfnamefont {L.}~\bibnamefont {Caldwell}}, \bibinfo {author}
  {\bibfnamefont {Y.}~\bibnamefont {Shagam}}, \bibinfo {author} {\bibfnamefont
  {A.~J.}\ \bibnamefont {Vigil}}, \bibinfo {author} {\bibfnamefont {E.~A.}\
  \bibnamefont {Cornell}},\ and\ \bibinfo {author} {\bibfnamefont
  {J.}~\bibnamefont {Ye}},\ }\href
  {https://doi.org/10.1103/PhysRevA.105.022823} {\bibfield  {journal} {\bibinfo
   {journal} {Phys. Rev. A}\ }\textbf {\bibinfo {volume} {105}},\ \bibinfo
  {pages} {022823} (\bibinfo {year} {2022})}\BibitemShut {NoStop}%
\bibitem [{\citenamefont {Taylor}\ \emph {et~al.}(2022)\citenamefont {Taylor},
  \citenamefont {Island},\ and\ \citenamefont {Zhou}}]{Taylor2022}%
  \BibitemOpen
  \bibfield  {author} {\bibinfo {author} {\bibfnamefont {T.~N.}\ \bibnamefont
  {Taylor}}, \bibinfo {author} {\bibfnamefont {J.~O.}\ \bibnamefont {Island}},\
  and\ \bibinfo {author} {\bibfnamefont {Y.}~\bibnamefont {Zhou}},\ }\href@noop
  {} {\bibinfo {title} {Quantum logic control and precision measurements of
  molecular ions in a ring trap -- a new approach for testing fundamental
  symmetries}} (\bibinfo {year} {2022}),\ \Eprint
  {https://arxiv.org/abs/2210.11613} {arXiv:2210.11613 [physics.atom-ph]}
  \BibitemShut {NoStop}%
\bibitem [{\citenamefont {Holland}\ \emph {et~al.}(2022)\citenamefont
  {Holland}, \citenamefont {Lu},\ and\ \citenamefont {Cheuk}}]{Holland2022}%
  \BibitemOpen
  \bibfield  {author} {\bibinfo {author} {\bibfnamefont {C.~M.}\ \bibnamefont
  {Holland}}, \bibinfo {author} {\bibfnamefont {Y.}~\bibnamefont {Lu}},\ and\
  \bibinfo {author} {\bibfnamefont {L.~W.}\ \bibnamefont {Cheuk}},\ }\href
  {https://doi.org/10.48550/ARXIV.2210.06309} {\bibinfo {title} {On-demand
  entanglement of molecules in a reconfigurable optical tweezer array}}
  (\bibinfo {year} {2022}),\ \Eprint {https://arxiv.org/abs/2210.06309}
  {arXiv:2210.06309} \BibitemShut {NoStop}%
\bibitem [{\citenamefont {Bao}\ \emph {et~al.}(2022)\citenamefont {Bao},
  \citenamefont {Yu}, \citenamefont {Anderegg}, \citenamefont {Chae},
  \citenamefont {Ketterle}, \citenamefont {Ni},\ and\ \citenamefont
  {Doyle}}]{Bao2022}%
  \BibitemOpen
  \bibfield  {author} {\bibinfo {author} {\bibfnamefont {Y.}~\bibnamefont
  {Bao}}, \bibinfo {author} {\bibfnamefont {S.~S.}\ \bibnamefont {Yu}},
  \bibinfo {author} {\bibfnamefont {L.}~\bibnamefont {Anderegg}}, \bibinfo
  {author} {\bibfnamefont {E.}~\bibnamefont {Chae}}, \bibinfo {author}
  {\bibfnamefont {W.}~\bibnamefont {Ketterle}}, \bibinfo {author}
  {\bibfnamefont {K.-K.}\ \bibnamefont {Ni}},\ and\ \bibinfo {author}
  {\bibfnamefont {J.~M.}\ \bibnamefont {Doyle}},\ }\href
  {https://doi.org/10.48550/ARXIV.2211.09780} {\bibinfo {title} {Dipolar
  spin-exchange and entanglement between molecules in an optical tweezer
  array}} (\bibinfo {year} {2022}),\ \Eprint {https://arxiv.org/abs/2211.09780}
  {arXiv:2211.09780} \BibitemShut {NoStop}%
\bibitem [{\citenamefont {Lin}\ \emph {et~al.}(2020)\citenamefont {Lin},
  \citenamefont {Leibrandt}, \citenamefont {Leibfried},\ and\ \citenamefont
  {Chou}}]{Lin2020}%
  \BibitemOpen
  \bibfield  {author} {\bibinfo {author} {\bibfnamefont {Y.}~\bibnamefont
  {Lin}}, \bibinfo {author} {\bibfnamefont {D.~R.}\ \bibnamefont {Leibrandt}},
  \bibinfo {author} {\bibfnamefont {D.}~\bibnamefont {Leibfried}},\ and\
  \bibinfo {author} {\bibfnamefont {C.-w.}\ \bibnamefont {Chou}},\ }\href
  {https://doi.org/10.1038/s41586-020-2257-1} {\bibfield  {journal} {\bibinfo
  {journal} {Nature}\ }\textbf {\bibinfo {volume} {581}},\ \bibinfo {pages}
  {273} (\bibinfo {year} {2020})}\BibitemShut {NoStop}%
\bibitem [{\citenamefont {Fan}\ \emph {et~al.}(2021)\citenamefont {Fan},
  \citenamefont {Holliman}, \citenamefont {Shi}, \citenamefont {Zhang},
  \citenamefont {Straus}, \citenamefont {Li}, \citenamefont {Buechele},\ and\
  \citenamefont {Jayich}}]{Fan2021}%
  \BibitemOpen
  \bibfield  {author} {\bibinfo {author} {\bibfnamefont {M.}~\bibnamefont
  {Fan}}, \bibinfo {author} {\bibfnamefont {C.~A.}\ \bibnamefont {Holliman}},
  \bibinfo {author} {\bibfnamefont {X.}~\bibnamefont {Shi}}, \bibinfo {author}
  {\bibfnamefont {H.}~\bibnamefont {Zhang}}, \bibinfo {author} {\bibfnamefont
  {M.~W.}\ \bibnamefont {Straus}}, \bibinfo {author} {\bibfnamefont
  {X.}~\bibnamefont {Li}}, \bibinfo {author} {\bibfnamefont {S.~W.}\
  \bibnamefont {Buechele}},\ and\ \bibinfo {author} {\bibfnamefont {A.~M.}\
  \bibnamefont {Jayich}},\ }\href
  {https://doi.org/10.1103/PhysRevLett.126.023002} {\bibfield  {journal}
  {\bibinfo  {journal} {Phys. Rev. Lett.}\ }\textbf {\bibinfo {volume} {126}},\
  \bibinfo {pages} {023002} (\bibinfo {year} {2021})}\BibitemShut {NoStop}%
\bibitem [{\citenamefont {Brewer}\ \emph {et~al.}(2019)\citenamefont {Brewer},
  \citenamefont {Chen}, \citenamefont {Hankin}, \citenamefont {Clements},
  \citenamefont {Chou}, \citenamefont {Wineland}, \citenamefont {Hume},\ and\
  \citenamefont {Leibrandt}}]{Brewer2019}%
  \BibitemOpen
  \bibfield  {author} {\bibinfo {author} {\bibfnamefont {S.~M.}\ \bibnamefont
  {Brewer}}, \bibinfo {author} {\bibfnamefont {J.-S.}\ \bibnamefont {Chen}},
  \bibinfo {author} {\bibfnamefont {A.~M.}\ \bibnamefont {Hankin}}, \bibinfo
  {author} {\bibfnamefont {E.~R.}\ \bibnamefont {Clements}}, \bibinfo {author}
  {\bibfnamefont {C.~W.}\ \bibnamefont {Chou}}, \bibinfo {author}
  {\bibfnamefont {D.~J.}\ \bibnamefont {Wineland}}, \bibinfo {author}
  {\bibfnamefont {D.~B.}\ \bibnamefont {Hume}},\ and\ \bibinfo {author}
  {\bibfnamefont {D.~R.}\ \bibnamefont {Leibrandt}},\ }\href
  {https://doi.org/10.1103/PhysRevLett.123.033201} {\bibfield  {journal}
  {\bibinfo  {journal} {Phys. Rev. Lett.}\ }\textbf {\bibinfo {volume} {123}},\
  \bibinfo {pages} {033201} (\bibinfo {year} {2019})}\BibitemShut {NoStop}%
\bibitem [{\citenamefont {Sanner}\ \emph {et~al.}(2019)\citenamefont {Sanner},
  \citenamefont {Huntemann}, \citenamefont {Lange}, \citenamefont {Tamm},
  \citenamefont {Peik}, \citenamefont {Safronova},\ and\ \citenamefont
  {Porsev}}]{Sanner2019}%
  \BibitemOpen
  \bibfield  {author} {\bibinfo {author} {\bibfnamefont {C.}~\bibnamefont
  {Sanner}}, \bibinfo {author} {\bibfnamefont {N.}~\bibnamefont {Huntemann}},
  \bibinfo {author} {\bibfnamefont {R.}~\bibnamefont {Lange}}, \bibinfo
  {author} {\bibfnamefont {C.}~\bibnamefont {Tamm}}, \bibinfo {author}
  {\bibfnamefont {E.}~\bibnamefont {Peik}}, \bibinfo {author} {\bibfnamefont
  {M.~S.}\ \bibnamefont {Safronova}},\ and\ \bibinfo {author} {\bibfnamefont
  {S.~G.}\ \bibnamefont {Porsev}},\ }\href
  {https://doi.org/10.1038/s41586-019-0972-2} {\bibfield  {journal} {\bibinfo
  {journal} {Nature}\ }\textbf {\bibinfo {volume} {567}},\ \bibinfo {pages}
  {204} (\bibinfo {year} {2019})}\BibitemShut {NoStop}%
\bibitem [{\citenamefont {Hughes}\ \emph {et~al.}(2020)\citenamefont {Hughes},
  \citenamefont {Frye}, \citenamefont {Sawant}, \citenamefont {Bhole},
  \citenamefont {Jones}, \citenamefont {Cornish}, \citenamefont {Tarbutt},
  \citenamefont {Hutson}, \citenamefont {Jaksch},\ and\ \citenamefont
  {Mur-Petit}}]{Hughes2020}%
  \BibitemOpen
  \bibfield  {author} {\bibinfo {author} {\bibfnamefont {M.}~\bibnamefont
  {Hughes}}, \bibinfo {author} {\bibfnamefont {M.~D.}\ \bibnamefont {Frye}},
  \bibinfo {author} {\bibfnamefont {R.}~\bibnamefont {Sawant}}, \bibinfo
  {author} {\bibfnamefont {G.}~\bibnamefont {Bhole}}, \bibinfo {author}
  {\bibfnamefont {J.~A.}\ \bibnamefont {Jones}}, \bibinfo {author}
  {\bibfnamefont {S.~L.}\ \bibnamefont {Cornish}}, \bibinfo {author}
  {\bibfnamefont {M.~R.}\ \bibnamefont {Tarbutt}}, \bibinfo {author}
  {\bibfnamefont {J.~M.}\ \bibnamefont {Hutson}}, \bibinfo {author}
  {\bibfnamefont {D.}~\bibnamefont {Jaksch}},\ and\ \bibinfo {author}
  {\bibfnamefont {J.}~\bibnamefont {Mur-Petit}},\ }\href
  {https://doi.org/10.1103/PhysRevA.101.062308} {\bibfield  {journal} {\bibinfo
   {journal} {Phys. Rev. A}\ }\textbf {\bibinfo {volume} {101}},\ \bibinfo
  {pages} {062308} (\bibinfo {year} {2020})}\BibitemShut {NoStop}%
\bibitem [{\citenamefont {Hudson}\ and\ \citenamefont
  {Campbell}(2018)}]{Hudson2018}%
  \BibitemOpen
  \bibfield  {author} {\bibinfo {author} {\bibfnamefont {E.~R.}\ \bibnamefont
  {Hudson}}\ and\ \bibinfo {author} {\bibfnamefont {W.~C.}\ \bibnamefont
  {Campbell}},\ }\href {https://doi.org/10.1103/PhysRevA.98.040302} {\bibfield
  {journal} {\bibinfo  {journal} {Phys. Rev. A}\ }\textbf {\bibinfo {volume}
  {98}},\ \bibinfo {pages} {040302(R)} (\bibinfo {year} {2018})}\BibitemShut
  {NoStop}%
\bibitem [{\citenamefont {Ni}\ \emph {et~al.}(2018)\citenamefont {Ni},
  \citenamefont {Rosenband},\ and\ \citenamefont {Grimes}}]{Ni2018}%
  \BibitemOpen
  \bibfield  {author} {\bibinfo {author} {\bibfnamefont {K.-K.}\ \bibnamefont
  {Ni}}, \bibinfo {author} {\bibfnamefont {T.}~\bibnamefont {Rosenband}},\ and\
  \bibinfo {author} {\bibfnamefont {D.~D.}\ \bibnamefont {Grimes}},\ }\href
  {https://doi.org/10.1039/C8SC02355G} {\bibfield  {journal} {\bibinfo
  {journal} {Chem. Sci.}\ }\textbf {\bibinfo {volume} {9}},\ \bibinfo {pages}
  {6830} (\bibinfo {year} {2018})}\BibitemShut {NoStop}%
\bibitem [{\citenamefont {Yelin}\ \emph {et~al.}(2006)\citenamefont {Yelin},
  \citenamefont {Kirby},\ and\ \citenamefont {C{\^{o}}t{\'{e}}}}]{Yelin2006}%
  \BibitemOpen
  \bibfield  {author} {\bibinfo {author} {\bibfnamefont {S.~F.}\ \bibnamefont
  {Yelin}}, \bibinfo {author} {\bibfnamefont {K.}~\bibnamefont {Kirby}},\ and\
  \bibinfo {author} {\bibfnamefont {R.}~\bibnamefont {C{\^{o}}t{\'{e}}}},\
  }\href {https://doi.org/10.1103/physreva.74.050301} {\bibfield  {journal}
  {\bibinfo  {journal} {Phys. Rev. A}\ }\textbf {\bibinfo {volume} {74}},\
  \bibinfo {pages} {050301(R)} (\bibinfo {year} {2006})}\BibitemShut {NoStop}%
\bibitem [{\citenamefont {Zhang}\ and\ \citenamefont
  {Tarbutt}(2022)}]{Zhang2022b}%
  \BibitemOpen
  \bibfield  {author} {\bibinfo {author} {\bibfnamefont {C.}~\bibnamefont
  {Zhang}}\ and\ \bibinfo {author} {\bibfnamefont {M.}~\bibnamefont
  {Tarbutt}},\ }\href {https://doi.org/10.1103/PRXQuantum.3.030340} {\bibfield
  {journal} {\bibinfo  {journal} {PRX Quantum}\ }\textbf {\bibinfo {volume}
  {3}},\ \bibinfo {pages} {030340} (\bibinfo {year} {2022})}\BibitemShut
  {NoStop}%
\bibitem [{\citenamefont {Wang}\ \emph {et~al.}(2022)\citenamefont {Wang},
  \citenamefont {Williams}, \citenamefont {Picard}, \citenamefont {Yao},\ and\
  \citenamefont {Ni}}]{Wang2022}%
  \BibitemOpen
  \bibfield  {author} {\bibinfo {author} {\bibfnamefont {K.}~\bibnamefont
  {Wang}}, \bibinfo {author} {\bibfnamefont {C.~P.}\ \bibnamefont {Williams}},
  \bibinfo {author} {\bibfnamefont {L.~R.}\ \bibnamefont {Picard}}, \bibinfo
  {author} {\bibfnamefont {N.~Y.}\ \bibnamefont {Yao}},\ and\ \bibinfo {author}
  {\bibfnamefont {K.-K.}\ \bibnamefont {Ni}},\ }\href
  {https://doi.org/10.1103/PRXQuantum.3.030339} {\bibfield  {journal} {\bibinfo
   {journal} {PRX Quantum}\ }\textbf {\bibinfo {volume} {3}},\ \bibinfo {pages}
  {030339} (\bibinfo {year} {2022})}\BibitemShut {NoStop}%
\bibitem [{\citenamefont {Omran}\ \emph {et~al.}(2019)\citenamefont {Omran},
  \citenamefont {Levine}, \citenamefont {Keesling}, \citenamefont {Semeghini},
  \citenamefont {Wang}, \citenamefont {Ebadi}, \citenamefont {Bernien},
  \citenamefont {Zibrov}, \citenamefont {Pichler}, \citenamefont {Choi},
  \citenamefont {Cui}, \citenamefont {Rossignolo}, \citenamefont {Rembold},
  \citenamefont {Montangero}, \citenamefont {Calarco}, \citenamefont {Endres},
  \citenamefont {Greiner}, \citenamefont {Vuletić},\ and\ \citenamefont
  {Lukin}}]{Omran2019}%
  \BibitemOpen
  \bibfield  {author} {\bibinfo {author} {\bibfnamefont {A.}~\bibnamefont
  {Omran}}, \bibinfo {author} {\bibfnamefont {H.}~\bibnamefont {Levine}},
  \bibinfo {author} {\bibfnamefont {A.}~\bibnamefont {Keesling}}, \bibinfo
  {author} {\bibfnamefont {G.}~\bibnamefont {Semeghini}}, \bibinfo {author}
  {\bibfnamefont {T.~T.}\ \bibnamefont {Wang}}, \bibinfo {author}
  {\bibfnamefont {S.}~\bibnamefont {Ebadi}}, \bibinfo {author} {\bibfnamefont
  {H.}~\bibnamefont {Bernien}}, \bibinfo {author} {\bibfnamefont {A.~S.}\
  \bibnamefont {Zibrov}}, \bibinfo {author} {\bibfnamefont {H.}~\bibnamefont
  {Pichler}}, \bibinfo {author} {\bibfnamefont {S.}~\bibnamefont {Choi}},
  \bibinfo {author} {\bibfnamefont {J.}~\bibnamefont {Cui}}, \bibinfo {author}
  {\bibfnamefont {M.}~\bibnamefont {Rossignolo}}, \bibinfo {author}
  {\bibfnamefont {P.}~\bibnamefont {Rembold}}, \bibinfo {author} {\bibfnamefont
  {S.}~\bibnamefont {Montangero}}, \bibinfo {author} {\bibfnamefont
  {T.}~\bibnamefont {Calarco}}, \bibinfo {author} {\bibfnamefont
  {M.}~\bibnamefont {Endres}}, \bibinfo {author} {\bibfnamefont
  {M.}~\bibnamefont {Greiner}}, \bibinfo {author} {\bibfnamefont
  {V.}~\bibnamefont {Vuletić}},\ and\ \bibinfo {author} {\bibfnamefont
  {M.~D.}\ \bibnamefont {Lukin}},\ }\href
  {https://doi.org/10.1126/science.aax9743} {\bibfield  {journal} {\bibinfo
  {journal} {Science}\ }\textbf {\bibinfo {volume} {365}},\ \bibinfo {pages}
  {570} (\bibinfo {year} {2019})}\BibitemShut {NoStop}%
\bibitem [{\citenamefont {Park}\ \emph {et~al.}(2017)\citenamefont {Park},
  \citenamefont {Yan}, \citenamefont {Loh}, \citenamefont {Will},\ and\
  \citenamefont {Zwierlein}}]{Park2017}%
  \BibitemOpen
  \bibfield  {author} {\bibinfo {author} {\bibfnamefont {J.~W.}\ \bibnamefont
  {Park}}, \bibinfo {author} {\bibfnamefont {Z.~Z.}\ \bibnamefont {Yan}},
  \bibinfo {author} {\bibfnamefont {H.}~\bibnamefont {Loh}}, \bibinfo {author}
  {\bibfnamefont {S.~A.}\ \bibnamefont {Will}},\ and\ \bibinfo {author}
  {\bibfnamefont {M.~W.}\ \bibnamefont {Zwierlein}},\ }\href
  {https://doi.org/10.1126/SCIENCE.AAL5066} {\bibfield  {journal} {\bibinfo
  {journal} {Science}\ }\textbf {\bibinfo {volume} {357}},\ \bibinfo {pages}
  {372} (\bibinfo {year} {2017})}\BibitemShut {NoStop}%
\bibitem [{\citenamefont {Gregory}\ \emph {et~al.}(2021)\citenamefont
  {Gregory}, \citenamefont {Blackmore}, \citenamefont {Bromley}, \citenamefont
  {Hutson},\ and\ \citenamefont {Cornish}}]{Gregory2021}%
  \BibitemOpen
  \bibfield  {author} {\bibinfo {author} {\bibfnamefont {P.~D.}\ \bibnamefont
  {Gregory}}, \bibinfo {author} {\bibfnamefont {J.~A.}\ \bibnamefont
  {Blackmore}}, \bibinfo {author} {\bibfnamefont {S.~L.}\ \bibnamefont
  {Bromley}}, \bibinfo {author} {\bibfnamefont {J.~M.}\ \bibnamefont
  {Hutson}},\ and\ \bibinfo {author} {\bibfnamefont {S.~L.}\ \bibnamefont
  {Cornish}},\ }\href {https://doi.org/10.1038/s41567-021-01328-7} {\bibfield
  {journal} {\bibinfo  {journal} {Nature Physics}\ }\textbf {\bibinfo {volume}
  {17}},\ \bibinfo {pages} {1149} (\bibinfo {year} {2021})}\BibitemShut
  {NoStop}%
\bibitem [{\citenamefont {Graner}\ \emph {et~al.}(2016)\citenamefont {Graner},
  \citenamefont {Chen}, \citenamefont {Lindahl},\ and\ \citenamefont
  {Heckel}}]{Graner2016}%
  \BibitemOpen
  \bibfield  {author} {\bibinfo {author} {\bibfnamefont {B.}~\bibnamefont
  {Graner}}, \bibinfo {author} {\bibfnamefont {Y.}~\bibnamefont {Chen}},
  \bibinfo {author} {\bibfnamefont {E.~G.}\ \bibnamefont {Lindahl}},\ and\
  \bibinfo {author} {\bibfnamefont {B.~R.}\ \bibnamefont {Heckel}},\ }\href
  {https://doi.org/10.1103/PhysRevLett.116.161601} {\bibfield  {journal}
  {\bibinfo  {journal} {Phys. Rev. Lett.}\ }\textbf {\bibinfo {volume} {116}},\
  \bibinfo {pages} {161601} (\bibinfo {year} {2016})}\BibitemShut {NoStop}%
\bibitem [{\citenamefont {Flambaum}\ \emph {et~al.}(2014)\citenamefont
  {Flambaum}, \citenamefont {DeMille},\ and\ \citenamefont
  {Kozlov}}]{Flambaum2014}%
  \BibitemOpen
  \bibfield  {author} {\bibinfo {author} {\bibfnamefont {V.~V.}\ \bibnamefont
  {Flambaum}}, \bibinfo {author} {\bibfnamefont {D.}~\bibnamefont {DeMille}},\
  and\ \bibinfo {author} {\bibfnamefont {M.~G.}\ \bibnamefont {Kozlov}},\
  }\href {https://doi.org/10.1103/PhysRevLett.113.103003} {\bibfield  {journal}
  {\bibinfo  {journal} {Phys. Rev. Lett.}\ }\textbf {\bibinfo {volume} {113}},\
  \bibinfo {pages} {103003} (\bibinfo {year} {2014})}\BibitemShut {NoStop}%
\bibitem [{\citenamefont {Hudson}\ \emph {et~al.}(2011)\citenamefont {Hudson},
  \citenamefont {Kara}, \citenamefont {Smallman}, \citenamefont {Sauer},
  \citenamefont {Tarbutt},\ and\ \citenamefont {Hinds}}]{Hudson2011}%
  \BibitemOpen
  \bibfield  {author} {\bibinfo {author} {\bibfnamefont {J.~J.}\ \bibnamefont
  {Hudson}}, \bibinfo {author} {\bibfnamefont {D.~M.}\ \bibnamefont {Kara}},
  \bibinfo {author} {\bibfnamefont {I.~J.}\ \bibnamefont {Smallman}}, \bibinfo
  {author} {\bibfnamefont {B.~E.}\ \bibnamefont {Sauer}}, \bibinfo {author}
  {\bibfnamefont {M.~R.}\ \bibnamefont {Tarbutt}},\ and\ \bibinfo {author}
  {\bibfnamefont {E.~A.}\ \bibnamefont {Hinds}},\ }\href
  {https://doi.org/10.1038/nature10104} {\bibfield  {journal} {\bibinfo
  {journal} {Nature}\ }\textbf {\bibinfo {volume} {473}},\ \bibinfo {pages}
  {493} (\bibinfo {year} {2011})}\BibitemShut {NoStop}%
\bibitem [{\citenamefont {Baron}\ \emph {et~al.}(2014)\citenamefont {Baron},
  \citenamefont {Campbell}, \citenamefont {DeMille}, \citenamefont {Doyle},
  \citenamefont {Gabrielse}, \citenamefont {Gurevich}, \citenamefont {Hess},
  \citenamefont {Hutzler}, \citenamefont {Kirilov}, \citenamefont {Kozyryev},
  \citenamefont {O'Leary}, \citenamefont {Panda}, \citenamefont {Parsons},
  \citenamefont {Petrik}, \citenamefont {Spaun}, \citenamefont {Vutha},\ and\
  \citenamefont {West}}]{Baron2014}%
  \BibitemOpen
  \bibfield  {author} {\bibinfo {author} {\bibfnamefont {J.}~\bibnamefont
  {Baron}}, \bibinfo {author} {\bibfnamefont {W.~C.}\ \bibnamefont {Campbell}},
  \bibinfo {author} {\bibfnamefont {D.}~\bibnamefont {DeMille}}, \bibinfo
  {author} {\bibfnamefont {J.~M.}\ \bibnamefont {Doyle}}, \bibinfo {author}
  {\bibfnamefont {G.}~\bibnamefont {Gabrielse}}, \bibinfo {author}
  {\bibfnamefont {Y.~V.}\ \bibnamefont {Gurevich}}, \bibinfo {author}
  {\bibfnamefont {P.~W.}\ \bibnamefont {Hess}}, \bibinfo {author}
  {\bibfnamefont {N.~R.}\ \bibnamefont {Hutzler}}, \bibinfo {author}
  {\bibfnamefont {E.}~\bibnamefont {Kirilov}}, \bibinfo {author} {\bibfnamefont
  {I.}~\bibnamefont {Kozyryev}}, \bibinfo {author} {\bibfnamefont {B.~R.}\
  \bibnamefont {O'Leary}}, \bibinfo {author} {\bibfnamefont {C.~D.}\
  \bibnamefont {Panda}}, \bibinfo {author} {\bibfnamefont {M.~F.}\ \bibnamefont
  {Parsons}}, \bibinfo {author} {\bibfnamefont {E.~S.}\ \bibnamefont {Petrik}},
  \bibinfo {author} {\bibfnamefont {B.}~\bibnamefont {Spaun}}, \bibinfo
  {author} {\bibfnamefont {A.~C.}\ \bibnamefont {Vutha}},\ and\ \bibinfo
  {author} {\bibfnamefont {A.~D.}\ \bibnamefont {West}},\ }\href
  {https://doi.org/10.1126/science.1248213} {\bibfield  {journal} {\bibinfo
  {journal} {Science}\ }\textbf {\bibinfo {volume} {343}},\ \bibinfo {pages}
  {269} (\bibinfo {year} {2014})}\BibitemShut {NoStop}%
\bibitem [{\citenamefont {Wineland}\ \emph {et~al.}(1992)\citenamefont
  {Wineland}, \citenamefont {Bollinger}, \citenamefont {Itano}, \citenamefont
  {Moore},\ and\ \citenamefont {Heinzen}}]{Wineland1992}%
  \BibitemOpen
  \bibfield  {author} {\bibinfo {author} {\bibfnamefont {D.~J.}\ \bibnamefont
  {Wineland}}, \bibinfo {author} {\bibfnamefont {J.~J.}\ \bibnamefont
  {Bollinger}}, \bibinfo {author} {\bibfnamefont {W.~M.}\ \bibnamefont
  {Itano}}, \bibinfo {author} {\bibfnamefont {F.~L.}\ \bibnamefont {Moore}},\
  and\ \bibinfo {author} {\bibfnamefont {D.~J.}\ \bibnamefont {Heinzen}},\
  }\href {https://doi.org/10.1103/PhysRevA.46.R6797} {\bibfield  {journal}
  {\bibinfo  {journal} {Phys. Rev. A}\ }\textbf {\bibinfo {volume} {46}},\
  \bibinfo {pages} {R6797} (\bibinfo {year} {1992})}\BibitemShut {NoStop}%
\bibitem [{\citenamefont {Wineland}\ \emph {et~al.}(1994)\citenamefont
  {Wineland}, \citenamefont {Bollinger}, \citenamefont {Itano},\ and\
  \citenamefont {Heinzen}}]{Wineland1994}%
  \BibitemOpen
  \bibfield  {author} {\bibinfo {author} {\bibfnamefont {D.~J.}\ \bibnamefont
  {Wineland}}, \bibinfo {author} {\bibfnamefont {J.~J.}\ \bibnamefont
  {Bollinger}}, \bibinfo {author} {\bibfnamefont {W.~M.}\ \bibnamefont
  {Itano}},\ and\ \bibinfo {author} {\bibfnamefont {D.~J.}\ \bibnamefont
  {Heinzen}},\ }\href {https://doi.org/10.1103/PhysRevA.50.67} {\bibfield
  {journal} {\bibinfo  {journal} {Phys. Rev. A}\ }\textbf {\bibinfo {volume}
  {50}},\ \bibinfo {pages} {67} (\bibinfo {year} {1994})}\BibitemShut {NoStop}%
\bibitem [{\citenamefont {Perlin}\ \emph {et~al.}(2020)\citenamefont {Perlin},
  \citenamefont {Qu},\ and\ \citenamefont {Rey}}]{Perlin2020}%
  \BibitemOpen
  \bibfield  {author} {\bibinfo {author} {\bibfnamefont {M.~A.}\ \bibnamefont
  {Perlin}}, \bibinfo {author} {\bibfnamefont {C.}~\bibnamefont {Qu}},\ and\
  \bibinfo {author} {\bibfnamefont {A.~M.}\ \bibnamefont {Rey}},\ }\href
  {https://doi.org/10.1103/PhysRevLett.125.223401} {\bibfield  {journal}
  {\bibinfo  {journal} {Phys. Rev. Lett.}\ }\textbf {\bibinfo {volume} {125}},\
  \bibinfo {pages} {223401} (\bibinfo {year} {2020})}\BibitemShut {NoStop}%
\bibitem [{\citenamefont {Petrov}\ and\ \citenamefont
  {Zakharova}(2022)}]{Petrov2022}%
  \BibitemOpen
  \bibfield  {author} {\bibinfo {author} {\bibfnamefont {A.}~\bibnamefont
  {Petrov}}\ and\ \bibinfo {author} {\bibfnamefont {A.}~\bibnamefont
  {Zakharova}},\ }\href {https://doi.org/10.1103/PhysRevA.105.L050801}
  {\bibfield  {journal} {\bibinfo  {journal} {Phys. Rev. A}\ }\textbf {\bibinfo
  {volume} {105}},\ \bibinfo {pages} {L050801} (\bibinfo {year}
  {2022})}\BibitemShut {NoStop}%
\bibitem [{\citenamefont {Loh}\ \emph {et~al.}(2013)\citenamefont {Loh},
  \citenamefont {Cossel}, \citenamefont {Grau}, \citenamefont {Ni},
  \citenamefont {Meyer}, \citenamefont {Bohn}, \citenamefont {Ye},\ and\
  \citenamefont {Cornell}}]{Loh2013}%
  \BibitemOpen
  \bibfield  {author} {\bibinfo {author} {\bibfnamefont {H.}~\bibnamefont
  {Loh}}, \bibinfo {author} {\bibfnamefont {K.~C.}\ \bibnamefont {Cossel}},
  \bibinfo {author} {\bibfnamefont {M.~C.}\ \bibnamefont {Grau}}, \bibinfo
  {author} {\bibfnamefont {K.-K.}\ \bibnamefont {Ni}}, \bibinfo {author}
  {\bibfnamefont {E.~R.}\ \bibnamefont {Meyer}}, \bibinfo {author}
  {\bibfnamefont {J.~L.}\ \bibnamefont {Bohn}}, \bibinfo {author}
  {\bibfnamefont {J.}~\bibnamefont {Ye}},\ and\ \bibinfo {author}
  {\bibfnamefont {E.~A.}\ \bibnamefont {Cornell}},\ }\href
  {https://doi.org/10.1126/science.1243683} {\bibfield  {journal} {\bibinfo
  {journal} {Science}\ }\textbf {\bibinfo {volume} {342}},\ \bibinfo {pages}
  {1220} (\bibinfo {year} {2013})}\BibitemShut {NoStop}%
\bibitem [{\citenamefont {Verma}\ \emph {et~al.}(2020)\citenamefont {Verma},
  \citenamefont {Jayich},\ and\ \citenamefont {Vutha}}]{Verma2020}%
  \BibitemOpen
  \bibfield  {author} {\bibinfo {author} {\bibfnamefont {M.}~\bibnamefont
  {Verma}}, \bibinfo {author} {\bibfnamefont {A.~M.}\ \bibnamefont {Jayich}},\
  and\ \bibinfo {author} {\bibfnamefont {A.~C.}\ \bibnamefont {Vutha}},\ }\href
  {https://doi.org/10.1103/PhysRevLett.125.153201} {\bibfield  {journal}
  {\bibinfo  {journal} {Phys. Rev. Lett.}\ }\textbf {\bibinfo {volume} {125}},\
  \bibinfo {pages} {153201} (\bibinfo {year} {2020})}\BibitemShut {NoStop}%
\bibitem [{\citenamefont {Monz}\ \emph {et~al.}(2009)\citenamefont {Monz},
  \citenamefont {Kim}, \citenamefont {Villar}, \citenamefont {Schindler},
  \citenamefont {Chwalla}, \citenamefont {Riebe}, \citenamefont {Roos},
  \citenamefont {H\"affner}, \citenamefont {H\"ansel}, \citenamefont
  {Hennrich},\ and\ \citenamefont {Blatt}}]{Monz2009}%
  \BibitemOpen
  \bibfield  {author} {\bibinfo {author} {\bibfnamefont {T.}~\bibnamefont
  {Monz}}, \bibinfo {author} {\bibfnamefont {K.}~\bibnamefont {Kim}}, \bibinfo
  {author} {\bibfnamefont {A.~S.}\ \bibnamefont {Villar}}, \bibinfo {author}
  {\bibfnamefont {P.}~\bibnamefont {Schindler}}, \bibinfo {author}
  {\bibfnamefont {M.}~\bibnamefont {Chwalla}}, \bibinfo {author} {\bibfnamefont
  {M.}~\bibnamefont {Riebe}}, \bibinfo {author} {\bibfnamefont {C.~F.}\
  \bibnamefont {Roos}}, \bibinfo {author} {\bibfnamefont {H.}~\bibnamefont
  {H\"affner}}, \bibinfo {author} {\bibfnamefont {W.}~\bibnamefont {H\"ansel}},
  \bibinfo {author} {\bibfnamefont {M.}~\bibnamefont {Hennrich}},\ and\
  \bibinfo {author} {\bibfnamefont {R.}~\bibnamefont {Blatt}},\ }\href
  {https://doi.org/10.1103/PhysRevLett.103.200503} {\bibfield  {journal}
  {\bibinfo  {journal} {Phys. Rev. Lett.}\ }\textbf {\bibinfo {volume} {103}},\
  \bibinfo {pages} {200503} (\bibinfo {year} {2009})}\BibitemShut {NoStop}%
\bibitem [{\citenamefont {S\o{}rensen}\ and\ \citenamefont
  {M\o{}lmer}(1999)}]{Sorensen1999}%
  \BibitemOpen
  \bibfield  {author} {\bibinfo {author} {\bibfnamefont {A.}~\bibnamefont
  {S\o{}rensen}}\ and\ \bibinfo {author} {\bibfnamefont {K.}~\bibnamefont
  {M\o{}lmer}},\ }\href {https://doi.org/10.1103/PhysRevLett.82.1971}
  {\bibfield  {journal} {\bibinfo  {journal} {Phys. Rev. Lett.}\ }\textbf
  {\bibinfo {volume} {82}},\ \bibinfo {pages} {1971} (\bibinfo {year}
  {1999})}\BibitemShut {NoStop}%
\bibitem [{\citenamefont {Briegel}\ and\ \citenamefont
  {Raussendorf}(2001)}]{Briegel2001}%
  \BibitemOpen
  \bibfield  {author} {\bibinfo {author} {\bibfnamefont {H.~J.}\ \bibnamefont
  {Briegel}}\ and\ \bibinfo {author} {\bibfnamefont {R.}~\bibnamefont
  {Raussendorf}},\ }\href {https://doi.org/10.1103/PhysRevLett.86.910}
  {\bibfield  {journal} {\bibinfo  {journal} {Phys. Rev. Lett.}\ }\textbf
  {\bibinfo {volume} {86}},\ \bibinfo {pages} {910} (\bibinfo {year}
  {2001})}\BibitemShut {NoStop}%
\bibitem [{\citenamefont {Verresen}\ \emph {et~al.}(2022)\citenamefont
  {Verresen}, \citenamefont {Tantivasadakarn},\ and\ \citenamefont
  {Vishwanath}}]{Verresen2022}%
  \BibitemOpen
  \bibfield  {author} {\bibinfo {author} {\bibfnamefont {R.}~\bibnamefont
  {Verresen}}, \bibinfo {author} {\bibfnamefont {N.}~\bibnamefont
  {Tantivasadakarn}},\ and\ \bibinfo {author} {\bibfnamefont {A.}~\bibnamefont
  {Vishwanath}},\ }\href {http://arxiv.org/abs/2112.03061} {\bibinfo {title}
  {Efficiently preparing {Schr\"odinger's} cat, fractons and non-abelian
  topological order in quantum devices}} (\bibinfo {year} {2022}),\ \Eprint
  {https://arxiv.org/abs/2112.03061 [cond-mat, physics:physics,
  physics:quant-ph]} {2112.03061 [cond-mat, physics:physics, physics:quant-ph]}
  \BibitemShut {NoStop}%
\bibitem [{\citenamefont {Lee}\ \emph {et~al.}(2022)\citenamefont {Lee},
  \citenamefont {Ji}, \citenamefont {Bi},\ and\ \citenamefont
  {Fisher}}]{Lee2022}%
  \BibitemOpen
  \bibfield  {author} {\bibinfo {author} {\bibfnamefont {J.~Y.}\ \bibnamefont
  {Lee}}, \bibinfo {author} {\bibfnamefont {W.}~\bibnamefont {Ji}}, \bibinfo
  {author} {\bibfnamefont {Z.}~\bibnamefont {Bi}},\ and\ \bibinfo {author}
  {\bibfnamefont {M.~P.~A.}\ \bibnamefont {Fisher}},\ }\href
  {http://arxiv.org/abs/2208.11699} {\bibinfo {title} {Decoding
  measurement-prepared quantum phases and transitions: from ising model to
  gauge theory, and beyond}} (\bibinfo {year} {2022}),\ \Eprint
  {https://arxiv.org/abs/2208.11699 [cond-mat, physics:quant-ph]} {2208.11699
  [cond-mat, physics:quant-ph]} \BibitemShut {NoStop}%
\bibitem [{\citenamefont {Tscherbul}\ \emph {et~al.}(2023)\citenamefont
  {Tscherbul}, \citenamefont {Ye},\ and\ \citenamefont {Rey}}]{Tscherbul2023}%
  \BibitemOpen
  \bibfield  {author} {\bibinfo {author} {\bibfnamefont {T.~V.}\ \bibnamefont
  {Tscherbul}}, \bibinfo {author} {\bibfnamefont {J.}~\bibnamefont {Ye}},\ and\
  \bibinfo {author} {\bibfnamefont {A.~M.}\ \bibnamefont {Rey}},\ }\href
  {https://doi.org/10.1103/PhysRevLett.130.143002} {\bibfield  {journal}
  {\bibinfo  {journal} {Phys. Rev. Lett.}\ }\textbf {\bibinfo {volume} {130}},\
  \bibinfo {pages} {143002} (\bibinfo {year} {2023})}\BibitemShut {NoStop}%
\bibitem [{\citenamefont {Baron}\ \emph {et~al.}(2017)\citenamefont {Baron},
  \citenamefont {Campbell}, \citenamefont {DeMille}, \citenamefont {Doyle},
  \citenamefont {Gabrielse}, \citenamefont {Gurevich}, \citenamefont {Hess},
  \citenamefont {Hutzler}, \citenamefont {Kirilov}, \citenamefont {Kozyryev},
  \citenamefont {O'Leary}, \citenamefont {Panda}, \citenamefont {Parsons},
  \citenamefont {Spaun}, \citenamefont {Vutha}, \citenamefont {West},\ and\
  \citenamefont {West}}]{Baron2017}%
  \BibitemOpen
  \bibfield  {author} {\bibinfo {author} {\bibfnamefont {J.}~\bibnamefont
  {Baron}}, \bibinfo {author} {\bibfnamefont {W.~C.}\ \bibnamefont {Campbell}},
  \bibinfo {author} {\bibfnamefont {D.}~\bibnamefont {DeMille}}, \bibinfo
  {author} {\bibfnamefont {J.~M.}\ \bibnamefont {Doyle}}, \bibinfo {author}
  {\bibfnamefont {G.}~\bibnamefont {Gabrielse}}, \bibinfo {author}
  {\bibfnamefont {Y.~V.}\ \bibnamefont {Gurevich}}, \bibinfo {author}
  {\bibfnamefont {P.~W.}\ \bibnamefont {Hess}}, \bibinfo {author}
  {\bibfnamefont {N.~R.}\ \bibnamefont {Hutzler}}, \bibinfo {author}
  {\bibfnamefont {E.}~\bibnamefont {Kirilov}}, \bibinfo {author} {\bibfnamefont
  {I.}~\bibnamefont {Kozyryev}}, \bibinfo {author} {\bibfnamefont {B.~R.}\
  \bibnamefont {O'Leary}}, \bibinfo {author} {\bibfnamefont {C.~D.}\
  \bibnamefont {Panda}}, \bibinfo {author} {\bibfnamefont {M.~F.}\ \bibnamefont
  {Parsons}}, \bibinfo {author} {\bibfnamefont {B.}~\bibnamefont {Spaun}},
  \bibinfo {author} {\bibfnamefont {A.~C.}\ \bibnamefont {Vutha}}, \bibinfo
  {author} {\bibfnamefont {A.~D.}\ \bibnamefont {West}},\ and\ \bibinfo
  {author} {\bibfnamefont {E.~P.}\ \bibnamefont {West}},\ }\href@noop {}
  {\bibfield  {journal} {\bibinfo  {journal} {New J. Phys.}\ }\textbf {\bibinfo
  {volume} {19}},\ \bibinfo {pages} {073029} (\bibinfo {year}
  {2017})}\BibitemShut {NoStop}%
\bibitem [{\citenamefont {Leibfried}\ \emph {et~al.}(2005)\citenamefont
  {Leibfried}, \citenamefont {Knill}, \citenamefont {Seidelin}, \citenamefont
  {Britton}, \citenamefont {Blakestad}, \citenamefont {Chiaverini},
  \citenamefont {Hume}, \citenamefont {Itano}, \citenamefont {Jost},
  \citenamefont {Langer}, \citenamefont {Ozeri}, \citenamefont {Reichle},\ and\
  \citenamefont {Wineland}}]{leibfried2005}%
  \BibitemOpen
  \bibfield  {author} {\bibinfo {author} {\bibfnamefont {D.}~\bibnamefont
  {Leibfried}}, \bibinfo {author} {\bibfnamefont {E.}~\bibnamefont {Knill}},
  \bibinfo {author} {\bibfnamefont {S.}~\bibnamefont {Seidelin}}, \bibinfo
  {author} {\bibfnamefont {J.}~\bibnamefont {Britton}}, \bibinfo {author}
  {\bibfnamefont {R.~B.}\ \bibnamefont {Blakestad}}, \bibinfo {author}
  {\bibfnamefont {J.}~\bibnamefont {Chiaverini}}, \bibinfo {author}
  {\bibfnamefont {D.~B.}\ \bibnamefont {Hume}}, \bibinfo {author}
  {\bibfnamefont {W.~M.}\ \bibnamefont {Itano}}, \bibinfo {author}
  {\bibfnamefont {J.~D.}\ \bibnamefont {Jost}}, \bibinfo {author}
  {\bibfnamefont {C.}~\bibnamefont {Langer}}, \bibinfo {author} {\bibfnamefont
  {R.}~\bibnamefont {Ozeri}}, \bibinfo {author} {\bibfnamefont
  {R.}~\bibnamefont {Reichle}},\ and\ \bibinfo {author} {\bibfnamefont {D.~J.}\
  \bibnamefont {Wineland}},\ }\href {https://doi.org/10.1038/nature04251}
  {\bibfield  {journal} {\bibinfo  {journal} {Nature}\ }\textbf {\bibinfo
  {volume} {438}},\ \bibinfo {pages} {639} (\bibinfo {year}
  {2005})}\BibitemShut {NoStop}%
\bibitem [{\citenamefont {Higgins}\ \emph {et~al.}(2021)\citenamefont
  {Higgins}, \citenamefont {Salim}, \citenamefont {Zhang}, \citenamefont
  {Parke}, \citenamefont {Pokorny},\ and\ \citenamefont
  {Hennrich}}]{Higgins2021}%
  \BibitemOpen
  \bibfield  {author} {\bibinfo {author} {\bibfnamefont {G.}~\bibnamefont
  {Higgins}}, \bibinfo {author} {\bibfnamefont {S.}~\bibnamefont {Salim}},
  \bibinfo {author} {\bibfnamefont {C.}~\bibnamefont {Zhang}}, \bibinfo
  {author} {\bibfnamefont {H.}~\bibnamefont {Parke}}, \bibinfo {author}
  {\bibfnamefont {F.}~\bibnamefont {Pokorny}},\ and\ \bibinfo {author}
  {\bibfnamefont {M.}~\bibnamefont {Hennrich}},\ }\href
  {https://doi.org/10.1088/1367-2630/ac3db6} {\bibfield  {journal} {\bibinfo
  {journal} {New Journal of Physics}\ }\textbf {\bibinfo {volume} {23}},\
  \bibinfo {pages} {123028} (\bibinfo {year} {2021})}\BibitemShut {NoStop}%
\bibitem [{\citenamefont {Nadlinger}\ \emph {et~al.}(2021)\citenamefont
  {Nadlinger}, \citenamefont {Drmota}, \citenamefont {Main}, \citenamefont
  {Nichol}, \citenamefont {Araneda}, \citenamefont {Srinivas}, \citenamefont
  {Stephenson}, \citenamefont {Ballance},\ and\ \citenamefont
  {Lucas}}]{Nadlinger2021}%
  \BibitemOpen
  \bibfield  {author} {\bibinfo {author} {\bibfnamefont {D.~P.}\ \bibnamefont
  {Nadlinger}}, \bibinfo {author} {\bibfnamefont {P.}~\bibnamefont {Drmota}},
  \bibinfo {author} {\bibfnamefont {D.}~\bibnamefont {Main}}, \bibinfo {author}
  {\bibfnamefont {B.~C.}\ \bibnamefont {Nichol}}, \bibinfo {author}
  {\bibfnamefont {G.}~\bibnamefont {Araneda}}, \bibinfo {author} {\bibfnamefont
  {R.}~\bibnamefont {Srinivas}}, \bibinfo {author} {\bibfnamefont {L.~J.}\
  \bibnamefont {Stephenson}}, \bibinfo {author} {\bibfnamefont {C.~J.}\
  \bibnamefont {Ballance}},\ and\ \bibinfo {author} {\bibfnamefont {D.~M.}\
  \bibnamefont {Lucas}},\ }\href {https://doi.org/10.48550/arXiv.2107.00056}
  {\bibinfo {title} {Micromotion minimisation by synchronous detection of
  parametrically excited motion}} (\bibinfo {year} {2021}),\ \Eprint
  {https://arxiv.org/abs/2107.00056 [physics, physics:quant-ph]} {2107.00056
  [physics, physics:quant-ph]} \BibitemShut {NoStop}%
\bibitem [{\citenamefont {Anderegg}\ \emph {et~al.}(2017)\citenamefont
  {Anderegg}, \citenamefont {Augenbraun}, \citenamefont {Chae}, \citenamefont
  {Hemmerling}, \citenamefont {Hutzler}, \citenamefont {Ravi}, \citenamefont
  {Collopy}, \citenamefont {Ye}, \citenamefont {Ketterle},\ and\ \citenamefont
  {Doyle}}]{Anderegg2017}%
  \BibitemOpen
  \bibfield  {author} {\bibinfo {author} {\bibfnamefont {L.}~\bibnamefont
  {Anderegg}}, \bibinfo {author} {\bibfnamefont {B.~L.}\ \bibnamefont
  {Augenbraun}}, \bibinfo {author} {\bibfnamefont {E.}~\bibnamefont {Chae}},
  \bibinfo {author} {\bibfnamefont {B.}~\bibnamefont {Hemmerling}}, \bibinfo
  {author} {\bibfnamefont {N.~R.}\ \bibnamefont {Hutzler}}, \bibinfo {author}
  {\bibfnamefont {A.}~\bibnamefont {Ravi}}, \bibinfo {author} {\bibfnamefont
  {A.}~\bibnamefont {Collopy}}, \bibinfo {author} {\bibfnamefont
  {J.}~\bibnamefont {Ye}}, \bibinfo {author} {\bibfnamefont {W.}~\bibnamefont
  {Ketterle}},\ and\ \bibinfo {author} {\bibfnamefont {J.~M.}\ \bibnamefont
  {Doyle}},\ }\href {https://doi.org/10.1103/PhysRevLett.119.103201} {\bibfield
   {journal} {\bibinfo  {journal} {Phys. Rev. Lett.}\ }\textbf {\bibinfo
  {volume} {119}},\ \bibinfo {pages} {103201} (\bibinfo {year}
  {2017})}\BibitemShut {NoStop}%
\bibitem [{\citenamefont {Arrowsmith-Kron}\ \emph {et~al.}(2023)\citenamefont
  {Arrowsmith-Kron}, \citenamefont {Athanasakis-Kaklamanakis}, \citenamefont
  {Au}, \citenamefont {Ballof}, \citenamefont {Berger}, \citenamefont
  {Borschevsky}, \citenamefont {Breier}, \citenamefont {Buchinger},
  \citenamefont {Budker}, \citenamefont {Caldwell}, \citenamefont {Charles},
  \citenamefont {Dattani}, \citenamefont {de~Groote}, \citenamefont {DeMille},
  \citenamefont {Dickel}, \citenamefont {Dobaczewski}, \citenamefont
  {Düllmann}, \citenamefont {Eliav}, \citenamefont {Engel}, \citenamefont
  {Fan}, \citenamefont {Flambaum}, \citenamefont {Flanagan}, \citenamefont
  {Gaiser}, \citenamefont {Ruiz}, \citenamefont {Gaul}, \citenamefont {Giesen},
  \citenamefont {Ginges}, \citenamefont {Gottberg}, \citenamefont {Gwinner},
  \citenamefont {Heinke}, \citenamefont {Hoekstra}, \citenamefont {Holt},
  \citenamefont {Hutzler}, \citenamefont {Jayich}, \citenamefont {Karthein},
  \citenamefont {Leach}, \citenamefont {Madison}, \citenamefont
  {Malbrunot-Ettenauer}, \citenamefont {Miyagi}, \citenamefont {Moore},
  \citenamefont {Moroch}, \citenamefont {Navrátil}, \citenamefont
  {Nazarewicz}, \citenamefont {Neyens}, \citenamefont {Norrgard}, \citenamefont
  {Nusgart}, \citenamefont {Pašteka}, \citenamefont {Petrov}, \citenamefont
  {Plass}, \citenamefont {Ready}, \citenamefont {Reiter}, \citenamefont
  {Reponen}, \citenamefont {Rothe}, \citenamefont {Safronova}, \citenamefont
  {Scheidenberger}, \citenamefont {Shindler}, \citenamefont {Singh},
  \citenamefont {Skripnikov}, \citenamefont {Titov}, \citenamefont {Udrescu},
  \citenamefont {Wilkins},\ and\ \citenamefont {Yang}}]{Arrowsmithkron2023}%
  \BibitemOpen
  \bibfield  {author} {\bibinfo {author} {\bibfnamefont {G.}~\bibnamefont
  {Arrowsmith-Kron}}, \bibinfo {author} {\bibfnamefont {M.}~\bibnamefont
  {Athanasakis-Kaklamanakis}}, \bibinfo {author} {\bibfnamefont
  {M.}~\bibnamefont {Au}}, \bibinfo {author} {\bibfnamefont {J.}~\bibnamefont
  {Ballof}}, \bibinfo {author} {\bibfnamefont {R.}~\bibnamefont {Berger}},
  \bibinfo {author} {\bibfnamefont {A.}~\bibnamefont {Borschevsky}}, \bibinfo
  {author} {\bibfnamefont {A.~A.}\ \bibnamefont {Breier}}, \bibinfo {author}
  {\bibfnamefont {F.}~\bibnamefont {Buchinger}}, \bibinfo {author}
  {\bibfnamefont {D.}~\bibnamefont {Budker}}, \bibinfo {author} {\bibfnamefont
  {L.}~\bibnamefont {Caldwell}}, \bibinfo {author} {\bibfnamefont
  {C.}~\bibnamefont {Charles}}, \bibinfo {author} {\bibfnamefont
  {N.}~\bibnamefont {Dattani}}, \bibinfo {author} {\bibfnamefont {R.~P.}\
  \bibnamefont {de~Groote}}, \bibinfo {author} {\bibfnamefont {D.}~\bibnamefont
  {DeMille}}, \bibinfo {author} {\bibfnamefont {T.}~\bibnamefont {Dickel}},
  \bibinfo {author} {\bibfnamefont {J.}~\bibnamefont {Dobaczewski}}, \bibinfo
  {author} {\bibfnamefont {C.~E.}\ \bibnamefont {Düllmann}}, \bibinfo {author}
  {\bibfnamefont {E.}~\bibnamefont {Eliav}}, \bibinfo {author} {\bibfnamefont
  {J.}~\bibnamefont {Engel}}, \bibinfo {author} {\bibfnamefont
  {M.}~\bibnamefont {Fan}}, \bibinfo {author} {\bibfnamefont {V.}~\bibnamefont
  {Flambaum}}, \bibinfo {author} {\bibfnamefont {K.~T.}\ \bibnamefont
  {Flanagan}}, \bibinfo {author} {\bibfnamefont {A.}~\bibnamefont {Gaiser}},
  \bibinfo {author} {\bibfnamefont {R.~G.}\ \bibnamefont {Ruiz}}, \bibinfo
  {author} {\bibfnamefont {K.}~\bibnamefont {Gaul}}, \bibinfo {author}
  {\bibfnamefont {T.~F.}\ \bibnamefont {Giesen}}, \bibinfo {author}
  {\bibfnamefont {J.}~\bibnamefont {Ginges}}, \bibinfo {author} {\bibfnamefont
  {A.}~\bibnamefont {Gottberg}}, \bibinfo {author} {\bibfnamefont
  {G.}~\bibnamefont {Gwinner}}, \bibinfo {author} {\bibfnamefont
  {R.}~\bibnamefont {Heinke}}, \bibinfo {author} {\bibfnamefont
  {S.}~\bibnamefont {Hoekstra}}, \bibinfo {author} {\bibfnamefont {J.~D.}\
  \bibnamefont {Holt}}, \bibinfo {author} {\bibfnamefont {N.~R.}\ \bibnamefont
  {Hutzler}}, \bibinfo {author} {\bibfnamefont {A.}~\bibnamefont {Jayich}},
  \bibinfo {author} {\bibfnamefont {J.}~\bibnamefont {Karthein}}, \bibinfo
  {author} {\bibfnamefont {K.~G.}\ \bibnamefont {Leach}}, \bibinfo {author}
  {\bibfnamefont {K.}~\bibnamefont {Madison}}, \bibinfo {author} {\bibfnamefont
  {S.}~\bibnamefont {Malbrunot-Ettenauer}}, \bibinfo {author} {\bibfnamefont
  {T.}~\bibnamefont {Miyagi}}, \bibinfo {author} {\bibfnamefont {I.~D.}\
  \bibnamefont {Moore}}, \bibinfo {author} {\bibfnamefont {S.}~\bibnamefont
  {Moroch}}, \bibinfo {author} {\bibfnamefont {P.}~\bibnamefont {Navrátil}},
  \bibinfo {author} {\bibfnamefont {W.}~\bibnamefont {Nazarewicz}}, \bibinfo
  {author} {\bibfnamefont {G.}~\bibnamefont {Neyens}}, \bibinfo {author}
  {\bibfnamefont {E.}~\bibnamefont {Norrgard}}, \bibinfo {author}
  {\bibfnamefont {N.}~\bibnamefont {Nusgart}}, \bibinfo {author} {\bibfnamefont
  {L.~F.}\ \bibnamefont {Pašteka}}, \bibinfo {author} {\bibfnamefont {A.~N.}\
  \bibnamefont {Petrov}}, \bibinfo {author} {\bibfnamefont {W.}~\bibnamefont
  {Plass}}, \bibinfo {author} {\bibfnamefont {R.~A.}\ \bibnamefont {Ready}},
  \bibinfo {author} {\bibfnamefont {M.~P.}\ \bibnamefont {Reiter}}, \bibinfo
  {author} {\bibfnamefont {M.}~\bibnamefont {Reponen}}, \bibinfo {author}
  {\bibfnamefont {S.}~\bibnamefont {Rothe}}, \bibinfo {author} {\bibfnamefont
  {M.}~\bibnamefont {Safronova}}, \bibinfo {author} {\bibfnamefont
  {C.}~\bibnamefont {Scheidenberger}}, \bibinfo {author} {\bibfnamefont
  {A.}~\bibnamefont {Shindler}}, \bibinfo {author} {\bibfnamefont {J.~T.}\
  \bibnamefont {Singh}}, \bibinfo {author} {\bibfnamefont {L.~V.}\ \bibnamefont
  {Skripnikov}}, \bibinfo {author} {\bibfnamefont {A.~V.}\ \bibnamefont
  {Titov}}, \bibinfo {author} {\bibfnamefont {S.-M.}\ \bibnamefont {Udrescu}},
  \bibinfo {author} {\bibfnamefont {S.~G.}\ \bibnamefont {Wilkins}},\ and\
  \bibinfo {author} {\bibfnamefont {X.}~\bibnamefont {Yang}},\ }\href@noop {}
  {\bibinfo {title} {Opportunities for fundamental physics research with
  radioactive molecules}} (\bibinfo {year} {2023}),\ \Eprint
  {https://arxiv.org/abs/2302.02165} {arXiv:2302.02165 [nucl-ex]} \BibitemShut
  {NoStop}%
\bibitem [{\citenamefont {Barrett}\ \emph {et~al.}(2003)\citenamefont
  {Barrett}, \citenamefont {DeMarco}, \citenamefont {Schaetz}, \citenamefont
  {Meyer}, \citenamefont {Leibfried}, \citenamefont {Britton}, \citenamefont
  {Chiaverini}, \citenamefont {Itano}, \citenamefont
  {Jelenkovi\ifmmode~\acute{c}\else \'{c}\fi{}}, \citenamefont {Jost},
  \citenamefont {Langer}, \citenamefont {Rosenband},\ and\ \citenamefont
  {Wineland}}]{Barrett2003}%
  \BibitemOpen
  \bibfield  {author} {\bibinfo {author} {\bibfnamefont {M.~D.}\ \bibnamefont
  {Barrett}}, \bibinfo {author} {\bibfnamefont {B.}~\bibnamefont {DeMarco}},
  \bibinfo {author} {\bibfnamefont {T.}~\bibnamefont {Schaetz}}, \bibinfo
  {author} {\bibfnamefont {V.}~\bibnamefont {Meyer}}, \bibinfo {author}
  {\bibfnamefont {D.}~\bibnamefont {Leibfried}}, \bibinfo {author}
  {\bibfnamefont {J.}~\bibnamefont {Britton}}, \bibinfo {author} {\bibfnamefont
  {J.}~\bibnamefont {Chiaverini}}, \bibinfo {author} {\bibfnamefont {W.~M.}\
  \bibnamefont {Itano}}, \bibinfo {author} {\bibfnamefont {B.}~\bibnamefont
  {Jelenkovi\ifmmode~\acute{c}\else \'{c}\fi{}}}, \bibinfo {author}
  {\bibfnamefont {J.~D.}\ \bibnamefont {Jost}}, \bibinfo {author}
  {\bibfnamefont {C.}~\bibnamefont {Langer}}, \bibinfo {author} {\bibfnamefont
  {T.}~\bibnamefont {Rosenband}},\ and\ \bibinfo {author} {\bibfnamefont
  {D.~J.}\ \bibnamefont {Wineland}},\ }\href
  {https://doi.org/10.1103/PhysRevA.68.042302} {\bibfield  {journal} {\bibinfo
  {journal} {Phys. Rev. A}\ }\textbf {\bibinfo {volume} {68}},\ \bibinfo
  {pages} {042302} (\bibinfo {year} {2003})}\BibitemShut {NoStop}%
\bibitem [{\citenamefont {Tan}\ \emph {et~al.}(2015)\citenamefont {Tan},
  \citenamefont {Gaebler}, \citenamefont {Lin}, \citenamefont {Wan},
  \citenamefont {Bowler}, \citenamefont {Leibfried},\ and\ \citenamefont
  {Wineland}}]{Tan2015}%
  \BibitemOpen
  \bibfield  {author} {\bibinfo {author} {\bibfnamefont {T.~R.}\ \bibnamefont
  {Tan}}, \bibinfo {author} {\bibfnamefont {J.~P.}\ \bibnamefont {Gaebler}},
  \bibinfo {author} {\bibfnamefont {Y.}~\bibnamefont {Lin}}, \bibinfo {author}
  {\bibfnamefont {Y.}~\bibnamefont {Wan}}, \bibinfo {author} {\bibfnamefont
  {R.}~\bibnamefont {Bowler}}, \bibinfo {author} {\bibfnamefont
  {D.}~\bibnamefont {Leibfried}},\ and\ \bibinfo {author} {\bibfnamefont
  {D.~J.}\ \bibnamefont {Wineland}},\ }\href
  {https://doi.org/10.1038/nature16186} {\bibfield  {journal} {\bibinfo
  {journal} {Nature}\ }\textbf {\bibinfo {volume} {528}},\ \bibinfo {pages}
  {380} (\bibinfo {year} {2015})}\BibitemShut {NoStop}%
\bibitem [{\citenamefont {Kielpinski}\ \emph {et~al.}(2002)\citenamefont
  {Kielpinski}, \citenamefont {Monroe},\ and\ \citenamefont
  {Wineland}}]{Kielpinski2002}%
  \BibitemOpen
  \bibfield  {author} {\bibinfo {author} {\bibfnamefont {D.}~\bibnamefont
  {Kielpinski}}, \bibinfo {author} {\bibfnamefont {C.}~\bibnamefont {Monroe}},\
  and\ \bibinfo {author} {\bibfnamefont {D.~J.}\ \bibnamefont {Wineland}},\
  }\href {https://doi.org/10.1038/nature00784} {\bibfield  {journal} {\bibinfo
  {journal} {Nature}\ }\textbf {\bibinfo {volume} {417}},\ \bibinfo {pages}
  {709} (\bibinfo {year} {2002})}\BibitemShut {NoStop}%
\bibitem [{\citenamefont {Keller}\ \emph {et~al.}(2015)\citenamefont {Keller},
  \citenamefont {Partner}, \citenamefont {Burgermeister},\ and\ \citenamefont
  {Mehlstäubler}}]{Keller2015}%
  \BibitemOpen
  \bibfield  {author} {\bibinfo {author} {\bibfnamefont {J.}~\bibnamefont
  {Keller}}, \bibinfo {author} {\bibfnamefont {H.~L.}\ \bibnamefont {Partner}},
  \bibinfo {author} {\bibfnamefont {T.}~\bibnamefont {Burgermeister}},\ and\
  \bibinfo {author} {\bibfnamefont {T.~E.}\ \bibnamefont {Mehlstäubler}},\
  }\href {https://doi.org/10.1063/1.4930037} {\bibfield  {journal} {\bibinfo
  {journal} {Journal of Applied Physics}\ }\textbf {\bibinfo {volume} {118}},\
  \bibinfo {pages} {104501} (\bibinfo {year} {2015})}\BibitemShut {NoStop}%
\end{thebibliography}%


\begin{thebibliography}{30}%
\makeatletter
\providecommand \@ifxundefined [1]{%
 \@ifx{#1\undefined}
}%
\providecommand \@ifnum [1]{%
 \ifnum #1\expandafter \@firstoftwo
 \else \expandafter \@secondoftwo
 \fi
}%
\providecommand \@ifx [1]{%
 \ifx #1\expandafter \@firstoftwo
 \else \expandafter \@secondoftwo
 \fi
}%
\providecommand \natexlab [1]{#1}%
\providecommand \enquote  [1]{``#1''}%
\providecommand \bibnamefont  [1]{#1}%
\providecommand \bibfnamefont [1]{#1}%
\providecommand \citenamefont [1]{#1}%
\providecommand \href@noop [0]{\@secondoftwo}%
\providecommand \href [0]{\begingroup \@sanitize@url \@href}%
\providecommand \@href[1]{\@@startlink{#1}\@@href}%
\providecommand \@@href[1]{\endgroup#1\@@endlink}%
\providecommand \@sanitize@url [0]{\catcode `\\12\catcode `\$12\catcode
  `\&12\catcode `\#12\catcode `\^12\catcode `\_12\catcode `\%12\relax}%
\providecommand \@@startlink[1]{}%
\providecommand \@@endlink[0]{}%
\providecommand \url  [0]{\begingroup\@sanitize@url \@url }%
\providecommand \@url [1]{\endgroup\@href {#1}{\urlprefix }}%
\providecommand \urlprefix  [0]{URL }%
\providecommand \Eprint [0]{\href }%
\providecommand \doibase [0]{https://doi.org/}%
\providecommand \selectlanguage [0]{\@gobble}%
\providecommand \bibinfo  [0]{\@secondoftwo}%
\providecommand \bibfield  [0]{\@secondoftwo}%
\providecommand \translation [1]{[#1]}%
\providecommand \BibitemOpen [0]{}%
\providecommand \bibitemStop [0]{}%
\providecommand \bibitemNoStop [0]{.\EOS\space}%
\providecommand \EOS [0]{\spacefactor3000\relax}%
\providecommand \BibitemShut  [1]{\csname bibitem#1\endcsname}%
\let\auto@bib@innerbib\@empty
\bibitem [{\citenamefont {Petrov}\ and\ \citenamefont
  {Zakharova}(2022)}]{Petrov2022}%
  \BibitemOpen
  \bibfield  {author} {\bibinfo {author} {\bibfnamefont {A.}~\bibnamefont
  {Petrov}}\ and\ \bibinfo {author} {\bibfnamefont {A.}~\bibnamefont
  {Zakharova}},\ }\bibfield  {title} {\bibinfo {title} {Sensitivity of the yboh
  molecule to $\mathcal{P}\mathcal{T}$-odd effects in an external electric
  field},\ }\href {https://doi.org/10.1103/PhysRevA.105.L050801} {\bibfield
  {journal} {\bibinfo  {journal} {Phys. Rev. A}\ }\textbf {\bibinfo {volume}
  {105}},\ \bibinfo {pages} {L050801} (\bibinfo {year} {2022})}\BibitemShut
  {NoStop}%
\bibitem [{\citenamefont {Borkowski}\ \emph {et~al.}(2023)\citenamefont
  {Borkowski}, \citenamefont {Reichsöllner}, \citenamefont {Thekkeppatt},
  \citenamefont {Barbé}, \citenamefont {van Roon}, \citenamefont {van
  Druten},\ and\ \citenamefont {Schreck}}]{Borkowski2023}%
  \BibitemOpen
  \bibfield  {author} {\bibinfo {author} {\bibfnamefont {M.}~\bibnamefont
  {Borkowski}}, \bibinfo {author} {\bibfnamefont {L.}~\bibnamefont
  {Reichsöllner}}, \bibinfo {author} {\bibfnamefont {P.}~\bibnamefont
  {Thekkeppatt}}, \bibinfo {author} {\bibfnamefont {V.}~\bibnamefont {Barbé}},
  \bibinfo {author} {\bibfnamefont {T.}~\bibnamefont {van Roon}}, \bibinfo
  {author} {\bibfnamefont {K.}~\bibnamefont {van Druten}},\ and\ \bibinfo
  {author} {\bibfnamefont {F.}~\bibnamefont {Schreck}},\ }\href
  {https://doi.org/10.48550/arXiv.2303.13682} {\bibinfo {title} {Active
  stabilization of kilogauss magnetic fields to the ppm level for
  magnetoassociation on ultranarrow feshbach resonances}} (\bibinfo {year}
  {2023}),\ \Eprint {https://arxiv.org/abs/2303.13682 [physics]} {2303.13682
  [physics]} \BibitemShut {NoStop}%
\bibitem [{\citenamefont {Yu}\ and\ \citenamefont
  {Hutzler}(2021)}]{Yu2021RaOCH3}%
  \BibitemOpen
  \bibfield  {author} {\bibinfo {author} {\bibfnamefont {P.}~\bibnamefont
  {Yu}}\ and\ \bibinfo {author} {\bibfnamefont {N.~R.}\ \bibnamefont
  {Hutzler}},\ }\bibfield  {title} {\bibinfo {title} {{Probing Fundamental
  Symmetries of Deformed Nuclei in Symmetric Top Molecules}},\ }\href
  {https://doi.org/10.1103/PhysRevLett.126.023003} {\bibfield  {journal}
  {\bibinfo  {journal} {Phys. Rev. Lett.}\ }\textbf {\bibinfo {volume} {126}},\
  \bibinfo {pages} {023003} (\bibinfo {year} {2021})}\BibitemShut {NoStop}%
\bibitem [{\citenamefont {Verma}\ \emph {et~al.}(2020)\citenamefont {Verma},
  \citenamefont {Jayich},\ and\ \citenamefont {Vutha}}]{Verma2020}%
  \BibitemOpen
  \bibfield  {author} {\bibinfo {author} {\bibfnamefont {M.}~\bibnamefont
  {Verma}}, \bibinfo {author} {\bibfnamefont {A.~M.}\ \bibnamefont {Jayich}},\
  and\ \bibinfo {author} {\bibfnamefont {A.~C.}\ \bibnamefont {Vutha}},\
  }\bibfield  {title} {\bibinfo {title} {Electron electric dipole moment
  searches using clock transitions in ultracold molecules},\ }\href
  {https://doi.org/10.1103/PhysRevLett.125.153201} {\bibfield  {journal}
  {\bibinfo  {journal} {Phys. Rev. Lett.}\ }\textbf {\bibinfo {volume} {125}},\
  \bibinfo {pages} {153201} (\bibinfo {year} {2020})}\BibitemShut {NoStop}%
\bibitem [{\citenamefont {Kozyryev}\ and\ \citenamefont
  {Hutzler}(2017)}]{Kozyryev2017b}%
  \BibitemOpen
  \bibfield  {author} {\bibinfo {author} {\bibfnamefont {I.}~\bibnamefont
  {Kozyryev}}\ and\ \bibinfo {author} {\bibfnamefont {N.~R.}\ \bibnamefont
  {Hutzler}},\ }\bibfield  {title} {\bibinfo {title} {{Precision measurement of
  time-reversal symmetry violation with laser-cooled polyatomic molecules}},\
  }\href {https://doi.org/10.1103/PhysRevLett.119.133002} {\bibfield  {journal}
  {\bibinfo  {journal} {Phys. Rev. Lett.}\ }\textbf {\bibinfo {volume} {119}},\
  \bibinfo {pages} {133002} (\bibinfo {year} {2017})}\BibitemShut {NoStop}%
\bibitem [{\citenamefont {Jadbabaie}\ \emph {et~al.}(2023)\citenamefont
  {Jadbabaie}, \citenamefont {Takahashi}, \citenamefont {Pilgram},
  \citenamefont {Conn}, \citenamefont {Zeng}, \citenamefont {Zhang},\ and\
  \citenamefont {Hutzler}}]{Jadbabaie2023}%
  \BibitemOpen
  \bibfield  {author} {\bibinfo {author} {\bibfnamefont {A.}~\bibnamefont
  {Jadbabaie}}, \bibinfo {author} {\bibfnamefont {Y.}~\bibnamefont
  {Takahashi}}, \bibinfo {author} {\bibfnamefont {N.~H.}\ \bibnamefont
  {Pilgram}}, \bibinfo {author} {\bibfnamefont {C.~J.}\ \bibnamefont {Conn}},
  \bibinfo {author} {\bibfnamefont {Y.}~\bibnamefont {Zeng}}, \bibinfo {author}
  {\bibfnamefont {C.}~\bibnamefont {Zhang}},\ and\ \bibinfo {author}
  {\bibfnamefont {N.~R.}\ \bibnamefont {Hutzler}},\ }\href
  {https://doi.org/10.48550/arXiv.2301.04124} {\bibinfo {title} {Characterizing
  the fundamental bending vibration of a linear polyatomic molecule for
  symmetry violation searches}} (\bibinfo {year} {2023})\BibitemShut {NoStop}%
\bibitem [{\citenamefont {Baron}\ \emph {et~al.}(2017)\citenamefont {Baron},
  \citenamefont {Campbell}, \citenamefont {DeMille}, \citenamefont {Doyle},
  \citenamefont {Gabrielse}, \citenamefont {Gurevich}, \citenamefont {Hess},
  \citenamefont {Hutzler}, \citenamefont {Kirilov}, \citenamefont {Kozyryev},
  \citenamefont {O'Leary}, \citenamefont {Panda}, \citenamefont {Parsons},
  \citenamefont {Spaun}, \citenamefont {Vutha}, \citenamefont {West},\ and\
  \citenamefont {West}}]{Baron2017}%
  \BibitemOpen
  \bibfield  {author} {\bibinfo {author} {\bibfnamefont {J.}~\bibnamefont
  {Baron}}, \bibinfo {author} {\bibfnamefont {W.~C.}\ \bibnamefont {Campbell}},
  \bibinfo {author} {\bibfnamefont {D.}~\bibnamefont {DeMille}}, \bibinfo
  {author} {\bibfnamefont {J.~M.}\ \bibnamefont {Doyle}}, \bibinfo {author}
  {\bibfnamefont {G.}~\bibnamefont {Gabrielse}}, \bibinfo {author}
  {\bibfnamefont {Y.~V.}\ \bibnamefont {Gurevich}}, \bibinfo {author}
  {\bibfnamefont {P.~W.}\ \bibnamefont {Hess}}, \bibinfo {author}
  {\bibfnamefont {N.~R.}\ \bibnamefont {Hutzler}}, \bibinfo {author}
  {\bibfnamefont {E.}~\bibnamefont {Kirilov}}, \bibinfo {author} {\bibfnamefont
  {I.}~\bibnamefont {Kozyryev}}, \bibinfo {author} {\bibfnamefont {B.~R.}\
  \bibnamefont {O'Leary}}, \bibinfo {author} {\bibfnamefont {C.~D.}\
  \bibnamefont {Panda}}, \bibinfo {author} {\bibfnamefont {M.~F.}\ \bibnamefont
  {Parsons}}, \bibinfo {author} {\bibfnamefont {B.}~\bibnamefont {Spaun}},
  \bibinfo {author} {\bibfnamefont {A.~C.}\ \bibnamefont {Vutha}}, \bibinfo
  {author} {\bibfnamefont {A.~D.}\ \bibnamefont {West}},\ and\ \bibinfo
  {author} {\bibfnamefont {E.~P.}\ \bibnamefont {West}},\ }\bibfield  {title}
  {\bibinfo {title} {{Methods, analysis, and the treatment of systematic errors
  for the electron electric dipole moment search in thorium monoxide}},\
  }\href@noop {} {\bibfield  {journal} {\bibinfo  {journal} {New J. Phys.}\
  }\textbf {\bibinfo {volume} {19}},\ \bibinfo {pages} {073029} (\bibinfo
  {year} {2017})}\BibitemShut {NoStop}%
\bibitem [{\citenamefont {Caldwell}\ and\ \citenamefont
  {Tarbutt}(2021)}]{Caldwell2021}%
  \BibitemOpen
  \bibfield  {author} {\bibinfo {author} {\bibfnamefont {L.}~\bibnamefont
  {Caldwell}}\ and\ \bibinfo {author} {\bibfnamefont {M.~R.}\ \bibnamefont
  {Tarbutt}},\ }\bibfield  {title} {\bibinfo {title} {General approach to
  state-dependent optical-tweezer traps for polar molecules},\ }\href
  {https://doi.org/10.1103/PhysRevResearch.3.013291} {\bibfield  {journal}
  {\bibinfo  {journal} {Phys. Rev. Res.}\ }\textbf {\bibinfo {volume} {3}},\
  \bibinfo {pages} {013291} (\bibinfo {year} {2021})}\BibitemShut {NoStop}%
\bibitem [{\citenamefont {Omran}\ \emph {et~al.}(2019)\citenamefont {Omran},
  \citenamefont {Levine}, \citenamefont {Keesling}, \citenamefont {Semeghini},
  \citenamefont {Wang}, \citenamefont {Ebadi}, \citenamefont {Bernien},
  \citenamefont {Zibrov}, \citenamefont {Pichler}, \citenamefont {Choi},
  \citenamefont {Cui}, \citenamefont {Rossignolo}, \citenamefont {Rembold},
  \citenamefont {Montangero}, \citenamefont {Calarco}, \citenamefont {Endres},
  \citenamefont {Greiner}, \citenamefont {Vuletić},\ and\ \citenamefont
  {Lukin}}]{Omran2019}%
  \BibitemOpen
  \bibfield  {author} {\bibinfo {author} {\bibfnamefont {A.}~\bibnamefont
  {Omran}}, \bibinfo {author} {\bibfnamefont {H.}~\bibnamefont {Levine}},
  \bibinfo {author} {\bibfnamefont {A.}~\bibnamefont {Keesling}}, \bibinfo
  {author} {\bibfnamefont {G.}~\bibnamefont {Semeghini}}, \bibinfo {author}
  {\bibfnamefont {T.~T.}\ \bibnamefont {Wang}}, \bibinfo {author}
  {\bibfnamefont {S.}~\bibnamefont {Ebadi}}, \bibinfo {author} {\bibfnamefont
  {H.}~\bibnamefont {Bernien}}, \bibinfo {author} {\bibfnamefont {A.~S.}\
  \bibnamefont {Zibrov}}, \bibinfo {author} {\bibfnamefont {H.}~\bibnamefont
  {Pichler}}, \bibinfo {author} {\bibfnamefont {S.}~\bibnamefont {Choi}},
  \bibinfo {author} {\bibfnamefont {J.}~\bibnamefont {Cui}}, \bibinfo {author}
  {\bibfnamefont {M.}~\bibnamefont {Rossignolo}}, \bibinfo {author}
  {\bibfnamefont {P.}~\bibnamefont {Rembold}}, \bibinfo {author} {\bibfnamefont
  {S.}~\bibnamefont {Montangero}}, \bibinfo {author} {\bibfnamefont
  {T.}~\bibnamefont {Calarco}}, \bibinfo {author} {\bibfnamefont
  {M.}~\bibnamefont {Endres}}, \bibinfo {author} {\bibfnamefont
  {M.}~\bibnamefont {Greiner}}, \bibinfo {author} {\bibfnamefont
  {V.}~\bibnamefont {Vuletić}},\ and\ \bibinfo {author} {\bibfnamefont
  {M.~D.}\ \bibnamefont {Lukin}},\ }\bibfield  {title} {\bibinfo {title}
  {Generation and manipulation of schrödinger cat states in rydberg atom
  arrays},\ }\href {https://doi.org/10.1126/science.aax9743} {\bibfield
  {journal} {\bibinfo  {journal} {Science}\ }\textbf {\bibinfo {volume}
  {365}},\ \bibinfo {pages} {570} (\bibinfo {year} {2019})}\BibitemShut
  {NoStop}%
\bibitem [{\citenamefont {Briegel}\ and\ \citenamefont
  {Raussendorf}(2001)}]{Briegel2001}%
  \BibitemOpen
  \bibfield  {author} {\bibinfo {author} {\bibfnamefont {H.~J.}\ \bibnamefont
  {Briegel}}\ and\ \bibinfo {author} {\bibfnamefont {R.}~\bibnamefont
  {Raussendorf}},\ }\bibfield  {title} {\bibinfo {title} {Persistent
  entanglement in arrays of interacting particles},\ }\href
  {https://doi.org/10.1103/PhysRevLett.86.910} {\bibfield  {journal} {\bibinfo
  {journal} {Phys. Rev. Lett.}\ }\textbf {\bibinfo {volume} {86}},\ \bibinfo
  {pages} {910} (\bibinfo {year} {2001})}\BibitemShut {NoStop}%
\bibitem [{\citenamefont {Verresen}\ \emph {et~al.}(2022)\citenamefont
  {Verresen}, \citenamefont {Tantivasadakarn},\ and\ \citenamefont
  {Vishwanath}}]{Verresen2022}%
  \BibitemOpen
  \bibfield  {author} {\bibinfo {author} {\bibfnamefont {R.}~\bibnamefont
  {Verresen}}, \bibinfo {author} {\bibfnamefont {N.}~\bibnamefont
  {Tantivasadakarn}},\ and\ \bibinfo {author} {\bibfnamefont {A.}~\bibnamefont
  {Vishwanath}},\ }\href {http://arxiv.org/abs/2112.03061} {\bibinfo {title}
  {Efficiently preparing {Schr\"odinger's} cat, fractons and non-abelian
  topological order in quantum devices}} (\bibinfo {year} {2022}),\ \Eprint
  {https://arxiv.org/abs/2112.03061 [cond-mat, physics:physics,
  physics:quant-ph]} {2112.03061 [cond-mat, physics:physics, physics:quant-ph]}
  \BibitemShut {NoStop}%
\bibitem [{\citenamefont {Lee}\ \emph {et~al.}(2022)\citenamefont {Lee},
  \citenamefont {Ji}, \citenamefont {Bi},\ and\ \citenamefont
  {Fisher}}]{Lee2022}%
  \BibitemOpen
  \bibfield  {author} {\bibinfo {author} {\bibfnamefont {J.~Y.}\ \bibnamefont
  {Lee}}, \bibinfo {author} {\bibfnamefont {W.}~\bibnamefont {Ji}}, \bibinfo
  {author} {\bibfnamefont {Z.}~\bibnamefont {Bi}},\ and\ \bibinfo {author}
  {\bibfnamefont {M.~P.~A.}\ \bibnamefont {Fisher}},\ }\href
  {http://arxiv.org/abs/2208.11699} {\bibinfo {title} {Decoding
  measurement-prepared quantum phases and transitions: from ising model to
  gauge theory, and beyond}} (\bibinfo {year} {2022}),\ \Eprint
  {https://arxiv.org/abs/2208.11699 [cond-mat, physics:quant-ph]} {2208.11699
  [cond-mat, physics:quant-ph]} \BibitemShut {NoStop}%
\bibitem [{\citenamefont {Tscherbul}\ \emph {et~al.}(2023)\citenamefont
  {Tscherbul}, \citenamefont {Ye},\ and\ \citenamefont {Rey}}]{Tscherbul2023}%
  \BibitemOpen
  \bibfield  {author} {\bibinfo {author} {\bibfnamefont {T.~V.}\ \bibnamefont
  {Tscherbul}}, \bibinfo {author} {\bibfnamefont {J.}~\bibnamefont {Ye}},\ and\
  \bibinfo {author} {\bibfnamefont {A.~M.}\ \bibnamefont {Rey}},\ }\bibfield
  {title} {\bibinfo {title} {Robust nuclear spin entanglement via dipolar
  interactions in polar molecules},\ }\href
  {https://doi.org/10.1103/PhysRevLett.130.143002} {\bibfield  {journal}
  {\bibinfo  {journal} {Phys. Rev. Lett.}\ }\textbf {\bibinfo {volume} {130}},\
  \bibinfo {pages} {143002} (\bibinfo {year} {2023})}\BibitemShut {NoStop}%
\bibitem [{\citenamefont {Sunaga}\ \emph {et~al.}(2019)\citenamefont {Sunaga},
  \citenamefont {Abe}, \citenamefont {Hada},\ and\ \citenamefont
  {Das}}]{sunaga_merits_2019}%
  \BibitemOpen
  \bibfield  {author} {\bibinfo {author} {\bibfnamefont {A.}~\bibnamefont
  {Sunaga}}, \bibinfo {author} {\bibfnamefont {M.}~\bibnamefont {Abe}},
  \bibinfo {author} {\bibfnamefont {M.}~\bibnamefont {Hada}},\ and\ \bibinfo
  {author} {\bibfnamefont {B.~P.}\ \bibnamefont {Das}},\ }\bibfield  {title}
  {\bibinfo {title} {Merits of heavy-heavy diatomic molecules for electron
  electric-dipole-moment searches},\ }\href
  {https://doi.org/10.1103/PhysRevA.99.062506} {\bibfield  {journal} {\bibinfo
  {journal} {Physical Review A}\ }\textbf {\bibinfo {volume} {99}},\ \bibinfo
  {pages} {062506} (\bibinfo {year} {2019})}\BibitemShut {NoStop}%
\bibitem [{\citenamefont {Fleig}\ and\ \citenamefont
  {DeMille}(2021)}]{fleig_theoretical_2021}%
  \BibitemOpen
  \bibfield  {author} {\bibinfo {author} {\bibfnamefont {T.}~\bibnamefont
  {Fleig}}\ and\ \bibinfo {author} {\bibfnamefont {D.}~\bibnamefont
  {DeMille}},\ }\bibfield  {title} {\bibinfo {title} {Theoretical aspects of
  radium-containing molecules amenable to assembly from laser-cooled atoms for
  new physics searches},\ }\href {https://doi.org/10.1088/1367-2630/ac3619}
  {\bibfield  {journal} {\bibinfo  {journal} {New Journal of Physics}\ }\textbf
  {\bibinfo {volume} {23}},\ \bibinfo {pages} {113039} (\bibinfo {year}
  {2021})}\BibitemShut {NoStop}%
\bibitem [{\citenamefont {Śmiałkowski}\ and\ \citenamefont
  {Tomza}(2021)}]{smialkowski_highly_2021}%
  \BibitemOpen
  \bibfield  {author} {\bibinfo {author} {\bibfnamefont {M.}~\bibnamefont
  {Śmiałkowski}}\ and\ \bibinfo {author} {\bibfnamefont {M.}~\bibnamefont
  {Tomza}},\ }\bibfield  {title} {\bibinfo {title} {Highly polar molecules
  consisting of a copper or silver atom interacting with an alkali-metal or
  alkaline-earth-metal atom},\ }\href
  {https://doi.org/10.1103/PhysRevA.103.022802} {\bibfield  {journal} {\bibinfo
   {journal} {Physical Review A}\ }\textbf {\bibinfo {volume} {103}},\ \bibinfo
  {pages} {022802} (\bibinfo {year} {2021})}\BibitemShut {NoStop}%
\bibitem [{\citenamefont {Andreev}\ \emph {et~al.}(2018)\citenamefont
  {Andreev}, \citenamefont {Ang}, \citenamefont {DeMille}, \citenamefont
  {Doyle}, \citenamefont {Gabrielse}, \citenamefont {Haefner}, \citenamefont
  {Hutzler}, \citenamefont {Lasner}, \citenamefont {Meisenhelder},
  \citenamefont {O’Leary}, \citenamefont {Panda}, \citenamefont {West},
  \citenamefont {West},\ and\ \citenamefont {Wu}}]{Andreev2018}%
  \BibitemOpen
  \bibfield  {author} {\bibinfo {author} {\bibfnamefont {V.}~\bibnamefont
  {Andreev}}, \bibinfo {author} {\bibfnamefont {D.~G.}\ \bibnamefont {Ang}},
  \bibinfo {author} {\bibfnamefont {D.}~\bibnamefont {DeMille}}, \bibinfo
  {author} {\bibfnamefont {J.~M.}\ \bibnamefont {Doyle}}, \bibinfo {author}
  {\bibfnamefont {G.}~\bibnamefont {Gabrielse}}, \bibinfo {author}
  {\bibfnamefont {J.}~\bibnamefont {Haefner}}, \bibinfo {author} {\bibfnamefont
  {N.~R.}\ \bibnamefont {Hutzler}}, \bibinfo {author} {\bibfnamefont
  {Z.}~\bibnamefont {Lasner}}, \bibinfo {author} {\bibfnamefont
  {C.}~\bibnamefont {Meisenhelder}}, \bibinfo {author} {\bibfnamefont {B.~R.}\
  \bibnamefont {O’Leary}}, \bibinfo {author} {\bibfnamefont {C.~D.}\
  \bibnamefont {Panda}}, \bibinfo {author} {\bibfnamefont {A.~D.}\ \bibnamefont
  {West}}, \bibinfo {author} {\bibfnamefont {E.~P.}\ \bibnamefont {West}},\
  and\ \bibinfo {author} {\bibfnamefont {X.}~\bibnamefont {Wu}},\ }\bibfield
  {title} {\bibinfo {title} {{Improved limit on the electric dipole moment of
  the electron}},\ }\href {https://doi.org/10.1038/s41586-018-0599-8}
  {\bibfield  {journal} {\bibinfo  {journal} {Nature}\ }\textbf {\bibinfo
  {volume} {562}},\ \bibinfo {pages} {355} (\bibinfo {year}
  {2018})}\BibitemShut {NoStop}%
\bibitem [{\citenamefont {Roussy}\ \emph {et~al.}(2023)\citenamefont {Roussy},
  \citenamefont {Caldwell}, \citenamefont {Wright}, \citenamefont {Cairncross},
  \citenamefont {Shagam}, \citenamefont {Ng}, \citenamefont {Schlossberger},
  \citenamefont {Park}, \citenamefont {Wang}, \citenamefont {Ye},\ and\
  \citenamefont {Cornell}}]{Roussy2023}%
  \BibitemOpen
  \bibfield  {author} {\bibinfo {author} {\bibfnamefont {T.~S.}\ \bibnamefont
  {Roussy}}, \bibinfo {author} {\bibfnamefont {L.}~\bibnamefont {Caldwell}},
  \bibinfo {author} {\bibfnamefont {T.}~\bibnamefont {Wright}}, \bibinfo
  {author} {\bibfnamefont {W.~B.}\ \bibnamefont {Cairncross}}, \bibinfo
  {author} {\bibfnamefont {Y.}~\bibnamefont {Shagam}}, \bibinfo {author}
  {\bibfnamefont {K.~B.}\ \bibnamefont {Ng}}, \bibinfo {author} {\bibfnamefont
  {N.}~\bibnamefont {Schlossberger}}, \bibinfo {author} {\bibfnamefont {S.~Y.}\
  \bibnamefont {Park}}, \bibinfo {author} {\bibfnamefont {A.}~\bibnamefont
  {Wang}}, \bibinfo {author} {\bibfnamefont {J.}~\bibnamefont {Ye}},\ and\
  \bibinfo {author} {\bibfnamefont {E.~A.}\ \bibnamefont {Cornell}},\ }\href
  {https://doi.org/10.1126/science.adg4084} {\bibinfo {title} {An improved
  bound on the electron’s electric dipole moment}} (\bibinfo {year}
  {2023})\BibitemShut {NoStop}%
\bibitem [{\citenamefont {Augenbraun}\ \emph
  {et~al.}(2020{\natexlab{a}})\citenamefont {Augenbraun}, \citenamefont
  {Doyle}, \citenamefont {Zelevinsky},\ and\ \citenamefont
  {Kozyryev}}]{Augenbraun2020b}%
  \BibitemOpen
  \bibfield  {author} {\bibinfo {author} {\bibfnamefont {B.~L.}\ \bibnamefont
  {Augenbraun}}, \bibinfo {author} {\bibfnamefont {J.~M.}\ \bibnamefont
  {Doyle}}, \bibinfo {author} {\bibfnamefont {T.}~\bibnamefont {Zelevinsky}},\
  and\ \bibinfo {author} {\bibfnamefont {I.}~\bibnamefont {Kozyryev}},\
  }\bibfield  {title} {\bibinfo {title} {Molecular asymmetry and optical
  cycling: laser cooling asymmetric top molecules},\ }\href
  {https://doi.org/10.1103/PhysRevX.10.031022} {\bibfield  {journal} {\bibinfo
  {journal} {Phys. Rev. X}\ }\textbf {\bibinfo {volume} {10}},\ \bibinfo
  {pages} {031022} (\bibinfo {year} {2020}{\natexlab{a}})}\BibitemShut
  {NoStop}%
\bibitem [{\citenamefont {Augenbraun}\ \emph
  {et~al.}(2020{\natexlab{b}})\citenamefont {Augenbraun}, \citenamefont
  {Lasner}, \citenamefont {Frenett}, \citenamefont {Sawaoka}, \citenamefont
  {Le}, \citenamefont {Doyle},\ and\ \citenamefont
  {Steimle}}]{Augenbraun2020c}%
  \BibitemOpen
  \bibfield  {author} {\bibinfo {author} {\bibfnamefont {B.~L.}\ \bibnamefont
  {Augenbraun}}, \bibinfo {author} {\bibfnamefont {Z.~D.}\ \bibnamefont
  {Lasner}}, \bibinfo {author} {\bibfnamefont {A.}~\bibnamefont {Frenett}},
  \bibinfo {author} {\bibfnamefont {H.}~\bibnamefont {Sawaoka}}, \bibinfo
  {author} {\bibfnamefont {A.~T.}\ \bibnamefont {Le}}, \bibinfo {author}
  {\bibfnamefont {J.~M.}\ \bibnamefont {Doyle}},\ and\ \bibinfo {author}
  {\bibfnamefont {T.~C.}\ \bibnamefont {Steimle}},\ }\bibfield  {title}
  {\bibinfo {title} {Observation and laser spectroscopy of ytterbium
  monomethoxide, {YbOCH$_3$}},\ }\href@noop {} {\bibfield  {journal} {\bibinfo
  {journal} {Phys. Rev. A}\ }\textbf {\bibinfo {volume} {103}},\ \bibinfo
  {pages} {022814} (\bibinfo {year} {2020}{\natexlab{b}})}\BibitemShut
  {NoStop}%
\bibitem [{\citenamefont {Hudson}\ \emph {et~al.}(2002)\citenamefont {Hudson},
  \citenamefont {Sauer}, \citenamefont {Tarbutt},\ and\ \citenamefont
  {Hinds}}]{Hudson2002}%
  \BibitemOpen
  \bibfield  {author} {\bibinfo {author} {\bibfnamefont {J.~J.}\ \bibnamefont
  {Hudson}}, \bibinfo {author} {\bibfnamefont {B.~E.}\ \bibnamefont {Sauer}},
  \bibinfo {author} {\bibfnamefont {M.~R.}\ \bibnamefont {Tarbutt}},\ and\
  \bibinfo {author} {\bibfnamefont {E.~A.}\ \bibnamefont {Hinds}},\ }\bibfield
  {title} {\bibinfo {title} {Measurement of the electron electric dipole moment
  using ybf molecules},\ }\href@noop {} {\bibfield  {journal} {\bibinfo
  {journal} {Phys. Rev. Lett.}\ }\textbf {\bibinfo {volume} {89}},\ \bibinfo
  {pages} {023003} (\bibinfo {year} {2002})}\BibitemShut {NoStop}%
\bibitem [{\citenamefont {Lim}\ \emph {et~al.}(2017)\citenamefont {Lim},
  \citenamefont {Almond}, \citenamefont {Tarbutt}, \citenamefont {Nguyen},\
  and\ \citenamefont {Steimle}}]{Lim2017}%
  \BibitemOpen
  \bibfield  {author} {\bibinfo {author} {\bibfnamefont {J.}~\bibnamefont
  {Lim}}, \bibinfo {author} {\bibfnamefont {J.~R.}\ \bibnamefont {Almond}},
  \bibinfo {author} {\bibfnamefont {M.~R.}\ \bibnamefont {Tarbutt}}, \bibinfo
  {author} {\bibfnamefont {D.~T.}\ \bibnamefont {Nguyen}},\ and\ \bibinfo
  {author} {\bibfnamefont {T.~C.}\ \bibnamefont {Steimle}},\ }\bibfield
  {title} {\bibinfo {title} {{The [557]-X$^2\Sigma^{+}$ and
  [561]-X$^2\Sigma^{+}$ bands of ytterbium fluoride, $^{174}$YbF}},\ }\href
  {https://doi.org/10.1016/j.jms.2017.06.007} {\bibfield  {journal} {\bibinfo
  {journal} {J. Mol. Spectrosc.}\ }\textbf {\bibinfo {volume} {338}},\ \bibinfo
  {pages} {81} (\bibinfo {year} {2017})}\BibitemShut {NoStop}%
\bibitem [{\citenamefont {Aggarwal}\ \emph {et~al.}(2018)\citenamefont
  {Aggarwal}, \citenamefont {Bethlem}, \citenamefont {Borschevsky},
  \citenamefont {Denis}, \citenamefont {Esajas}, \citenamefont {Haase},
  \citenamefont {Hao}, \citenamefont {Hoekstra}, \citenamefont {Jungmann},
  \citenamefont {Meijknecht}, \citenamefont {Mooij}, \citenamefont
  {Timmermans}, \citenamefont {Ubachs}, \citenamefont {Willmann},\ and\
  \citenamefont {Zapara}}]{Aggarwal2018}%
  \BibitemOpen
  \bibfield  {author} {\bibinfo {author} {\bibfnamefont {P.}~\bibnamefont
  {Aggarwal}}, \bibinfo {author} {\bibfnamefont {H.~L.}\ \bibnamefont
  {Bethlem}}, \bibinfo {author} {\bibfnamefont {A.}~\bibnamefont
  {Borschevsky}}, \bibinfo {author} {\bibfnamefont {M.}~\bibnamefont {Denis}},
  \bibinfo {author} {\bibfnamefont {K.}~\bibnamefont {Esajas}}, \bibinfo
  {author} {\bibfnamefont {P.~A.~B.}\ \bibnamefont {Haase}}, \bibinfo {author}
  {\bibfnamefont {Y.}~\bibnamefont {Hao}}, \bibinfo {author} {\bibfnamefont
  {S.}~\bibnamefont {Hoekstra}}, \bibinfo {author} {\bibfnamefont
  {K.}~\bibnamefont {Jungmann}}, \bibinfo {author} {\bibfnamefont {T.~B.}\
  \bibnamefont {Meijknecht}}, \bibinfo {author} {\bibfnamefont {M.~C.}\
  \bibnamefont {Mooij}}, \bibinfo {author} {\bibfnamefont {R.~G.~E.}\
  \bibnamefont {Timmermans}}, \bibinfo {author} {\bibfnamefont
  {W.}~\bibnamefont {Ubachs}}, \bibinfo {author} {\bibfnamefont
  {L.}~\bibnamefont {Willmann}},\ and\ \bibinfo {author} {\bibfnamefont
  {A.}~\bibnamefont {Zapara}},\ }\bibfield  {title} {\bibinfo {title}
  {{Measuring the electric dipole moment of the electron in BaF}},\ }\href
  {https://doi.org/10.1140/epjd/e2018-90192-9} {\bibfield  {journal} {\bibinfo
  {journal} {Eur. Phys. J. D}\ }\textbf {\bibinfo {volume} {72}},\ \bibinfo
  {pages} {197} (\bibinfo {year} {2018})}\BibitemShut {NoStop}%
\bibitem [{\citenamefont {Steimle}\ \emph {et~al.}(2011)\citenamefont
  {Steimle}, \citenamefont {Frey}, \citenamefont {Le}, \citenamefont {DeMille},
  \citenamefont {Rahmlow},\ and\ \citenamefont
  {Linton}}]{steimle_molecular-beam_2011}%
  \BibitemOpen
  \bibfield  {author} {\bibinfo {author} {\bibfnamefont {T.~C.}\ \bibnamefont
  {Steimle}}, \bibinfo {author} {\bibfnamefont {S.}~\bibnamefont {Frey}},
  \bibinfo {author} {\bibfnamefont {A.}~\bibnamefont {Le}}, \bibinfo {author}
  {\bibfnamefont {D.}~\bibnamefont {DeMille}}, \bibinfo {author} {\bibfnamefont
  {D.~A.}\ \bibnamefont {Rahmlow}},\ and\ \bibinfo {author} {\bibfnamefont
  {C.}~\bibnamefont {Linton}},\ }\bibfield  {title} {\bibinfo {title}
  {Molecular-beam optical {Stark} and {Zeeman} study of the {A}$^2\pi$- {X}
  $^2\sigma^+$(0,0) band system of {BaF}},\ }\href
  {https://doi.org/10.1103/PhysRevA.84.012508} {\bibfield  {journal} {\bibinfo
  {journal} {Physical Review A}\ }\textbf {\bibinfo {volume} {84}},\ \bibinfo
  {pages} {012508} (\bibinfo {year} {2011})}\BibitemShut {NoStop}%
\bibitem [{\citenamefont {Isaev}\ \emph {et~al.}(2010)\citenamefont {Isaev},
  \citenamefont {Hoekstra},\ and\ \citenamefont {Berger}}]{Isaev2010}%
  \BibitemOpen
  \bibfield  {author} {\bibinfo {author} {\bibfnamefont {T.~A.}\ \bibnamefont
  {Isaev}}, \bibinfo {author} {\bibfnamefont {S.}~\bibnamefont {Hoekstra}},\
  and\ \bibinfo {author} {\bibfnamefont {R.}~\bibnamefont {Berger}},\
  }\bibfield  {title} {\bibinfo {title} {Laser-cooled raf as a promising
  candidate to measure molecular parity violation},\ }\href
  {https://doi.org/10.1103/PhysRevA.82.052521} {\bibfield  {journal} {\bibinfo
  {journal} {Phys. Rev. A}\ }\textbf {\bibinfo {volume} {82}},\ \bibinfo
  {pages} {052521} (\bibinfo {year} {2010})}\BibitemShut {NoStop}%
\bibitem [{\citenamefont {Ruiz}\ \emph {et~al.}(2020)\citenamefont {Ruiz},
  \citenamefont {Berger}, \citenamefont {Billowes}, \citenamefont {Binnersley},
  \citenamefont {Bissell}, \citenamefont {Breier}, \citenamefont {Brinson},
  \citenamefont {Chrysalidis}, \citenamefont {Cocolios}, \citenamefont
  {Cooper}, \citenamefont {Flanagan}, \citenamefont {Giesen}, \citenamefont
  {de~Groote}, \citenamefont {Franchoo}, \citenamefont {Gustafsson},
  \citenamefont {Isaev}, \citenamefont {Koszorus}, \citenamefont {Neyens},
  \citenamefont {Perrett}, \citenamefont {Ricketts}, \citenamefont {Rothe},
  \citenamefont {Schweikhard}, \citenamefont {Vernon}, \citenamefont {Wendt},
  \citenamefont {Wienholtz}, \citenamefont {Wilkins},\ and\ \citenamefont
  {Yang}}]{garcia_ruiz_spectroscopy_2020}%
  \BibitemOpen
  \bibfield  {author} {\bibinfo {author} {\bibfnamefont {R.~F.~G.}\
  \bibnamefont {Ruiz}}, \bibinfo {author} {\bibfnamefont {R.}~\bibnamefont
  {Berger}}, \bibinfo {author} {\bibfnamefont {J.}~\bibnamefont {Billowes}},
  \bibinfo {author} {\bibfnamefont {C.~L.}\ \bibnamefont {Binnersley}},
  \bibinfo {author} {\bibfnamefont {M.~L.}\ \bibnamefont {Bissell}}, \bibinfo
  {author} {\bibfnamefont {A.~A.}\ \bibnamefont {Breier}}, \bibinfo {author}
  {\bibfnamefont {A.~J.}\ \bibnamefont {Brinson}}, \bibinfo {author}
  {\bibfnamefont {K.}~\bibnamefont {Chrysalidis}}, \bibinfo {author}
  {\bibfnamefont {T.~E.}\ \bibnamefont {Cocolios}}, \bibinfo {author}
  {\bibfnamefont {B.~S.}\ \bibnamefont {Cooper}}, \bibinfo {author}
  {\bibfnamefont {K.~T.}\ \bibnamefont {Flanagan}}, \bibinfo {author}
  {\bibfnamefont {T.~F.}\ \bibnamefont {Giesen}}, \bibinfo {author}
  {\bibfnamefont {R.~P.}\ \bibnamefont {de~Groote}}, \bibinfo {author}
  {\bibfnamefont {S.}~\bibnamefont {Franchoo}}, \bibinfo {author}
  {\bibfnamefont {F.~P.}\ \bibnamefont {Gustafsson}}, \bibinfo {author}
  {\bibfnamefont {T.~A.}\ \bibnamefont {Isaev}}, \bibinfo {author}
  {\bibfnamefont {A.}~\bibnamefont {Koszorus}}, \bibinfo {author}
  {\bibfnamefont {G.}~\bibnamefont {Neyens}}, \bibinfo {author} {\bibfnamefont
  {H.~A.}\ \bibnamefont {Perrett}}, \bibinfo {author} {\bibfnamefont {C.~M.}\
  \bibnamefont {Ricketts}}, \bibinfo {author} {\bibfnamefont {S.}~\bibnamefont
  {Rothe}}, \bibinfo {author} {\bibfnamefont {L.}~\bibnamefont {Schweikhard}},
  \bibinfo {author} {\bibfnamefont {A.~R.}\ \bibnamefont {Vernon}}, \bibinfo
  {author} {\bibfnamefont {K.~D.~A.}\ \bibnamefont {Wendt}}, \bibinfo {author}
  {\bibfnamefont {F.}~\bibnamefont {Wienholtz}}, \bibinfo {author}
  {\bibfnamefont {S.~G.}\ \bibnamefont {Wilkins}},\ and\ \bibinfo {author}
  {\bibfnamefont {X.~F.}\ \bibnamefont {Yang}},\ }\bibfield  {title} {\bibinfo
  {title} {Spectroscopy of short-lived radioactive molecules},\ }\href
  {https://doi.org/10.1038/s41586-020-2299-4} {\bibfield  {journal} {\bibinfo
  {journal} {Nature}\ }\textbf {\bibinfo {volume} {581}},\ \bibinfo {pages}
  {396} (\bibinfo {year} {2020})}\BibitemShut {NoStop}%
\bibitem [{\citenamefont {Udrescu}\ \emph {et~al.}(2023)\citenamefont
  {Udrescu}, \citenamefont {Wilkins}, \citenamefont {Breier}, \citenamefont
  {Ruiz}, \citenamefont {Athanasakis-Kaklamanakis}, \citenamefont {Au},
  \citenamefont {Belošević}, \citenamefont {Berger}, \citenamefont {Bissell},
  \citenamefont {Chrysalidis}, \citenamefont {Cocolios}, \citenamefont
  {Groote}, \citenamefont {Dorne}, \citenamefont {Flanagan}, \citenamefont
  {Franchoo}, \citenamefont {Gaul}, \citenamefont {Geldhof}, \citenamefont
  {Giesen}, \citenamefont {Hanstorp}, \citenamefont {Heinke}, \citenamefont
  {Koszorús}, \citenamefont {Kujanpää}, \citenamefont {Lalanne},
  \citenamefont {Neyens}, \citenamefont {Nichols}, \citenamefont {Perrett},
  \citenamefont {Reilly}, \citenamefont {Rothe}, \citenamefont {Borne},
  \citenamefont {Wang}, \citenamefont {Wessolek}, \citenamefont {Yang},\ and\
  \citenamefont {Zülch}}]{udrescu_precision_2023}%
  \BibitemOpen
  \bibfield  {author} {\bibinfo {author} {\bibfnamefont {S.-M.}\ \bibnamefont
  {Udrescu}}, \bibinfo {author} {\bibfnamefont {S.}~\bibnamefont {Wilkins}},
  \bibinfo {author} {\bibfnamefont {A.}~\bibnamefont {Breier}}, \bibinfo
  {author} {\bibfnamefont {R.~F.~G.}\ \bibnamefont {Ruiz}}, \bibinfo {author}
  {\bibfnamefont {M.}~\bibnamefont {Athanasakis-Kaklamanakis}}, \bibinfo
  {author} {\bibfnamefont {M.}~\bibnamefont {Au}}, \bibinfo {author}
  {\bibfnamefont {I.}~\bibnamefont {Belošević}}, \bibinfo {author}
  {\bibfnamefont {R.}~\bibnamefont {Berger}}, \bibinfo {author} {\bibfnamefont
  {M.}~\bibnamefont {Bissell}}, \bibinfo {author} {\bibfnamefont
  {K.}~\bibnamefont {Chrysalidis}}, \bibinfo {author} {\bibfnamefont
  {T.}~\bibnamefont {Cocolios}}, \bibinfo {author} {\bibfnamefont
  {R.}~\bibnamefont {Groote}}, \bibinfo {author} {\bibfnamefont
  {A.}~\bibnamefont {Dorne}}, \bibinfo {author} {\bibfnamefont
  {K.}~\bibnamefont {Flanagan}}, \bibinfo {author} {\bibfnamefont
  {S.}~\bibnamefont {Franchoo}}, \bibinfo {author} {\bibfnamefont
  {K.}~\bibnamefont {Gaul}}, \bibinfo {author} {\bibfnamefont {S.}~\bibnamefont
  {Geldhof}}, \bibinfo {author} {\bibfnamefont {T.}~\bibnamefont {Giesen}},
  \bibinfo {author} {\bibfnamefont {D.}~\bibnamefont {Hanstorp}}, \bibinfo
  {author} {\bibfnamefont {R.}~\bibnamefont {Heinke}}, \bibinfo {author}
  {\bibfnamefont {A.}~\bibnamefont {Koszorús}}, \bibinfo {author}
  {\bibfnamefont {S.}~\bibnamefont {Kujanpää}}, \bibinfo {author}
  {\bibfnamefont {L.}~\bibnamefont {Lalanne}}, \bibinfo {author} {\bibfnamefont
  {G.}~\bibnamefont {Neyens}}, \bibinfo {author} {\bibfnamefont
  {M.}~\bibnamefont {Nichols}}, \bibinfo {author} {\bibfnamefont
  {H.}~\bibnamefont {Perrett}}, \bibinfo {author} {\bibfnamefont
  {J.}~\bibnamefont {Reilly}}, \bibinfo {author} {\bibfnamefont
  {S.}~\bibnamefont {Rothe}}, \bibinfo {author} {\bibfnamefont {B.~V.~D.}\
  \bibnamefont {Borne}}, \bibinfo {author} {\bibfnamefont {Q.}~\bibnamefont
  {Wang}}, \bibinfo {author} {\bibfnamefont {J.}~\bibnamefont {Wessolek}},
  \bibinfo {author} {\bibfnamefont {X.}~\bibnamefont {Yang}},\ and\ \bibinfo
  {author} {\bibfnamefont {C.}~\bibnamefont {Zülch}},\ }\href
  {https://doi.org/10.21203/rs.3.rs-2648482/v1} {\bibinfo {title} {Precision
  spectroscopy and laser cooling scheme of a radium-containing molecule}}
  (\bibinfo {year} {2023})\BibitemShut {NoStop}%
\bibitem [{\citenamefont {Shafer-Ray}(2006)}]{shafer-ray_possibility_2006}%
  \BibitemOpen
  \bibfield  {author} {\bibinfo {author} {\bibfnamefont {N.~E.}\ \bibnamefont
  {Shafer-Ray}},\ }\bibfield  {title} {\bibinfo {title} {Possibility of
  0-$g$-factor paramagnetic molecules for measurement of the electron's
  electric dipole moment},\ }\href {https://doi.org/10.1103/PhysRevA.73.034102}
  {\bibfield  {journal} {\bibinfo  {journal} {Physical Review A}\ }\textbf
  {\bibinfo {volume} {73}},\ \bibinfo {pages} {034102} (\bibinfo {year}
  {2006})},\ \bibinfo {note} {publisher: American Physical Society}\BibitemShut
  {NoStop}%
\bibitem [{\citenamefont {Mawhorter}\ \emph {et~al.}(2011)\citenamefont
  {Mawhorter}, \citenamefont {Murphy}, \citenamefont {Baum}, \citenamefont
  {Sears}, \citenamefont {Yang}, \citenamefont {Rupasinghe}, \citenamefont
  {McRaven}, \citenamefont {Shafer-Ray}, \citenamefont {Alphei},\ and\
  \citenamefont {Grabow}}]{mawhorter_characterization_2011}%
  \BibitemOpen
  \bibfield  {author} {\bibinfo {author} {\bibfnamefont {R.~J.}\ \bibnamefont
  {Mawhorter}}, \bibinfo {author} {\bibfnamefont {B.~S.}\ \bibnamefont
  {Murphy}}, \bibinfo {author} {\bibfnamefont {A.~L.}\ \bibnamefont {Baum}},
  \bibinfo {author} {\bibfnamefont {T.~J.}\ \bibnamefont {Sears}}, \bibinfo
  {author} {\bibfnamefont {T.}~\bibnamefont {Yang}}, \bibinfo {author}
  {\bibfnamefont {P.~M.}\ \bibnamefont {Rupasinghe}}, \bibinfo {author}
  {\bibfnamefont {C.~P.}\ \bibnamefont {McRaven}}, \bibinfo {author}
  {\bibfnamefont {N.~E.}\ \bibnamefont {Shafer-Ray}}, \bibinfo {author}
  {\bibfnamefont {L.~D.}\ \bibnamefont {Alphei}},\ and\ \bibinfo {author}
  {\bibfnamefont {J.-U.}\ \bibnamefont {Grabow}},\ }\bibfield  {title}
  {\bibinfo {title} {Characterization of the ground {X}$_1$ state of
  $^{204}${Pb}$^{19}${F}, $^{206}${Pb}$^{19}${F}, $^{207}${Pb}$^{19}${F}, and
  $^{208}${Pb}$^{19}${F}},\ }\href {https://doi.org/10.1103/PhysRevA.84.022508}
  {\bibfield  {journal} {\bibinfo  {journal} {Physical Review A}\ }\textbf
  {\bibinfo {volume} {84}},\ \bibinfo {pages} {022508} (\bibinfo {year}
  {2011})}\BibitemShut {NoStop}%
\bibitem [{\citenamefont {Zhu}\ \emph {et~al.}(2022)\citenamefont {Zhu},
  \citenamefont {Wang}, \citenamefont {Chen}, \citenamefont {Chen},
  \citenamefont {Yang}, \citenamefont {Yin},\ and\ \citenamefont
  {Liu}}]{zhu_fine_2022}%
  \BibitemOpen
  \bibfield  {author} {\bibinfo {author} {\bibfnamefont {C.}~\bibnamefont
  {Zhu}}, \bibinfo {author} {\bibfnamefont {H.}~\bibnamefont {Wang}}, \bibinfo
  {author} {\bibfnamefont {B.}~\bibnamefont {Chen}}, \bibinfo {author}
  {\bibfnamefont {Y.}~\bibnamefont {Chen}}, \bibinfo {author} {\bibfnamefont
  {T.}~\bibnamefont {Yang}}, \bibinfo {author} {\bibfnamefont {J.}~\bibnamefont
  {Yin}},\ and\ \bibinfo {author} {\bibfnamefont {J.}~\bibnamefont {Liu}},\
  }\bibfield  {title} {\bibinfo {title} {Fine and hyperfine interactions of
  {PbF} studied by laser-induced fluorescence spectroscopy},\ }\href
  {https://doi.org/10.1063/5.0099716} {\bibfield  {journal} {\bibinfo
  {journal} {The Journal of Chemical Physics}\ }\textbf {\bibinfo {volume}
  {157}},\ \bibinfo {pages} {084307} (\bibinfo {year} {2022})}\BibitemShut
  {NoStop}%
\end{thebibliography}%

\end{document}


\title{Supplemental Material: Quantum-Enhanced Metrology for\\ Molecular Symmetry Violation using Decoherence-Free Subspaces}

\author{Chi Zhang}
\email[]{chizhang@caltech.edu}
\author{Phelan Yu}
\author{Arian Jadbabaie}
\author{Nicholas R. Hutzler}
\affiliation{California Institute of Technology, Division of Physics, Mathematics, and Astronomy.  Pasadena, CA 91125}

\newcommand{\py}{\textcolor{purple}}

\renewcommand{\thefigure}{S\arabic{figure}}
\renewcommand{\thetable}{S\arabic{table}}

\maketitle

The structure of the Supplemental Material is as follows: 

\begin{enumerate}
\item In Section I, we discuss the details of the eEDM coupling in a single molecule. When the molecule is prepared in a superposition of the opposite parity doublet states it has an electric dipole moment oscillating at the splitting frequency, and an associated oscillating effective electric field. When the electron spin is polarized along a transverse direction, the interaction of the eEDM with the effective electric field will result in a spin precession which oscillates back and forth because the direction of $\Eeff$ oscillates back and forth. It time-averages to zero. If a transverse static magnetic field is added and its magnitude tuned to match the parity splitting, the spin will stay in phase with the oscillation of $\Eeff$ and there will be a net precession due to the eEDM. This requires an extremely stable magnetic field which is probably not feasible unless the doublet splitting is very small. Instead, an rf magnetic field can be applied along the quantization axis whose frequency matches the oscillation frequency of $\Eeff$. If the amplitude is strong enough, the spin follow the rf field adiabatically and there will be a net precession in the rotating frame. This method suffers from noise on the amplitude of the rf magnetic field which will wash out the eEDM signal unless the noise can be made extremely small.

\item In Section II, we show that the method using an rf magnetic field can be extended to a two-molecule system, in which the unwanted shifts from the magnetic field cancel out but the eEDM spin precession adds up linearly to molecule number (Heisenberg scaling). We also present a detailed example of an experimental sequence and show what the observables are in the lab frame.

\item In Section III, we present an example of entangling two molecules using an existing entanglement protocol.

\item In Section IV, we briefly discuss the Heisenberg scaling of the eEDM sensitivity in larger entangled systems, as well as the possible ways to prepare a large entangled system.

\item In Section V, we summarize the requirements for choice of molecule species for our scheme, and list some suitable molecule species.
\end{enumerate}

\section{Details of the eEDM coupling}

\begin{figure*}
	\includegraphics[width=\textwidth]{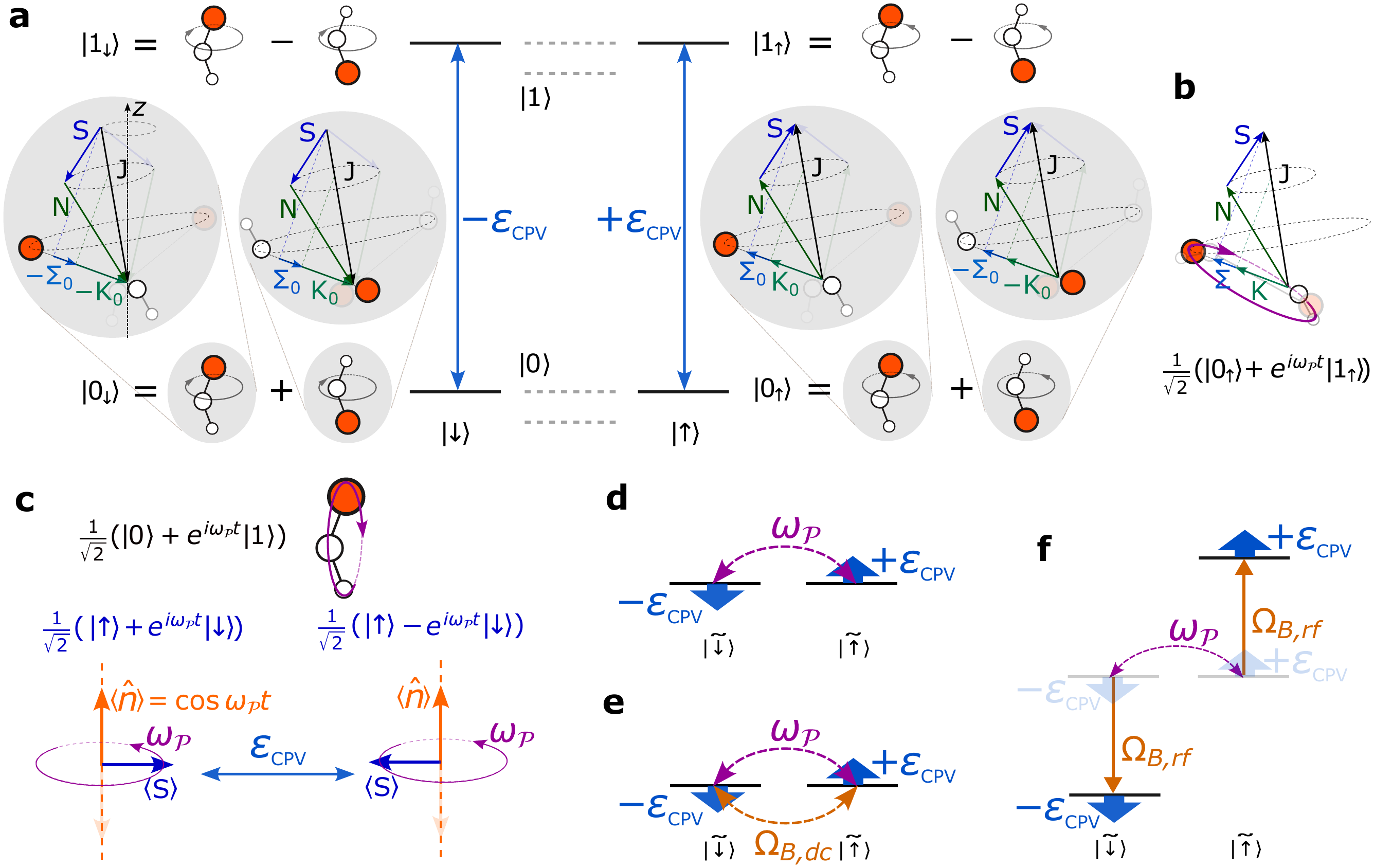}
	\caption{(a) A more general and detailed coupling diagram of angular momenta in a molecule, similar to Fig.~1 in the main text. The arrows represent angular momenta or their projections (see text for details), the dashed circles stand for precession, i.e., the angular momentum or molecule axis is in a superposition around another axis or angular momentum and has zero expectation value perpendicular to that axis. (b) In a superposition of two opposite parity states, the molecule axis is rotating (purple arrow) while all the angular momenta are stationary. Note that the rotational axis is in a superposition (precessing) about the total angular momentum $J$ and $J$ precess about the lab $z$ axis, as a result, the dipole moment vanishes in the $x,y$ plane and is oscillating along $z$ direction. (c) In a superposition of the opposite parity states, the eEDM spin precession only happens between the spin states that rotate in phase with the molecule axis. (d) In the rotating frame, the molecule rotation couples $\Tilde{\ket{\uparrow}}$ and $\Tilde{\ket{\downarrow}}$ and suppresses eEDM spin precession. (e) The non-adiabatic method. A dc magnetic field can be used to cancel the coupling of the rotation. (f) The adiabatic method. An rf magnetic field can be used to split $\Tilde{\ket{\uparrow}}$ and $\Tilde{\ket{\downarrow}}$ and thus suppress the coupling of the rotation. The problems of (e) and (f) are explained in the text.}
	\label{FigS1}
\end{figure*}

In this section we present a more general and detailed description about the eEDM coupling in one molecule.  Fig.~\ref{FigS1} shows the total angular momentum excluding spins labeled as $N$, and its projection on the molecule axis $K=N\cdot\hat{n}$, which can receive contributions from orbital angular momentum and rotation about the molecule axis. For molecules with nonzero $K$, the good parity states are superpositions of equal and opposite $K$ states, i.e., $\ket{N,K,\pm}=\frac{1}{\sqrt{2}}\left( \ket{N,K} \pm (-1)^{N-K} \ket{N,-K} \right)$. The degeneracy between opposite parity states (i.e., a parity doublet) is lifted by high-order interactions such as the interaction with the end-to-end rotation of the molecule. For molecules with $K=0$, the good parity states are rotational states $\ket{N}$ and are split by the rotational energy. Next, $S$ is aligned, or partially aligned, to $N$ by spin-orbit or spin-rotation interactions with the sub-components of $N$. The total angular momentum formed by $S$ and $N$ is labeled as $J$. Molecule eigenstates have well-defined $J$ and are superpositions of states of the same parity. $S$ precess about $J$ and the averaged $S$ projection on $\hat{n}$ is $\Sigma_0$~\cite{Petrov2022}. Molecule eigenstates are superpositions of $\pm \Sigma_0$. In addition, nuclear spins ($I$) may interact with $J$ to form $F$. Here we consider the extreme $M$ (magnetic quantum number) states with $M=\pm F$, where the nuclear spins are separable.

As in the main text, we label the positive and negative parity states as $\ket{0}$ and $\ket{1}$, and the superposition $\ket{\Uparrow} = \frac{1}{\sqrt{2}}(\ket{0}+\ket{1})=  \ket{N,K}$ and $\ket{\Downarrow} = \frac{1}{\sqrt{2}}(\ket{0}-\ket{1})=  \ket{N,-K}$. We label the spin states in the lab frame as $\ket{\uparrow}$ and $\ket{\downarrow}$. The eEDM oppositely shifts the energies of the spins aligned and anti-aligned with the dipole. In the $\{\ket{\Uparrow},\ket{\Downarrow}\}\otimes\{\ket{\uparrow},\ket{\downarrow}\}$ basis (the quantization axis is along $z$), the molecular Hamiltonian including the eEDM coupling is:
\begin{equation}
    H_\mathrm{pol} =
    \begin{pmatrix}
        \varepsilon_\mathrm{CPV} & 0 & -\omega_\mathcal{P}/2 & 0 \\
        0 & -\varepsilon_\mathrm{CPV} & 0 & -\omega_\mathcal{P}/2 \\
        -\omega_\mathcal{P}/2 & 0 & -\varepsilon_\mathrm{CPV} & 0 \\
        0 & -\omega_\mathcal{P}/2 & 0 & \varepsilon_\mathrm{CPV}
    \end{pmatrix},
\end{equation}
$\omega_\mathcal{P}$ is from the higher order coupling between $\ket{\Uparrow} \leftrightarrow \ket{\Downarrow}$. The eEDM interaction causes opposite spin precession in the subspaces of $\{\ket{\Uparrow_\uparrow},\ket{\Uparrow_\downarrow}\}$ and $\{\ket{\Downarrow_\uparrow},\ket{\Downarrow_\downarrow}\}$. However, a molecule initially prepared in $\Uparrow$ oscillates between $\Uparrow \leftrightarrow \Downarrow$ at the frequency of the parity doubling $\omega_\mathcal{P}$. As a result, the eEDM spin precession oscillates and averages to zero.

$H_\mathrm{pol}$ can be transformed to the molecule eigenbasis ($\{\ket{0},\ket{1}\}\otimes\{\ket{\uparrow},\ket{\downarrow}\}$) as
\begin{equation}
    H_\mathrm{mol} =
    \begin{pmatrix}
        0 & 0 & \varepsilon_\mathrm{CPV} & 0 \\
        0 & 0 & 0 & -\varepsilon_\mathrm{CPV} \\
        \varepsilon_\mathrm{CPV} & 0 & \omega_\mathcal{P} & 0 \\
        0 & -\varepsilon_\mathrm{CPV} & 0 & \omega_\mathcal{P}
    \end{pmatrix}.
\end{equation}
This is the coupling shown in Fig.~\ref{FigS1}, where the eEDM only has a vanishing second order effect. In the frame rotating at $\omega_\mathcal{P}$ frequency about the $x$-axis, the Hamiltonian is transformed by $\ket{0}\bra{0} + e^{i \omega_\mathcal{P} t} \ket{1}\bra{1}$ and $\ket{+}\bra{+} + e^{i \omega_\mathcal{P} t} \ket{-}\bra{-}$, with $\ket{\pm} = \frac{1}{\sqrt{2}}(\ket{\uparrow} \pm \ket{\downarrow})$. After neglecting the small and fast-oscillating terms proportional to $\varepsilon_\mathrm{CPV} e^{\pm i \omega_\mathcal{P} t}$, the Hamiltonian is:
\begin{equation}
    \widetilde{H}_\mathrm{mol} = \frac{1}{2}
    \begin{pmatrix}
        0 & \omega_\mathcal{P} & \varepsilon_\mathrm{CPV} & \varepsilon_\mathrm{CPV} \\
        \omega_\mathcal{P} & 0 & -\varepsilon_\mathrm{CPV} & -\varepsilon_\mathrm{CPV} \\
        \varepsilon_\mathrm{CPV} & -\varepsilon_\mathrm{CPV} & 0 & \omega_\mathcal{P} \\
        \varepsilon_\mathrm{CPV} & -\varepsilon_\mathrm{CPV} & \omega_\mathcal{P} & 0
    \end{pmatrix}.
\end{equation}

We define the rotating frame basis $\widetilde{\ket{\Uparrow}} = \frac{1}{2}(\ket{0}+ e^{-i \omega_\mathcal{P} t} \ket{1})$, $\widetilde{\ket{\Downarrow}} = \frac{1}{2}(\ket{0} - e^{-i \omega_\mathcal{P} t} \ket{1})$, $\widetilde{\ket{\uparrow}} = \frac{1}{2}(\ket{+}+ e^{-i \omega_\mathcal{P} t} \ket{-})$ and $\widetilde{\ket{\downarrow}} = \frac{1}{2}(\ket{+}- e^{-i \omega_\mathcal{P} t} \ket{-})$. Note that this basis consists of states which are oscillating in the lab frame. $\widetilde{\ket{\Uparrow}}$ and $\widetilde{\ket{\Downarrow}}$ are eigenstates of $\widetilde{H}_\mathrm{mol}$ while $\widetilde{\ket{\uparrow}}$ and $\widetilde{\ket{\downarrow}}$ are not. In the rotating frame basis $\{\widetilde{\ket{\Uparrow}},\widetilde{\ket{\Downarrow}}\}\otimes\{\widetilde{\ket{\uparrow}},\widetilde{\ket{\downarrow}}\}$, the Hamiltonian is simply
\begin{equation}
    \widetilde{H}^\prime_\mathrm{pol} = \frac{1}{2}
    \begin{pmatrix}
        \varepsilon_\mathrm{CPV} & \omega_\mathcal{P} & 0 & -\varepsilon_\mathrm{CPV} \\
        \omega_\mathcal{P} & -\varepsilon_\mathrm{CPV} & \varepsilon_\mathrm{CPV} & 0 \\
        0 & \varepsilon_\mathrm{CPV} & -\varepsilon_\mathrm{CPV} & \omega_\mathcal{P} \\
        -\varepsilon_\mathrm{CPV} & 0 & \omega_\mathcal{P} & \varepsilon_\mathrm{CPV}
    \end{pmatrix}.
\end{equation}
In this two-by-two block matrix (in the basis of $\{\widetilde{\ket{\Uparrow}},\widetilde{\ket{\Downarrow}}\}$), the diagonal parts are the eEDM shifts on the spin states ($\pm\varepsilon_\mathrm{CPV}\sigma_z$) and the coupling between the spin states by the rotation ($\omega_\mathcal{P} \sigma_x$). So far we have simplified the molecule orientation as a two-level systems for which there are some coupling terms between $\widetilde{\ket{\Uparrow}}$ and $\widetilde{\ket{\Downarrow}}$ subspaces. For two interacting spin-$1/2$ degrees of freedom (if molecule orientation was spin-$1/2$), these couplings indicate transverse interactions ($XY$ interaction). However, because the real molecule orientation is not a two-level system, and it does not have a transverse dipole moment in superpositions of $\widetilde{\ket{\Uparrow}}$ and $\widetilde{\ket{\Downarrow}}$, these couplings are not physical and need to be removed. As a result, the Hamiltonian in the basis of $\{\widetilde{\ket{\Uparrow}},\widetilde{\ket{\Downarrow}}\}\otimes\{\widetilde{\ket{\uparrow}},\widetilde{\ket{\downarrow}}\}$ is
\begin{equation}
\label{eqDiagEDM}
    \widetilde{H}_\mathrm{pol} = \frac{1}{2}
    \begin{pmatrix}
        \varepsilon_\mathrm{CPV} & \omega_\mathcal{P} & 0 & 0 \\
        \omega_\mathcal{P} & -\varepsilon_\mathrm{CPV} & 0 & 0 \\
        0 & 0 & -\varepsilon_\mathrm{CPV} & \omega_\mathcal{P} \\
        0 & 0 & \omega_\mathcal{P} & \varepsilon_\mathrm{CPV}
    \end{pmatrix}.
\end{equation}

Now we have two decoupled subspaces, which correspond to $\widetilde{\ket{\Uparrow}}$ and $\widetilde{\ket{\Downarrow}}$, where the spin can precess oppositely. However, the spin precession is still suppressed by the coupling caused by the rotation of the frame. As shown in Fig.~\ref{FigS1}, in a magnetic field, we find two types of schemes: a non-adiabatic one and an adiabatic one, of spin-precession between the rotating spin states. Here we explain the schemes, and discuss why they won't work for the case of a single molecule.

The non-adiabatic method is to apply a static magnetic field along the $x$ direction that cancels the coupling of $\omega_\mathcal{P}$ exactly, as shown in Fig.~\ref{FigS1}(e). More specifically, the Hamiltonian of the magnetic field is $\widetilde{H}_{B,dc} = I_2 \otimes \frac{\Omega_B}{2}\widetilde{\sigma_x}$, where $I_2$ is the two-dimensional identity matrix for the molecule alignment. The total Hamiltonian is therefore
\begin{equation}
    \widetilde{H}_\mathrm{pol,dcB} = \frac{1}{2}
    \begin{pmatrix}
        \varepsilon_\mathrm{CPV} & \omega_\mathcal{P}+\Omega_B & 0 & 0 \\
        \omega_\mathcal{P}+\Omega_B & -\varepsilon_\mathrm{CPV} & 0 & 0 \\
        0 & 0 & -\varepsilon_\mathrm{CPV} & \omega_\mathcal{P}+\Omega_B \\
        0 & 0 & \omega_\mathcal{P}+\Omega_B & \varepsilon_\mathrm{CPV}
    \end{pmatrix}.
\end{equation}

This requires fine tuning and stabilization of the magnetic field strength $\Omega_B$ to $\omega_\mathcal{P}$ ($\gtrsim 100~\mathrm{kHz}$) within a fluctuation less than the decoherence rate (typically $\lesssim \mathrm{Hz}$). This is challenging, although magnetic field stabilization to ppm level has been achieved \cite{Borkowski2023} and this scheme may work for molecules with very small parity doubling \cite{Yu2021RaOCH3}.

The adiabatic method is to apply an oscillating or rotating magnetic field in phase with the oscillating dipole. The spin states follow the magnetic field adiabatically and rotate in phase with the molecule axis. Note that this only requires tuning the magnetic field frequency to the parity doubling frequency, which is achievable. Equivalently, in the rotating frame, as Fig.~\ref{FigS1}(f) shows, the Hamiltonian is $\widetilde{H}_{B,rf} = I_2 \otimes \frac{\Omega_B}{2}\widetilde{\sigma_z}$, and the total Hamiltonian is
\begin{equation}
\label{eqSingleHam}
    \widetilde{H}_\mathrm{pol,rfB} = \frac{1}{2}
    \begin{pmatrix}
        \Omega_B+\varepsilon_\mathrm{CPV} & \omega_\mathcal{P} & 0 & 0 \\
        \omega_\mathcal{P} & -\Omega_B-\varepsilon_\mathrm{CPV} & 0 & 0 \\
        0 & 0 & \Omega_B-\varepsilon_\mathrm{CPV} & \omega_\mathcal{P} \\
        0 & 0 & \omega_\mathcal{P} & -\Omega_B+\varepsilon_\mathrm{CPV}
    \end{pmatrix}.
\end{equation}

For $\Omega_B \gtrsim \omega_\mathcal{P}$, a total spin precession caused by the magnetic field and the eEDM may be observed. This can also be understood equivalently as that the dressed eigenstates $\ket{\pm}$ are superpositions with non-equal populations in $\Tilde{\ket{\uparrow}}$ and $\Tilde{\ket{\downarrow}}$, $\ket{+} = \sin{\theta} \Tilde{\ket{\uparrow}} +\cos{\theta}\Tilde{\ket{\downarrow}}$ and $\ket{-} = \cos{\theta} \Tilde{\ket{\uparrow}} -\sin{\theta}\Tilde{\ket{\downarrow}}$, with the mixing angle $\theta$ given by $\tan \theta = \Omega_B / \omega_\mathcal{P}$. Here $\Omega_B$ does not need to match $\omega_\mathcal{P}$. The eEDM interaction, which splits $\Tilde{\ket{\uparrow}}$ and $\Tilde{\ket{\downarrow}}$, causes spin precession in the dressed eigenstates $\ket{\pm}$ since they have non-equal $\Tilde{\ket{\uparrow}}$ and $\Tilde{\ket{\downarrow}}$ components. The problem with this scheme is that the magnetic field contributes to the same spin precession as the eEDM interaction. As a consequence, magnetic field fluctuations need to be reduced to below the eEDM shift, otherwise the eEDM spin precession phase will be washed out in the magnetic field noise.  Note that this is conceptually similar to the approach proposed in \cite{Verma2020}, where this problem is avoided by using magnetically-insensitive $M=0$ states.

However, the adiabatic method can be extended to two entangled molecules. For two non-interacting molecules, the total Hamiltonian is $H = H_1 \otimes I + I \otimes H_2$, where $H_i$ ($i=1,2$) is the single molecule Hamiltonian $\widetilde{H}_\mathrm{pol,rfB}$ in Eq.~\ref{eqSingleHam} and $I$ is the identity operator. The eEDM coupling, as well as the detailed experimental sequence, for two molecules is discussed in the next section.

\begin{figure*}
	\includegraphics[width=\textwidth]{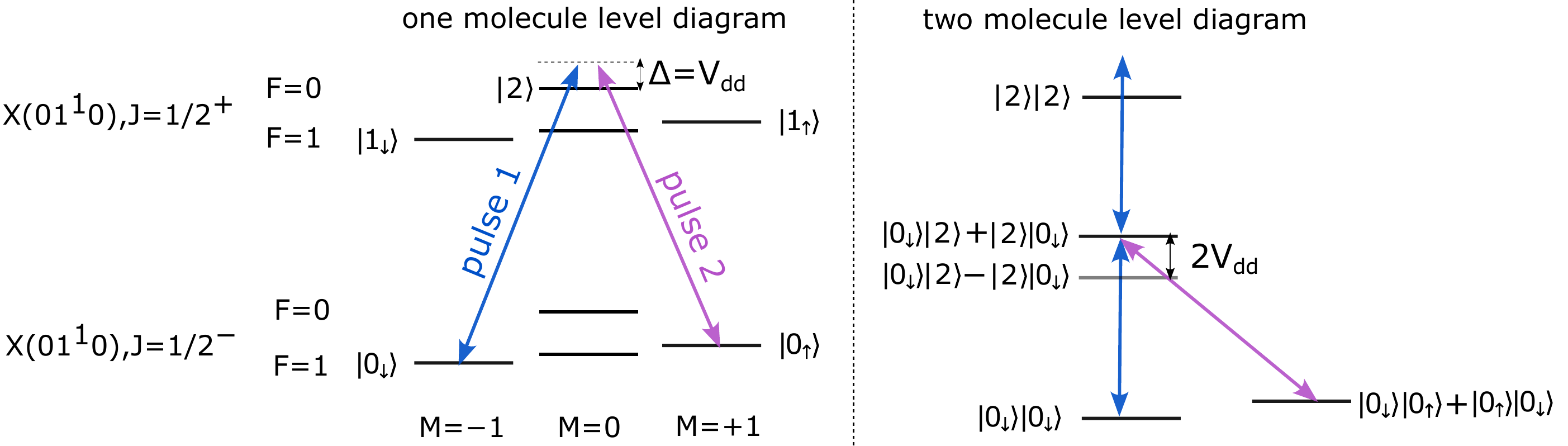}
	\caption{Level diagram of a bending mode in a linear $^2\Sigma$ molecule, for example YbOH~\cite{Kozyryev2017b,Jadbabaie2023}. We choose $\ket{X(01^10),J=1/2-,F=1,M=\pm 1} = \ket{0_{\uparrow,\downarrow}}$, $\ket{X(01^10),J=1/2+,F=1,M=\pm 1} = \ket{1_{\uparrow,\downarrow}}$, and $\ket{X(01^10),J=1/2+,F=0,M=0} = \ket{2}$. $\ket{0_\uparrow}$ and $\ket{2}$ are connected by an electric dipole transition and the pair states $\frac{1}{\sqrt{2}}(\ket{0_\downarrow 2}\pm\ket{2 0_\downarrow})$ and they are split by $2V_{dd}$. Rf pulse 1 (blue arrows) couples $\ket{0_\downarrow} \leftrightarrow \ket{2}$ and pulse 2 (purple arrows) couples $\ket{2} \leftrightarrow \ket{0_\uparrow}$ with the same detuning $\Delta=V_{dd}$. A magnetic field is applied to split the Zeeman sublevels during entanglement generation.}
	\label{FigS3}
\end{figure*}

\section{Details about the experimental sequence and detection scheme}

In this section we explain the experimental sequence by an example of an ideal experiment. The molecules are initialized in $\ket{0_\downarrow 0_\downarrow}$ via optical pumping and then entangled to $\frac{1}{\sqrt{2}} (\ket{0_\uparrow 0_\downarrow} - \ket{0_\downarrow 0_\uparrow})$.  An example protocol for entanglement generation is discussed in the next section. Next the molecule orientation is prepared in $\widetilde{\ket{\Uparrow\Downarrow}}$ by a global $\pi/2$-pulse and a scalar light shift on one of the molecules, as described in the main text.

Since the total Hamiltonian for one molecule $\widetilde{H}_\mathrm{pol,rfB}$ (Eq.~\ref{eqSingleHam}) can be separated into two decoupled subspaces, we can reduce it to a $2\times 2$ matrix for each molecule orientation. For two molecules in $\widetilde{\ket{\Uparrow\Downarrow}}$, the total Hamiltonian in the basis of $\{\widetilde{\ket{\uparrow\uparrow}}, \widetilde{\ket{\uparrow\downarrow}}, \widetilde{\ket{\downarrow\uparrow}}, \widetilde{\ket{\downarrow\downarrow}}\}$ is 
\begin{equation}
    H_{\Uparrow\Downarrow} = \frac{1}{2}
    \begin{pmatrix}
        2\Omega_B & \omega_\mathcal{P} & \omega_\mathcal{P} & 0 \\
        \omega_\mathcal{P} & 2\varepsilon_\mathrm{CPV} & 0 & \omega_\mathcal{P} \\
        \omega_\mathcal{P} & 0 & -2\varepsilon_\mathrm{CPV} & \omega_\mathcal{P} \\
        0 & \omega_\mathcal{P} & \omega_\mathcal{P} & -2\Omega_B
    \end{pmatrix}.
\end{equation}

The initial spin state in this basis is
\begin{equation}
    \ket{\Psi^-} = \frac{1}{\sqrt{2}}
    \begin{pmatrix}
        0 \\
        +1 \\
        -1 \\
        0
    \end{pmatrix}.
\end{equation}

Before turning on the magnetic field ($\Omega_B=0$), the eEDM spin precession is suppressed by the $\omega_\mathcal{P}$ coupling in the triplet subspace. After turning on the magnetic field, as described in the main text, $\ket{\Psi^-}$ is coupled by the eEDM interaction resonantly to the unshifted state
\begin{equation}
\label{eqDarkstate}
    \ket{u} = \sin{\theta} \ket{\Psi^+} + \cos{\theta} \ket{\Phi^+} = 
    \frac{1}{\sqrt{2}} 
    \begin{pmatrix}
        \cos{\theta} \\
        \sin{\theta} \\
        \sin{\theta} \\
        \cos{\theta}
    \end{pmatrix},
\end{equation}
with the mixing angle $\theta$ given by $\tan{\theta} = \Omega_B/\omega_\mathcal{P}$, and
\begin{equation}
    \ket{\Psi^+} = \frac{1}{\sqrt{2}}
    \begin{pmatrix}
        0 \\
        1 \\
        1 \\
        0
    \end{pmatrix},
    \ket{\Phi^\pm} = \frac{1}{\sqrt{2}}
    \begin{pmatrix}
        1 \\
        0 \\
        0 \\
        \pm 1
    \end{pmatrix}.
\end{equation}
The coupling strength is reduced to $\varepsilon_u =4 \varepsilon_\mathrm{CPV}\sin{\theta}$.  

After spin precession time $T$, the spin state is
\begin{equation}
\label{eqpsi}
\begin{split}
    \ket{\psi} = & \cos{\varepsilon_u T} \ket{\Psi^-} + i\sin{\varepsilon_u T} \ket{u} \\
    = & \cos{\varepsilon_u T} \ket{\Psi^-} + i\sin{\varepsilon_u T} (\sin{\theta} \ket{\Psi^+} + \cos{\theta} \ket{\Phi^+}) = 
    \frac{1}{\sqrt{2}}
    \begin{pmatrix}
        i\cos{\theta}\sin{\varepsilon_u T} \\
        \cos{\varepsilon_u T} + i\sin{\theta}\sin{\varepsilon_u T} \\
        -\cos{\varepsilon_u T} + i\sin{\theta}\sin{\varepsilon_u T} \\
        i\cos{\theta}\sin{\varepsilon_u T}
    \end{pmatrix}.
\end{split}
\end{equation}

The magnetic field is turned off after an integer cycles of oscillations, when the lab basis coincides with the rotating frame basis, the spin state freezes in the lab frame (but starts to oscillate between $\ket{\Psi^+}\leftrightarrow\ket{\Phi^-}$ in the rotating frame). After rotating the molecule orientation back to $\ket{00}$, the spin state remains the same. As a result, $\ket{\psi}$ is the final spin state in the lab frame. $\ket{\psi}$ is mostly $\ket{\Psi^-}$, because the eEDM spin precession phase is small, with a small admixture of $\ket{u}$ (Eq.~\ref{eqDarkstate}). The small difference between the initial ($\ket{\Psi^-}$) and final ($\ket{\psi}$) spin states in the lab frame indicates the eEDM spin precession phase.

To maximize the sensitivity, we need to measure in the $\frac{1}{\sqrt{2}} (\ket{\uparrow \downarrow} \pm i \ket{\downarrow \uparrow})$ basis, because $\ket{\Psi^-}$ has equal projection on this set of basis states and the spin precession is in the same plane as the basis states. This is conceptually similar to rotating the phase of the spin or rotating the measurement basis by $\pm\pi/4$ between spin initialization and measurement in conventional eEDM measurements to maximize the sensitivity to the spin precession~\cite{Baron2017}. Here, it can be achieved by two similar methods. 

The first method works as follows. To start, we add an extra $\pm\pi/2$ phase between $\ket{\uparrow}$ and $\ket{\downarrow}$ by a vector Stark shift \cite{Caldwell2021} from an addressing beam on one of the molecules. The addressing beam on the second molecule shifting $\ket{\downarrow}$ by $\delta$, as an example, is described by $I\otimes \delta \ket{\downarrow}\bra{\downarrow}$. After a pulse time $t$ with $\delta t =\pi/2$, the state $\ket{\psi}$ becomes
\begin{equation}
\label{eqpsiprime}
\begin{split}
    \ket{\psi ^\prime} = & \cos{\left(\frac{\pi}{4} + \varepsilon_u T\right)} \ket{\Psi^-} + 
    i\sin{\left(\frac{\pi}{4} + \varepsilon_u T\right)} \sin{\theta} \ket{\Psi^+} + i \frac{1}{\sqrt{2}} \sin{\varepsilon_u T} \cos{\theta} \left(\ket{\Phi^+} + i \ket{\Phi^-}\right) \\
    = & \frac{1}{\sqrt{2}}
    \begin{pmatrix}
        i\cos{\theta}\sin{\varepsilon_u T} \\
        i\cos{\varepsilon_u T} - \sin{\theta}\sin{\varepsilon_u T} \\
        -\cos{\varepsilon_u T} + i\sin{\theta}\sin{\varepsilon_u T} \\
        -\cos{\theta}\sin{\varepsilon_u T}
    \end{pmatrix}.
\end{split}
\end{equation}
$\ket{\psi ^\prime}$ now is roughly an equal superposition of $\ket{\Psi^-}$ and a state in the triplet subspace. Next, we apply a global $\pi$ rotation between $\ket{\uparrow} \leftrightarrow \ket{\downarrow}$. The singlet $\ket{\Psi^-}$ is not coupled by global rotations. The $\ket{\Psi^+}$ state is coupled to a superposition $\cos{\phi} \ket{\Phi^+} + i \sin{\phi} \ket{\Phi^-})$, where $\phi$ is the phase of the global $\pi$-pulse, and the other superposition $\cos{\phi} \ket{\Phi^+} - i \sin{\phi} \ket{\Phi^-})$ is a dark state. We choose a phase of $-\pi/2$ (i.e., $-\sigma_y$ rotation), as a result, for the components in $\ket{\psi ^\prime}$ (Eq.~\ref{eqpsiprime}), $\frac{1}{\sqrt{2}} (\ket{\Phi^+} + i \ket{\Phi^-})$ is a dark state, $\ket{\Phi^+}$ is mapped to $\frac{1}{\sqrt{2}} (\ket{\Phi^+} - i \ket{\Phi^-})$, and $\ket{\Psi^-}$ is uncoupled. The state after rotation for a small $\varepsilon_u T$ ($\sin{\varepsilon_u T} \approx \varepsilon_u T$ and $\cos{\varepsilon_u T} \approx 1$) is
\begin{equation}
\begin{split}
    \ket{\psi ^{\prime\prime}} = & \frac{1}{\sqrt{2}} (1 - \varepsilon_u T) \ket{\Psi^-} + 
    i\frac{1}{2\sqrt{2}} (1+(\sin{\theta}+\cos{\theta}) \varepsilon_u T) \ket{\Phi^+}
    + i\frac{1}{2\sqrt{2}} (1+(\sin{\theta}-\cos{\theta}) \varepsilon_u T) \ket{\Phi^-} \\
    = &
    \begin{pmatrix}
        i\frac{1}{\sqrt{2}} (1+\varepsilon_u T\sin{\theta} ) \\
        \frac{1}{2} (1 - \varepsilon_u T) \\
        -\frac{1}{2} (1 - \varepsilon_u T) \\
        i\frac{1}{\sqrt{2}} \varepsilon_u T\cos{\theta} 
    \end{pmatrix}.
\end{split}
\end{equation}

Finally, the populations are measured by fluorescence in $\ket{\uparrow}$ and $\ket{\downarrow}$. The eEDM phase information is mapped to the parity of the population and we do not need single molecule resolved imaging. The population in the even and odd parity states are 
\begin{equation}
\begin{split}
    P_{\uparrow\uparrow, \downarrow\downarrow} = & \frac{1}{2} (1+\varepsilon_u T) \\
    P_{\uparrow\downarrow, \downarrow\uparrow} = & \frac{1}{2} (1-\varepsilon_u T)
\end{split}
\end{equation}

An alternative detection method, similar to the one described above, is to apply a $\pi$ pulse on two molecules with different phases on the state $\ket{\psi}$ (Eq.~\ref{eqpsi}) and measure the parity. For instance, if we apply $-\sigma_y \otimes I + I \otimes \sigma_x$, the $\frac{1}{\sqrt{2}} (\ket{\uparrow \downarrow} + i \ket{\downarrow \uparrow})$ state is mapped to the even parity states ($\ket{\uparrow\uparrow}, \ket{\downarrow\downarrow}$) while the $\frac{1}{\sqrt{2}} (\ket{\uparrow \downarrow} + i \ket{\downarrow \uparrow})$ remains in the odd parity states ($\ket{\uparrow\downarrow}, \ket{\downarrow\uparrow}$), which can be distinguished by fluorescence detection. This method requires the ability to perform single molecule-resolved rotation, which can be achieved by a two-photon transition with focused lasers. An advantage compared to the first method is that the phase of the measurement basis is set by the phase of the laser field, but not the intensity of the addressing beam. Both methods together may be used for checking systematic effects, and even in parallel in systems with multiple pairs of molecules.

\section{An example of entanglement generation}

As mentioned in the main text, the spin entangled initial state $\frac{1}{\sqrt{2}} (\ket{0_\uparrow 0_\downarrow} - \ket{0_\downarrow 0_\uparrow})$ can be prepared by existing entanglement protocols together with single molecule rotations. Here we present an example for the YbOH molecule, which is the only molecule for which the parity-doubled bending mode has been completely mapped out \cite{Jadbabaie2023}. Note that the level structures for other metal hydroxide molecules (SrOH, CaOH, RaOH, etc) are similar and therefore the same experimental sequence can be applied. More generally, a similar entanglement sequence can be found for any polar molecules using the dipole-dipole interaction or other types of interactions.

The level diagram of the YbOH molecule is shown in Fig.~\ref{FigS3}. We propose to use the $X(01^10),J=1/2-,F=1,M=\pm 1$ states as $\ket{0_\uparrow}$ and $\ket{0_\downarrow}$ states, and use $X(01^10),J=1/2+,F=1,M=\pm 1$ states as $\ket{1_\uparrow}$ and $\ket{1_\downarrow}$ states. The $\ket{0}$ and $\ket{1}$ states are separated by $\sim 35~\mathrm{MHz}$ for YbOH \cite{Jadbabaie2023}, and similarly for other metal hydroxide molecules. We list the parity splitting of other types of molecules in Sec.~\ref{sec_molecule}. A magnetic field is applied to split the Zeeman sublevels during entanglement generation. We choose another state $X(01^10),J=1/2+,F=0,M=0$, labeled as $\ket{2}$, as an ancillary state for the entanglement generation. $\ket{2}$ can be any state that is connected with $\ket{0_\downarrow}$ by an electric dipole transition. The dipole-dipole interaction between $\ket{0_\downarrow 2} \leftrightarrow \ket{2 0_\downarrow}$ is $V_{dd}$, which depends on the transition dipole moment and the distance between two molecules. For molecules with $\sim 2~\mathrm{Debye}$ molecule frame dipole moment and $\sim \mathrm{\mu m}$ separation, $V_{dd}$ is around $100~\mathrm{kHz}$. The eigenstates of the dipole-dipole interaction are $\frac{1}{\sqrt{2}}(\ket{0_\downarrow 2}\pm\ket{2 0_\downarrow})$ and they are split by $2V_{dd}$.

Two molecules are initialized in $\ket{0_\downarrow 0_\downarrow}$ by optical pumping. An rf pulse coupling $\ket{0_\downarrow} \leftrightarrow \ket{2}$ with a detuning $\Delta=V_{dd}$ is applied. This pulse resonantly couples the pair states $\ket{0_\downarrow 0_\downarrow} \leftrightarrow \frac{1}{\sqrt{2}}(\ket{0_\downarrow 2}+\ket{2 0_\downarrow})$, and off-resonantly couples $\frac{1}{\sqrt{2}}(\ket{0_\downarrow 2}+\ket{2 0_\downarrow}) \leftrightarrow \ket{22}$. If the coupling Rabi frequency is much less than $V_{dd}$, only the entangled state $\frac{1}{\sqrt{2}}(\ket{0_\downarrow 2}+\ket{2 0_\downarrow})$ is populated after a $\pi$ pulse (the pulse area is $\pi/\sqrt{2}$ for a single molecule). Next, another pulse coupling $\ket{2} \leftrightarrow \ket{0_\uparrow}$ with a detuning $\Delta=V_{dd}$ is applied. Note that the first and second pulses can be different in polarization or frequency, so the first pulse does not drive the $\ket{2} \leftrightarrow \ket{0_\uparrow}$ transition. After a $\pi$ pulse (the pulse area is $\pi$ for a single molecule), the population in $\ket{2}$ is mapped to $\ket{0_\uparrow}$ for each molecule and the pair state is $\frac{1}{\sqrt{2}} (\ket{0_\uparrow 0_\downarrow} + \ket{0_\downarrow 0_\uparrow})$. Subsequently, a $\pi$ phase can be added on the $\ket{0_\downarrow 0_\uparrow}$ component by the vector Stark shift of an addressing beam focusing on one of the molecules. The entangled state $\frac{1}{\sqrt{2}} (\ket{0_\uparrow 0_\downarrow} - \ket{0_\downarrow 0_\uparrow})$ is prepared.

Next, the DC magnetic field is switched off, and as described in the main text, a $\pi/2$ pulse $\ket{0}\leftrightarrow \ket{1}$ is applied to both molecules. They are prepared in $\Uparrow\Uparrow$, and an AC Stark shift by an addressing beam focusing on one of the molecules is applied to shift the phase of $\ket{1_{\uparrow,\downarrow}}$ by $\pi$. The state is prepared in $\Uparrow\Downarrow$. Then an rf magnetic field is turned on and spin precession starts. After spin precession, the eEDM phase shift is measured by the sequence described in the previous section.

After spin precession, we use the methods described in the previous section to measure the phase shift from the eEDM interaction.

\section{Scaling up to $2N$ molecules}

\begin{figure*}[ht]
	\includegraphics[width=\textwidth]{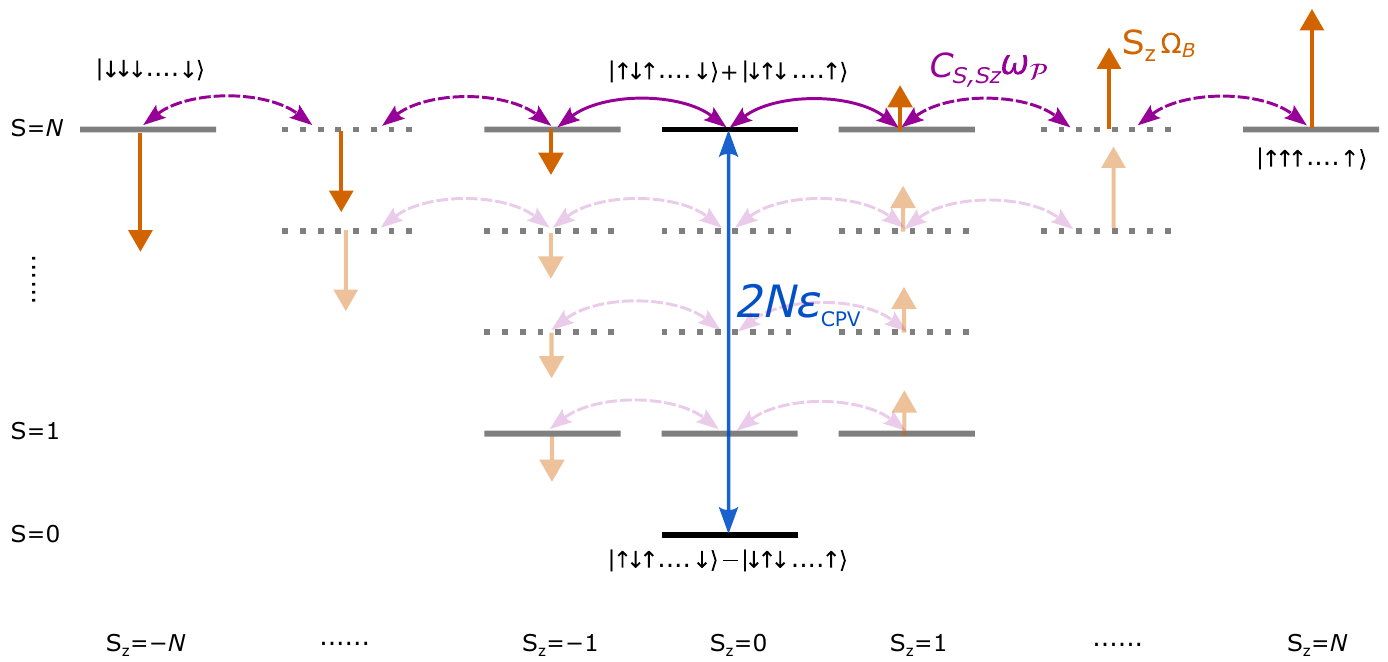}
	\caption{For $2N$ molecules with oppositely aligned molecule orientations, the spin states can be described by the Dicke ladder. The vertical dimension is ordered by total spin and the horizontal dimension is ordered by the spin projection on the quantization axis. In the rotating frame, the eEDM interaction couples the $\ket{S=0,S_z=0} \leftrightarrow \ket{S=N,S_z=0}$ (blue arrow). Similar to the two molecule case, the $S=N$ subspace is coupled by the rotation of the reference frame (purple arrows), the parameter $C_{S,S_z}$ is given by the Clebsch-Gordan coefficients. We apply an rf magnetic field, which shifts the energies of states with nonzero $S_z$. Similar to the two molecule case, the magnetic field and the rotation together gives an unshifted state with most population in $\ket{S=N,S_z=0}$.}
	\label{FigS4}
\end{figure*}

For $2N$ molecules, the eEDM sensitivity increases linearly as the molecule number (Heisenberg scaling). We explain the scaling in Fig.~\ref{FigS4}. Since the eEDM interaction is diagonal in the $\otimes_{2N}\{\widetilde{\ket{\uparrow}},\widetilde{\ket{\downarrow}}\}$ basis (see Eq.~\ref{eqDiagEDM}), it does not flip spins and it shifts the two states $\widetilde{\ket{\uparrow\downarrow\uparrow ...}}$ and $\widetilde{\ket{\downarrow\uparrow\downarrow ...}}$ oppositely. As a result, the eEDM causes spin precession within the two dimensional subspace of $\{\widetilde{\ket{\uparrow\downarrow\uparrow ...}}, \widetilde{\ket{\downarrow\uparrow\downarrow ...}} \}$.  Since these two states form a two-dimensional subspace spanned by $\ket{S=N,S_z=0}$ and $\ket{S=0,S_z=0}$, the eEDM interaction will not couple them to any other states.  The entangled state of $2N$ molecules may be generated by adiabatic sweeping to the many-body ground state, which has been demonstrated in Rydberg atom systems \cite{Omran2019}, or using universal gate operations. These methods requires longer time for entangling larger systems. In addition, the entangled state may be generated by measurement and feedback on a cluster state \cite{Briegel2001,Verresen2022,Lee2022,Tscherbul2023}, which can be generated by parallel operations. This method has a constant circuit depth for arbitrary numbers of entangled molecules. Once the molecules are entangled, the qubit states can be mapped to the spin states.

\newpage
\section{Molecular Design Requirements}
\label{sec_molecule}

As discussed in the main text, the entangled-basis eEDM couplings are maximal when the molecular dipole oscillations -- set by the opposite-parity splitting $\omega_P$ -- are adiabatic relative to the RF magnetic field drive $\Omega_B$. Limiting RF $B$-fields to achievable amplitudes therefore imposes upper bounds on the size of the parity splitting $\omega_P$ relative to the magnetic tuning of the molecule. Conversely, stray-field considerations also impose a lower bound on the minimum parity doubling to avoid excess decoherence and accidental polarization. RF drives on trapped ions must additionally be well-separated from trap frequencies ($\sim 20$ MHz).

In rigid-rotor molecules, the most generic parity splitting scale is set by the end-to-end rotation, which is inversely proportional to the largest rotational moment of inertia. For small molecules, typical end-to-end rotational scales are several GHz.  While far-detuned from typical ion trapping frequencies, these splittings impose demanding requirements on RF $B$-field intensities (e.g. $> 500$ G for a paramagnetic $^2\Sigma$ molecule). Achieving sub-GHz end-to-end rotation is possible, but requires molecules with both heavy metal and ligand partners~\cite{sunaga_merits_2019, fleig_theoretical_2021, smialkowski_highly_2021}. 

Technical requirements on $B$-field amplitudes can be significantly relaxed, however, by using molecules with near-degenerate parity doubling, for which typical $\omega_P$ splittings are $< 100$ MHz. A standard approach is to utilize states with non-zero orbital angular momentum ($\Lambda > 0$), which form near-degenerate $\Omega$-doublets of combined electronic and rotational angular momenta. These states can be found in the electronic configurations of linear diatomic molecules, where the relevant quantum numbers include orbital angular momentum projection onto the molecular axis ($\Lambda$), electron spin angular momentum projection on the molecular axis ($\Sigma$), and the sum of these projection quantum numbers ($\Omega$). To facilitate the RF magnetic drive, it is furthermore desirable to utilize the stretched states with maximal $|\Omega|$, where there are no cancellations to magnetic sensitivity from mixed orientations of orbital and electron spin angular momenta.

Note that this is opposite to the design considerations in some contemporary eEDM experiments, where the non-stretched $^3\Delta_1$ configuration is utilized for measurements due to its suppressed $g$-factor~\cite{Andreev2018,Roussy2023}. In our scheme, the antiferromagnetic ordering of the entangled states already confers insensitivity to global magnetic field noise, which in combination with immunity to slow noise from the rotating frame, significantly reduces the technical need for a magnetically insensitive state. If local magnetic insensitivity (or a diamagnetic molecule) were desired, however, one could alternatively perform the effective RF $B$-field via two-photon E1 drives to a magnetic, excited electronic state or amplitude-modulated AC light shifts. However, this merely shifts the experimental susceptibility to magnetic noise onto laser power and polarization noise; whether this approach is indeed advantageous depends on details of the technical implementation.

\begin{table*}
    \centering
    \begin{ruledtabular}
    \begin{tabular}{cccccc}
        State & Interactions & Effective form & Scaling & Prefactors & $\omega_p$ (approx.) \\
        \hline
        $^2\Pi_{1/2}$ & $H_{L^+}\times H_{\text{so}}$ & $J_+S_+ + J_-S_-$ & $\frac{B(A)}{\Delta E}$ & $2\times(J+\frac{1}{2})$ & $\sim 1-10$ GHz \\
        $^2\Pi_{3/2}$ & $(H_{L^+})^2\times H_{S^+}$ & $J_+^3 + J_-^3$ & $\frac{B^3}{A(\Delta E)}$ & $6\times{\prod_{i=-1/2}^{3/2}(J+i)}$ & $\sim 10-100$ Hz\\
        $^3\Pi_{1}$ & $(H_{L^+})^2$ & $J_+^2 + J_-^2$ & $\frac{B^2}{\Delta E}$ & $2\times J(J+1)$ & $\sim 0.1-1$ MHz\\\
        $^3\Pi_{2}$ & $(H_{L^+})^2\times (H_{S^+})^2$ & $J_+^4 + J_-^4$ & $\frac{B^4}{A^2(\Delta E)}$ & $24\times{\prod_{i=-1}^{2}(J+i)}$ & $\sim 10-100$ mHz\\
        $^2\Delta_{3/2}$ & $(H_{L^+})^3\times H_{\text{so}}$ & $J_+^3S_+ + J_-^3S_-$ & $\frac{B^3A}{\Delta E^3}$ & $6\times{\prod_{i=-1/2}^{3/2}(J+i)}$ & $\sim 1-10$ Hz \\
        $^2\Delta_{5/2}$ & $(H_{L^+})^4\times H_{S^+}$ & $J_+^5 + J_-^5$ & $\frac{B^5}{A(\Delta E)^3}$ & $120\times {\prod_{i=-3/2}^{5/2}(J+i)}$ & $< 10$ mHz\\
        $^3\Delta_{1}$ & $(H_{L^+})^2\times(H_{\text{so}})^2$  & $J_+^2S_+^2 + J_-^2S_-^2$ & $\frac{B^2 A^2}{\Delta E^3}$ & $24\times J(J+1)$ & $\sim 0.1 -1 $ MHz \\
        $^3\Delta_{2}$ & $(H_{L^+})^4$ & $J_+^4 + J_-^4$ & $\frac{B^4}{\Delta E^3}$ & $24\times{\prod_{i=-1}^{2}(J+i)}$ & $\sim 1-10$ mHz \\
        $^3\Delta_{3}$ & $(H_{L^+})^4\times(H_{S^+})^2$ & $J_+^6 + J_-^6$ & $\frac{B^6}{A^2 (\Delta E)^3}$ & $720\times{\prod_{i=-2}^{3}(J+i)}$ & $< 10$ mHz\\
        $^4\Delta_{1/2}$ & $H_{L^+}\times(H_{\text{so}})^3$ & $J_+S_+^3 + J_-S_-^3$ & $\frac{A^3 B}{\Delta E^3}$ & $24\times (J+\frac{1}{2})$ & $\sim 0.1-1$ GHz\\
        $^2\Phi_  {5/2}$ &  $(H_{L^+})^5\times H_{\text{so}}$ & $J_+^5S_+ + J_-^5S_-$ & $\frac{B^5 A}{ \Delta E^5}$ & $720\times{\prod_{i=-3/2}^{5/2}(J+i)}$ & $< 10$ mHz  \\
        $^2\Phi_{7/2}$ &  $(H_{L^+})^6\times H_{S^+}$ & $J_+^7 + J_-^7$ & $\frac{B^6}{ A(\Delta E)^4}$ & $720\times{\prod_{i=-5/2}^{7/2}(J+i)}$ & $< 10$ mHz\\
        $^4\Phi_{3/2}$ &  $(H_{L^+})^3\times (H_{\text{so}})^3$ & $J_+^3S_+^3 + J_-^3S_-^3$ & $\frac{B^3 A^3}{\Delta E^5}$ & $720\times{\prod_{i=-1/2}^{3/2}(J+i)}$ & $\sim 1-100$ Hz
    \end{tabular}
    \end{ruledtabular}
    \caption{Orbital parity-doubling mechanisms and approximate $\omega_p$ scales for heavy, spin-orbit-coupled molecules. Listed are $\Omega$-doubling matrix elements (at the single-configuration level) of selected $C_{\infty v}$ electronic terms ($^{2S+1}\Lambda_\Omega$). The $H_{L^+}$, $H_{S^+}$, and $H_{so}$ interactions refer to $L$-uncoupling ($J\cdot L$), $S$-uncoupling ($J\cdot S$), and microscopic spin-orbit ($\sum_i l_i\cdot s_i$) terms, respectively. In the ``scaling" column, the terms $B$, $A$, and $\Delta E$ refer to the rotational constant, spin-orbit constant, and electronic bandgaps to the perturbing level. Numerical prefactors are given by the product of $n!$ coupling paths for an $n$-th order perturbation and factors of $\sqrt{J(J+1)}$ from evaluating $\hat{J}^{+/-}$ terms in the effective Hamiltonians. The size of $\omega_p$ for lowest-$J$ states are estimated assuming $A\sim 4\times 10^{3}$ cm$^{-1}$, $\Delta E\sim 2\times 10^{4}$ cm$^{-1}$, and $B\sim 0.2$ cm$^{-1}$, which provides rough values for the scale of typical single-configuration, single-perturber contributions to the $\Omega$-doubling. Full computation of the $\Omega$-doubling splittings is highly species-dependent and usually involves complicated sums over multiple perturbing channels and electronic state configurations, which can exhibit cancellations and contributions that are not accounted for in these simplified estimates.}\label{tab:omega}
    
        \vspace{8mm}
    \begin{ruledtabular}
    \begin{tabular}{ccccc}
        Type & Mechanism(s) & Doublet Quanta & $\omega_p$ (typ.)\\
        \hline 
        \begin{tabular}{@{}c@{}}rotation-vibration \end{tabular} & \begin{tabular}{@{}c@{}}centrifugal distortion\\inertial asymmetry\end{tabular} &  \begin{tabular}{@{}c@{}}$l$, $K$\\ $K_a$\end{tabular} & \begin{tabular}{@{}c@{}} $< 10$ MHz\end{tabular}\\
        \hline
        \begin{tabular}{@{}c@{}}\shortstack{anisotropic \\ electron hyperfine}\end{tabular} & \begin{tabular}{@{}c@{}}spin-dipolar ($S\cdot I$)\\spin-rotation ($S\cdot N$)\end{tabular} & \begin{tabular}{@{}c@{}}$l$, $K$, $K_a$\\ $l$, $K$, $K_a$\end{tabular} & \begin{tabular}{@{}c@{}} 1 - 10 MHz \end{tabular}\\
        \hline
        \begin{tabular}{@{}c@{}}\shortstack{anisotropic \\ nuclear hyperfine}\end{tabular} & \begin{tabular}{@{}c@{}}spin-dipolar ($I_i\cdot I_j$)\\spin-rotation ($I\cdot N$)\end{tabular} & \begin{tabular}{@{}c@{}}$l$, $K$, $K_a$\\ $l$, $K$, $K_a$\end{tabular} & \begin{tabular}{@{}c@{}} 1 - 10 kHz \end{tabular}\\
  \end{tabular}
    \end{ruledtabular}
    \caption{Common rovibrational parity-doubling mechanisms and splitting scales in polyatomic molecules.}\label{tab:rovibrational}
\end{table*}

Operating under the assumption that a magnetic state is desired, we note that states with larger values of $|\Omega|$ are coupled at progressively higher orders and therefore exhibit smaller $\Omega$-doubling and $\omega_p$. As discussed earlier, an excessively small or unresolved $\omega_p$ decreases the protection conferred by the rotating frame (because it is slow) and risks accidental polarization from stray electric fields. Table \ref{tab:omega} lists the leading $\Omega$-doubling mechanisms and matrix element scales with respect to electronic ($\Delta E$), spin-orbit ($A$), and rotational ($B$) splittings for a variety of open-shell, non-zero $\Lambda$ and $\Omega$ electronic configurations, which can be utilized for order-of-magnitude estimates of $\omega_p$ for molecules with $\Omega$-doubled electronic configurations. Imposing the additional constraint that eEDM-sensitive states must have non-zero spin projection on the molecular axis ($\Sigma\neq 0$), we find that $^2\Pi_{3/2}$ and $^4\Delta_{1/2}$ states are most likely to meet the requirements for $\sim\mu_B$ magnetic tuning and kHz to MHz-scale parity-splitting $\omega_p$. These electronic configurations can be found in a range of EDM-sensitive molecular ions, several of which are listed in table \ref{tab:candidates}. 

An even more flexible approach to engineering parity doublets is to rely on near-degeneracies that originate from ro-vibrational, rather than orbital electronic degrees of freedom, which are present universally in polyatomic (more than two atom) molecules. This provides the added advantage of decoupling polarization from the choice of metal center -- a feature which is particularly useful for neutral molecules, where the metal center can be designed to be compatible with optical cycling and laser cooling, as well as for integrating exotic rare isotopes (with arbitrary electronic structure) into parity-doubled neutral and ionic molecules.  Common structural motifs~\cite{Kozyryev2017b,Augenbraun2020b,Augenbraun2020c} for engineering rovibrational doublets include linear molecules with bending-induced $\ell$-doubling (e.g. linear MOH) as well as non-linear symmetric (e.g. MCH$_3$) and asymmetric tops (e.g. planar MNH$_2$, bent MSH) with rotationally induced $K$-doubling. Control and trapping of eEDM-sensitive polyatomics is being actively pursued across several experiments and molecular species. In table \ref{tab:rovibrational}, we list several common rovibrational doubling mechanisms and scales in polyatomic molecules. Specific polyatomic typologies are listed in Table \ref{tab:candidates}.

\begin{table}
    \begin{ruledtabular}
    \begin{tabular}{ccccc}
        Class & Species & Science State & $\omega_p$\\
        \hline 
        \multirow{2}{*}{rigid-rotor} &alkaline-earth monofluorides (e.g. YbF \cite{Hudson2002, Lim2017}, BaF \cite{Aggarwal2018, steimle_molecular-beam_2011}, RaF \cite{Isaev2010, garcia_ruiz_spectroscopy_2020, udrescu_precision_2023}) & \multirow{2}{*}{$^2\Sigma^+$} & $\sim 5$ GHz \\
        & assembled alkaline-earth coinage (e.g. RaAg, RaAu \cite{sunaga_merits_2019, fleig_theoretical_2021, smialkowski_highly_2021}) & & $\sim 500$ MHz\\
        \hline
        \multirow{2}{*}{$\Omega$-doubled}& $\Lambda = 1$ diatomics (e.g. PbF \cite{shafer-ray_possibility_2006, mawhorter_characterization_2011,zhu_fine_2022}, BiF$^+$) & $^2\Pi_{3/2}$ & $\sim 10$ - 100 Hz\\
        & $\Lambda = 2$ diatomics (e.g. IrF$^+$, PtO$^+$) & $^4\Delta_{1/2}$ & $\sim 100$ MHz \\
        \hline 
        \multirow{4}{*}{polyatomics} & $C_{\infty v}$ linear (e.g. MOH \cite{Jadbabaie2023}) & $^2\Sigma^+$($v_\text{bend}, \ell>0$) & $\sim 20$ MHz\\
        & $C_{(n\geq 3) v}$ symmetric (e.g. MOCH$_3$ \cite{Augenbraun2020c}) & $^2A_1$ ($K>0$) & $\sim 100$ kHz \\
        & $C_{2v}$ planar asymmetric (e.g. MNH$_2$ \cite{Augenbraun2020b}) & $^2A_1$ ($K_a>0$) & $\sim 1$ MHz\\
        & $C_{s}$, $C_1$ bent asymmetric/chiral (e.g. MSH \cite{Augenbraun2020b}) & $^2A'$ ($K_a>0$)  & $\sim 5$ MHz
  \end{tabular}
    \end{ruledtabular}
    \caption{Examples of paramagnetic EDM-sensitive molecules, science state configurations, and approximate parity splitting scales $\omega_p$, based on the scaling relations described in the text and figs. \ref{tab:omega} and \ref{tab:rovibrational}.  Molecules listed without references have not, to our knowledge, been previously considered in the literature. Electronic configurations are inferred from periodic trends and comparison to iso-electronic systems.}\label{tab:candidates}
\end{table}

\FloatBarrier

\bibliography{references}